\pdfoutput=1
\documentclass[prd,letterpaper,aps,notitlepage,superscriptaddress,nofootinbib,10pt]{revtex4-1}

\usepackage{bm}
\usepackage{amsmath,amssymb,graphicx,placeins,slashed}

\usepackage{hyperref}

\newcommand{\bss}[1]{\ensuremath{{\boldsymbol{#1}}}}
\def\sfrac#1#2{{\textstyle\frac{#1}{#2}}}

\newcommand{\nb}{\phantom{0}}
\newcommand{\wm}{\phantom{-}}

\begin{document}

\title{Charmed bottom baryon spectroscopy from lattice QCD}

\author{Zachary S.~Brown}
\affiliation{Department of Physics, College of William and Mary, Williamsburg, VA 23187, USA}
\affiliation{Thomas Jefferson National Accelerator Facility, Newport News, VA 23606, USA}
\author{William Detmold}
\affiliation{Center for Theoretical Physics, Massachusetts Institute of Technology, Cambridge, MA 02139, USA}
\author{Stefan Meinel}
\affiliation{Center for Theoretical Physics, Massachusetts Institute of Technology, Cambridge, MA 02139, USA}
\affiliation{Department of Physics, University of Arizona, Tucson, AZ 85721, USA}
\affiliation{RIKEN BNL Research Center, Brookhaven National Laboratory, Upton, NY 11973, USA}
\author{Kostas Orginos}
\affiliation{Department of Physics, College of William and Mary, Williamsburg, VA 23187, USA}
\affiliation{Thomas Jefferson National Accelerator Facility, Newport News, VA 23606, USA}

\date{September 1, 2014}

\begin{abstract}
We calculate the masses of baryons containing one, two, or three heavy quarks using lattice QCD. We consider
all possible combinations of charm and bottom quarks, and compute a total of 36 different states with $J^P = \frac12^+$ and $J^P = \frac32^+$.
We use domain-wall fermions for the up, down, and strange quarks, a relativistic heavy-quark action
for the charm quarks, and nonrelativistic QCD for the bottom quarks. Our analysis includes results from
two different lattice spacings and seven different pion masses. We perform extrapolations of the baryon masses to the continuum limit and to the physical
pion mass using $SU(4|2)$ heavy-hadron chiral perturbation theory including $1/m_Q$ and finite-volume effects.
For the 14 singly heavy baryons that have already been observed, our results agree with the experimental values within the uncertainties.
We compare our predictions for the hitherto unobserved states with other lattice calculations and
quark-model studies.
\end{abstract}

\maketitle

\FloatBarrier
\section{Introduction}
\FloatBarrier

Baryons containing heavy quarks are interesting both from the theoretical and experimental points of view.
Because the bottom and charm quark masses are greater than the intrinsic energy scale of QCD,
approximate heavy-quark flavor and spin symmetries constrain the spectrum and dynamics of heavy baryons \cite{Manohar:2000dt, Korner:1994nh}.
Singly charmed and singly bottom baryons exhibit a similar spectrum of excitations of the light degrees of freedom. Interactions with the
spin of the heavy quark, and hence the hyperfine splittings, are suppressed by $1/m_Q$.
A particularly interesting symmetry emerges for doubly heavy baryons: in the large-mass limit, the two heavy quarks
are expected to form a point-like diquark that acts like a single heavy antiquark, and the light degrees of freedom behave as in a
heavy-light meson \cite{Savage:1990di}. The ratio of hyperfine splittings of doubly heavy baryons (with two equal
heavy-quark flavors) and singly heavy mesons is predicted to approach the value $3/4$ in the heavy-quark limit \cite{Brambilla:2005yk}.
Finally, triply heavy baryons can be viewed as baryonic analogues of heavy quarkonia, making them very interesting systems to
study in effective field theories and perturbative QCD \cite{Brambilla:2005yk, Brambilla:2009cd, LlanesEstrada:2011kc, Brambilla:2013vx}.

The masses of all low-lying\footnote{Here and in the following, ``low-lying'' refers to the states that have zero orbital angular momentum and are not radially excited
in the quark model.}
singly charmed baryons with $J^P = \frac12^+$ and $J^P = \frac32^+$, and of most of their singly bottom partners,
are well known from experiments \cite{Beringer:1900zz}. In this sector, the most recent discoveries are the $\Omega_b$ \cite{Abazov:2008qm, Aaltonen:2009ny}, and a state that is likely
the $J^P = \frac32^+$ $\Xi_b^*$ \cite{Chatrchyan:2012ni}; the $\Omega_b^*$ an $\Xi_b^\prime$ remain to be found.
The $\Omega_b$ masses reported by D$\slashed{0}$ \cite{Abazov:2008qm} and CDF \cite{Aaltonen:2009ny}
are inconsistent with each other, but a recent more precise measurement by the LHCb collaboration \cite{Aaij:2013qja} agrees with the CDF result.
In contrast to the singly heavy baryons, the arena of doubly and triply heavy baryons remains experimentally
unexplored to a large extent, with the only possibly observed state being the $\Xi_{cc}^+$. The discovery of the $\Xi_{cc}^+$ was
reported by the SELEX collaboration \cite{Mattson:2002vu, Ocherashvili:2004hi}, but subsequent searches for this state by the FOCUS \cite{Ratti:2003ez}, BaBar \cite{Aubert:2006qw},
Belle \cite{Chistov:2006zj}, and LHCb collaborations \cite{Aaij:2013voa} returned negative results. Nevertheless, there is still potential
for discoveries of doubly and triply heavy baryons at the LHC \cite{Doncheski:1995ye, Chen:2011mb, Chen:2014hqa} and perhaps
also at the coming generation of spectroscopy experiments at BESIII \cite{Asner:2008nq}, Belle II \cite{Aushev:2010bq}, and PANDA \cite{Lutz:2009ff}.

Lattice QCD can predict the masses and other properties of heavy baryons from first principles, and can help resolve
experimental controversies such as those surrounding the $\Omega_b$ and $\Xi_{cc}$. For the doubly and triply heavy baryons, which
may remain beyond the reach of experiments at the present time, lattice QCD results can also serve as a benchmark for other
theoretical approaches, such as quark models and perturbative QCD. Complete control over all sources of systematic uncertainties,
including the nonzero lattice spacing and finite lattice volume, unphysical values used for the quark masses, any approximations made for the heavy quarks,
as well as excited-state contamination in the correlation functions, is essential in both of these contexts.
Most lattice calculations of heavy baryon masses that have been published to date are still lacking in some of these aspects.
The earliest studies \cite{Alexandrou:1994dm, Bowler:1996ws, Michael:1998sg, AliKhan:1999yb, Woloshyn:2000fe, Lewis:2001iz, Mathur:2002ce,
Flynn:2003vz, Chiu:2005zc} were performed in the quenched approximation, removing the effects of sea quarks to reduce the computational cost
but at the expense of connection to experiment. The first unquenched calculations were reported in Refs.~\cite{Na:2006qz, Na:2007pv, Na:2008hz}.
Since then, additional unquenched calculations have been performed with various choices of lattice actions for the light and heavy quarks \cite{Burch:2008qx,
Detmold:2008ww, Lewis:2008fu, Lin:2009rx, Liu:2009jc, Meinel:2010pw, Meinel:2012qz, Alexandrou:2012xk, Briceno:2012wt,
Namekawa:2013vu, Perez-Rubio:2013oha, Padmanath:2013zfa, Basak:2013oya, Alexandrou:2014sha}; reviews can be
found in Refs.~\cite{Orginos:2010lza, Lewis:2010xj, Lin:2011ti}.

In this paper, we present the first lattice QCD determination of singly, doubly, and triply heavy baryon masses that includes
both charm and bottom quarks in any combination, and also achieves good control over all major sources of systematic uncertainties. Our calculation
includes dynamical up, down, and strange quarks implemented with a domain-wall action \cite{Kaplan:1992bt, Furman:1994ky, Shamir:1993zy},
and is performed at two different lattice spacings and seven different values of the up/down quark mass corresponding to pion masses
as low as 227(3) MeV. Because the masses of the charm and bottom quarks are not small in units of the lattice spacing, special heavy-quark actions are needed
for them to avoid large discretization errors. We use a relativistic heavy-quark action
\cite{ElKhadra:1996mp, Chen:2000, Aoki:2001ra, Aoki:2003dg, Christ:2006us, Lin:2006ur, Aoki:2012xaa}
for the charm quarks and improved nonrelativistic QCD \cite{Thacker:1990bm, Lepage:1992tx} for the bottom quarks.
Details of the actions and parameters are given in Sec.~\ref{sec:actions}. The interpolating fields we use for the heavy baryons and
our methodology for fitting the two-point functions are described in Sec.~\ref{sec:2pt}. We extrapolate the results for all baryon masses
to the physical pion mass and the continuum limit as explained in Sec.~\ref{sec:extrap}. For the singly and
doubly heavy baryon masses, heavy-hadron chiral perturbation theory at next-to-leading order is used to fit the light-quark mass
dependence and to remove the leading finite-volume effects. Because some of our data sets use valence light-quark
masses lower than the sea-quark masses, we use the partially quenched extension of heavy-hadron chiral perturbation
theory \cite{Tiburzi:2004kd, Mehen:2006}. For the singly heavy baryons, we generalize the expressions given in Ref.~\cite{Tiburzi:2004kd}
to include hyperfine splittings. The final results for the baryon masses and mass splittings are presented in Sec.~\ref{sec:finalresults},
which also includes a detailed discussion of the systematic uncertainties. We conclude in Sec.~\ref{sec:conclusions}
with a comparison of our results to the literature.

\FloatBarrier
\section{\label{sec:actions}Lattice actions}
\FloatBarrier

\FloatBarrier
\subsection{\label{sec:lightquarks}Light quark and gluon actions}
\FloatBarrier

In this work, we performed the Euclidean path integral using ensembles of gauge field configurations generated by the
RBC and UKQCD collaborations \cite{Aoki:2010dy}. These ensembles include the effects of dynamical up-, down- and strange quarks, implemented
with a domain-wall action \cite{Kaplan:1992bt, Furman:1994ky, Shamir:1993zy}.
The quark fields in this action depend on an auxiliary fifth dimension with extent $L_5$. Four-dimensional quark fields,
for which the low-energy effective field theory obeys an exact lattice chiral symmetry in the limit $L_5\rightarrow \infty$,
are obtained in terms of the quark fields at the boundaries $x_5=a$ and $x_5=L_5$ \cite{Kaplan:1992bt, Furman:1994ky, Shamir:1993zy}.
The RBC and UKQCD collaborations chose the Iwasaki gauge action \cite{Iwasaki:1983ck, Iwasaki:1984cj},
which, compared to the standard Wilson or Symanzik gauge actions, reduces the residual chiral symmetry breaking of the
domain-wall action at finite $L_5$ \cite{Aoki:2002vt}. For hadron spectroscopy calculations, the primary benefit of approximate
chiral symmetry is the smallness of $\mathcal{O}(a)$ discretization
errors.

Here we selected four different ensembles of gauge fields: two ensemble with lattice size $24^3\times64$ and lattice spacing $a\approx 0.11$ fm
(in the following referred to as ``coarse''), and two ensembles with lattice size $32^3\times 64$ and lattice spacing $a\approx 0.085$ fm
(in the following referred to as ``fine'') \cite{Aoki:2010dy}. All ensembles have $L_5/a=16$, and the domain-wall height is $a M_5=1.8$. The residual chiral symmetry
breaking can be quantified by the residual additive quark-mass renormalization, $am_{\rm res}$. The coarse ensembles have $a m_{\rm res}\approx 0.003$,
while the fine ensembles have $a m_{\rm res}\approx 0.0007$ \cite{Aoki:2010dy}.

We work in the isospin limit $m_u=m_d$; this means that our results for the baryon masses should be considered as isospin-averaged values.
We computed domain-wall light and strange quark propagators with various quark
masses $am_{u,d}^{(\mathrm{val})}$ and $am_{s}^{(\mathrm{val})}$ as shown in Table \ref{tab:params}, leading to eight different
data sets in total. For the four data sets \texttt{C104}, \texttt{C54}, \texttt{F43}, and \texttt{F63}, the valence quark masses are equal
to the sea quark masses. The data sets \texttt{C14}, \texttt{C24}, and \texttt{F23} have $am_{u,d}^{(\mathrm{val})} < am_{u,d}^{(\mathrm{sea})}$
in order to achieve lighter valence pion masses, and the data set \texttt{C53} has $am_{s}^{(\mathrm{val})} < am_{s}^{(\mathrm{sea})}$
to enable interpolations to the physical strange quark mass. We used about 200 gauge configurations from each ensemble,
and computed domain-wall propagators for multiple source locations on each configuration. The resulting total
numbers of light/strange propagator pairs in each data set are given in the last column of Table \ref{tab:params}. To further increase
the statistical precision of our calculations, we computed hadron two-point functions propagating both forward and backward in Euclidean time and averaged over these.

\begin{table}
\begin{tabular}{cccccccccccccccccccccc}
\hline\hline
Set & \hspace{1ex} & $N_s^3\times N_t$ & \hspace{1ex} & $\beta$ & \hspace{1ex} & $am_{u,d}^{(\mathrm{sea})}$
& \hspace{1ex} & $am_{s}^{(\mathrm{sea})}$   & \hspace{1ex} & $a$ (fm) & \hspace{1ex} & $am_{u,d}^{(\mathrm{val})}$
& \hspace{1ex} & $am_{s}^{(\mathrm{val})}$  & \hspace{1ex} & $m_\pi^{(\mathrm{vv})}$ (MeV)
& \hspace{1ex} & $m_{\eta_s}^{(\mathrm{vv})}$ (MeV)  & \hspace{1ex} & $N_{\rm meas}$ \\
\hline
\texttt{C104} && $24^3\times64$ && 2.13 && $0.01\nb$  && $0.04$ && $0.1139(18)$ && $0.01\nb$  && $0.04$ && 419(7) && 752(12)  && 2554 \\
\texttt{C14}  && $24^3\times64$ && 2.13 && $0.005$    && $0.04$ && $0.1119(17)$ && $0.001$    && $0.04$ && 245(4) && 761(12)  && 2705 \\
\texttt{C24}  && $24^3\times64$ && 2.13 && $0.005$    && $0.04$ && $0.1119(17)$ && $0.002$    && $0.04$ && 270(4) && 761(12)  && 2683 \\
\texttt{C54}  && $24^3\times64$ && 2.13 && $0.005$    && $0.04$ && $0.1119(17)$ && $0.005$    && $0.04$ && 336(5) && 761(12)  && 2780 \\
\texttt{C53}  && $24^3\times64$ && 2.13 && $0.005$    && $0.04$ && $0.1119(17)$ && $0.005$    && $0.03$ && 336(5) && 665(10)  && 1192 \\
\texttt{F23}  && $32^3\times64$ && 2.25 && $0.004$    && $0.03$ && $0.0849(12)$ && $0.002$    && $0.03$ && 227(3) && 747(10)  && 1918 \\
\texttt{F43}  && $32^3\times64$ && 2.25 && $0.004$    && $0.03$ && $0.0849(12)$ && $0.004$    && $0.03$ && 295(4) && 747(10)  && 1919 \\
\texttt{F63}  && $32^3\times64$ && 2.25 && $0.006$    && $0.03$ && $0.0848(17)$ && $0.006$    && $0.03$ && 352(7) && 749(14)  && 2785 \\
\hline\hline
\end{tabular}
\caption{\label{tab:params}Properties of the gauge field ensembles \protect\cite{Aoki:2010dy} and of the light/strange quark propagators
we computed on them. Here, $N_s$ and $N_t$ are the numbers of lattice 
points in the spatial and temporal directions, $\beta=6/g^2$ is the gauge coupling, $am_{u,d}^{(\mathrm{sea})}$ and $am_{s}^{(\mathrm{sea})}$ are the
light and strange sea quark masses, and $a$ is the lattice spacing (determined in Ref.~\protect\cite{Meinel:2010pv}). The valence 
quark masses used for the calculation of the light and strange quark propagators are denoted by $am_{u,d}^{(\mathrm{val})}$ and $am_{s}^{(\mathrm{val})}$. The corresponding
valence pion and $\eta_s$ masses are denoted as $m_\pi^{(\mathrm{vv})}$ and $m_{\eta_s}^{(\mathrm{vv})}$. The $\eta_s$ is an artificial
$s\bar{s}$ state that is defined by treating the $s$ and $\bar{s}$ as different, but mass-degenerate flavors. This state is useful
as an intermediate quantity to tune the strange-quark mass \cite{Davies:2009tsa}; its mass at the physical point has
been computed precisely by the HPQCD collaboration and is $m_{\eta_s}^{(\mathrm{phys})}=689.3(1.2)\:\:{\rm MeV}$ \cite{Dowdall:2011wh}. In the last column
of the table, $N_{\rm meas}$ is the number of pairs of light and strange quark propagators computed in each data set.}
\end{table}

\FloatBarrier
\subsection{Bottom quark action}
\FloatBarrier

The typical momentum of a bottom quark inside a hadron at rest is much smaller than the bottom-quark mass. For hadrons
containing only a single bottom quark and no charm quarks, one expects $\langle |\mathbf{p}_b| \rangle \sim \Lambda \sim 500$ MeV \cite{Shifman:1987rj, Manohar:2000dt}. For bottomonium
and triply bottom baryons, one expects $\langle |\mathbf{p}_b| \rangle \sim m_b v \sim 1.5$ GeV, corresponding to $v^2\sim 0.1$ \cite{Thacker:1990bm}. For hadrons containing both
bottom and charm quarks, the typical momentum of the $b$ quark is between these extremes. In all
cases, the separation of scales, $\langle |\mathbf{p}_b| \rangle \ll m_b$, allows the treatment of the $b$ quarks with nonrelativistic effective field theory.
Here we used improved lattice NRQCD, which was introduced in Refs.~\cite{Thacker:1990bm, Lepage:1992tx}. The $b$-quark is described by a two-component spinor field $\psi$,
with Euclidean lattice action
\begin{equation}
  S_{\psi}=a^3\sum_{\mathbf{x},t}\psi^\dagger(\mathbf{x},t)\big[{\psi}(\mathbf{x},t)
  -K(t) \: {\psi}(\mathbf{x},t-a) \big], \label{eq:latact}
\end{equation}
where
\begin{eqnarray}
  K(t)&=&\left(1-\frac{a\:\delta H|_t}{2}\right)
  \left(1-\frac{a H_0|_t}{2n} \right)^n U_0^\dag(t-a)
  \hspace*{1cm}
  \nonumber\\
  &&\hspace*{0.5cm}
 \times\left(1-\frac{a H_0|_{t-a}}{2n} \right)^n
  \left(1-\frac{a\:\delta H|_{t-a}}{2}\right)\,.
  \label{eq:mNRQCD_action_kernel}
\end{eqnarray}
In Eq.~(\ref{eq:mNRQCD_action_kernel}), $U_0(t-a)$ denotes a temporal gauge link, and $H_0$ and $\delta H$ are given by
\begin{eqnarray}
  H_0 &=& -\frac{\Delta^{(2)}}{2 m_b}, \label{eq:H0} \\
 \nonumber
  \delta H&=&-c_1\:\frac{\left(\Delta^{(2)}\right)^2}{8 m_b^3}+c_2\:\frac{ig}{8 m_b^2}\:\Big(\bss{\nabla}\cdot\mathbf{\widetilde{E}}
  -\mathbf{\widetilde{E}}\cdot\bss{\nabla}\Big)\\
  \nonumber&&-c_3\:\frac{g}{8 m_b^2}\:\bss{\sigma}\cdot
  \left(\bss{\widetilde{\nabla}}\times\mathbf{\widetilde{E}}
    -\mathbf{\widetilde{E}}\times\bss{\widetilde{\nabla}} \right)-c_4\:\frac{g}{2 m_b}\:\bss{\sigma}\cdot\mathbf{\widetilde{B}}\\
  && + c_5\:\frac{a^2\Delta^{(4)}}{24m_b}
  -c_6\:\frac{a\left(\Delta^{(2)}\right)^2}{16n\:m_b^2}.
  \label{eq:dH_full}
\end{eqnarray}
This action was originally introduced for heavy quarkonium, for which $H_0$ is the leading-order
term (order $v^2$), and the terms with coefficients $c_1$ through $c_4$ in $\delta H$ are of order $v^4$ \cite{Thacker:1990bm, Lepage:1992tx}.
The parameter $n\geq 1$ was introduced to avoid numerical instabilities occurring at small $a m_b$ \cite{Lepage:1992tx}; here we set $n=2$.
The operators with coefficients $c_5$ and $c_6$ correct discretization errors associated with $H_0$ and with
the time derivative. We performed tadpole-improvement of the action using the Landau-gauge mean link, $u_{0L}$ \cite{Lepage:1992xa},
and set the matching coefficients $c_1$ through $c_3$ to their tree-level values ($c_i=1$). The matching coefficient $c_4$
was computed to one-loop in perturbation theory \cite{Hammant:2011bt}. We tuned the bare $b$-quark mass by requiring that the spin-averaged
bottomonium kinetic mass agrees with experiment (see Ref.~\cite{Meinel:2010pv} for details). The resulting values of $a m_b$, as well as the values of $u_{0L}$
and $c_4$ are given in Table \ref{tab:NRQCDparams}. The values of $c_4$ are specific for our gauge action (the Iwasaki action),
and were computed for us by Tom Hammant.

\begin{table}
  \begin{tabular}{lcccccc}
    \hline\hline
    Data sets & \hspace{2ex} &  $a m_b$  & \hspace{2ex} & $u_{0L}$ & \hspace{2ex} & $c_4$ \\
    \hline
    \texttt{C104}, \texttt{C14}, \texttt{C24}, \texttt{C54}, \texttt{C53} &&  2.52  &&  0.8439  &&  1.09389 \\
    \texttt{F23}, \texttt{F43}, \texttt{F63}               &&  1.85  &&  0.8609  &&  1.07887 \\
    \hline\hline
  \end{tabular}
  \caption{\label{tab:NRQCDparams}Parameters used in the NRQCD action for the bottom quarks.}
\end{table}

When applied to hadrons containing only a single $b$-quark and no charm quarks, the power counting for the NRQCD action
is different. In this case, the expansion parameter is $\Lambda/m_b$, and the action shown above is complete
through order $(\Lambda/m_b)^2$. For singly bottom hadrons, the operator $-c_4\:\frac{g}{2 m_b}\:\bss{\sigma}\cdot\mathbf{\widetilde{B}}$
in $\delta H$ is of the same order in the power counting as the operator $H_0$, while all other operators are of higher order. This means that
the one-loop matching used for $c_4$ is especially important for heavy-light hadrons.

\FloatBarrier
\subsection{\label{sec:RHQ}Charm quark action}
\FloatBarrier

Because the nonrelativistic expansion converges poorly for charm quarks (and because lattice NRQCD requires $a m>1$, which is
not satisfied for the charm quark on the present lattices), we used instead a relativistic heavy quark action
\cite{ElKhadra:1996mp, Chen:2000, Aoki:2001ra, Aoki:2003dg, Christ:2006us, Lin:2006ur, Aoki:2012xaa}.
Beginning with a clover fermion action, separate coefficients are introduced for the spatial and temporal components of the operators, so
that the action becomes
\begin{equation}
S_Q = a^4 \sum_x \bar{Q} \left[ m_Q + \gamma_0 \nabla_0 - \frac{a}{2} \nabla^{(2)}_0 + \nu\sum_{i=1}^3\left(\gamma_i \nabla_i - \frac{a}{2} \nabla^{(2)}_i\right)
- c_E  \frac{a}{2} \sum_{i=1}^3 \sigma_{0i}F_{0i} - c_B \frac{a}{4} \sum_{i,\, j=1}^3 \sigma_{ij}F_{ij}  \right] Q \, . \label{eq:RHQ}
\end{equation}
The (bare) parameters are the mass $m_Q$, the anisotropy $\nu$, and the chromoelectric and chromomagnetic 
coefficients $c_E$, $c_B$. Discretization errors proportional to powers of the heavy-quark mass can then be removed to all orders by allowing the
coefficients $\nu$, $c_E$, and $c_B$ to depend on $am_Q$ and tuning them. The remaining discretization errors are of order $a^2|\mathbf{p}|^2$, where $|\mathbf{p}|$ is
the typical magnitude of the spatial momentum of the heavy quark inside the hadron. The standard clover action with $\nu=1$ and $c_E=c_B=c_{\rm SW}$
is recovered in the continuum limit.

Several different approaches have been suggested for determining the parameters $m_Q$, $\nu$, $c_B$, and $c_E$
\cite{ElKhadra:1996mp, Chen:2000, Aoki:2001ra, Aoki:2003dg, Christ:2006us, Lin:2006ur, Aoki:2012xaa}. Here we followed Ref.~\cite{Liu:2009jc} and
tuned the two parameters $m_Q$ and $\nu$ nonperturbatively while setting the coefficients $c_E$, $c_B$
equal to the values predicted by tadpole-improved tree-level perturbation theory \cite{Chen:2000},
\begin{equation}
c_E = \frac{\left(1 + \nu \right)}{2u_0^3}, \hspace{4ex} c_B = \frac{\nu}{u_0^3}\,. \label{eq:cEcB}
\end{equation}
We set the tadpole improvement parameter $u_0$ equal to the fourth root of the average plaquette.
In order to tune the parameters $m_Q$ and $\nu$, we nonperturbatively computed the energies of
the charmonium states $\eta_c$ and $J/\psi$ at zero and nonzero momentum, and extracted the ``speed of light''
in the $J/\psi$ dispersion relation,
\begin{equation}
c^2(\mathbf{p}) = \frac{E^2_{J/\psi}(\mathbf{p}) - E^2_{J/\psi}(0)}{\mathbf{p}^2}, \label{eq:speedoflight}
\end{equation}
as well as the spin-averaged mass
\begin{equation}
\overline{M} = \frac{3}{4}E_{J/\psi}(0) + \frac{1}{4}E_{\eta_c}(0). \label{eq:spinavccbar}
\end{equation}
The parameters $m_Q$ and $\nu$ need to be adjusted such that $\overline{M}$ agrees with the experimental value
and the relativistic continuum dispersion relation is restored, i.e., $c=1$.

We obtained the energies $E_{J/\psi}$, $E_{\eta_c}$ from single-exponential fits at large Euclidean time to the two-point functions
\begin{equation}
C(\mathbf{p},t) = \sum_{\mathbf{x}} e^{-i\mathbf{p} \cdot (\mathbf{x}-\mathbf{x}_{\rm src})}  \left\langle O(\mathbf{x},t_{\rm src}+t)
\:\:\:  \overline{O}(\mathbf{x}_{\rm src},t_{\rm src}) \right\rangle, \label{eq:ccbar2pt}
\end{equation}
where $O = \bar{c}\gamma_5 c$ for the $\eta_c$ and $O=\bar{c}\gamma_i c$ for the $J/\psi$. For the extraction of the speed of light using
Eq.~(\ref{eq:speedoflight}), we used the smallest nonzero momentum allowed by the periodic boundary conditions, $|\mathbf{p}|=2\pi/L$ with $L=N_s a$.
We generated data points $\left\{ c, \overline{M} \right\}$ for a few good initial guesses of $\left\{ m_{Q}, \nu \right\}$ and performed linear fits using
the functions
\begin{eqnarray}
f^{\overline{M}}\left(\nu, m_Q\right) &=& \delta^{\overline{M}} + C^{\overline{M}}_{\nu}\: \nu + C^{\overline{M}}_{m_Q}\: m_Q, \label{eq:tuninglinearfit1} \\
f^c\left(\nu, m_Q\right) &=& \delta^c + C^c_{\nu}\: \nu + C^c_{m_Q}\: m_Q,  \label{eq:tuninglinearfit2}
\end{eqnarray}
with parameters $\delta^{\overline{M}}$, $C^{\overline{M}}_{\nu}$, $C^{\overline{M}}_{m_Q}$, $\delta^c$, $C^c_{\nu}$, and $C^c_{m_Q}$.
We then solved the equations
\begin{eqnarray}
f^{\overline{M}}\left(\nu, m_Q\right) &=& \overline{M}_{\rm phys}, \\
f^c\left(\nu, m_Q\right) &=& 1,
\end{eqnarray}
for $m_Q$ and $\nu$, and recomputed the actual values of $c$, $\overline{M}$ using Eqs.~(\ref{eq:speedoflight}), (\ref{eq:spinavccbar}) with $m_Q$ and $\nu$
set equal to the solution [the values of $c_E$ and $c_B$ were updated for each new choice of $\nu$ according to Eq.~(\ref{eq:cEcB})].
If the result was consistent with $c=1$ and $\overline{M}=\overline{M}_{\rm phys}$, the procedure was stopped; otherwise,
the new data point was added to the linear fit (\ref{eq:tuninglinearfit1}), (\ref{eq:tuninglinearfit2}) and the procedure iterated.

The final tuned values of the parameters for the coarse and fine lattices are given in Table \ref{tab:c_tuning}, and the resulting values
of the spin-averaged charmonium mass and speed of light for the data sets \texttt{C54} and \texttt{F43} are given in Table \ref{tab:c_masses_speedoflight}.
There, we also show the lattice results for the hyperfine splittings
\begin{equation}
 M_{J/\psi}-M_{\eta_c}\, ,
\end{equation}
which are the first predictions from our charm quark action, and serve as a stringent test of the relativistic heavy-quark formalism
adopted here (hyperfine splittings are highly sensitive to discretization errors). Note that we did not include the disconnected quark
contractions when evaluating the two-point functions (\ref{eq:ccbar2pt}); we neglect the possible annihilation of the $\eta_c$ and $\Upsilon$ to
light hadrons. This affects mainly the $\eta_c$, which can annihilate through two gluons. At leading order in perturbation theory,
the resulting mass shift of the $\eta_c$ can be expressed in terms of its hadronic width \cite{Bodwin:1994jh, Follana:2006rc},
\begin{equation}
 \Delta M_{\eta_c} = \Gamma(\eta_c \to {\rm hadrons}) \left(\frac{\ln(2)-1}{\pi} + \mathcal{O}(\alpha_s) \right). \label{eq:annihilation}
\end{equation}
Using $\Gamma(\eta_c \to {\rm hadrons})=32.0(0.9)$ MeV \cite{Beringer:1900zz}, this gives $\Delta M_{\eta_c}\approx -3$ MeV,
corresponding to a 3 MeV increase in the hyperfine splitting. After adding this correction to our lattice data, we obtain
agreement with the experimental result for the hyperfine splitting for both the coarse and the fine lattice spacings.

\begin{table}
\centering
\begin{tabular}{lcccccccc}
\hline\hline
Data sets & & $a m_Q$ & & $\nu$ & &  $c_E$ & & $c_B$ \\
\hline
\texttt{C104}, \texttt{C14}, \texttt{C24}, \texttt{C54}, \texttt{C53}  && $\wm0.1214$ && $1.2362$ && $1.6650$ && $1.8409$ \\
\texttt{F23},  \texttt{F43}, \texttt{F63}                              && $-0.0045$   && $1.1281$ && $1.5311$ && $1.6232$ \\
\hline\hline
\end{tabular}
\caption{Parameters used in the relativistic heavy-quark action for the charm quarks (a negative bare mass parameter is not unusual because
of the additive quark-mass renormalization for Wilson-type actions).}
\label{tab:c_tuning}
\end{table}

\begin{table}
\centering
\begin{tabular}{lcccccc}
\hline\hline
Quantity                       & & \texttt{C54} & & \texttt{F43} & & Experiment   \\
\hline
$\overline{M}$ (MeV)           & & 3062(43)     & & 3065(42)     & & 3068.6(0.2)  \\
$M_{J/\psi}-M_{\eta_c}$ (MeV)  & & 108.5(1.5)   & & 109.0(1.5)   & & 113.2(0.7)   \\
$c$                            & & 1.010(15)    & & 1.000(30)    & & 1             \\
\hline\hline
\end{tabular}
\caption{Charmonium spin-averaged mass, hyperfine splitting, and ``speed of light'', computed with the tuned RHQ parameters from Table \protect\ref{tab:c_tuning}.
Charm annihilation effects have not been included in the calculation; perturbation theory predicts that these effects would increase the
hyperfine splitting by about 3 MeV \cite{Bodwin:1994jh, Follana:2006rc}. The experimental values are from Ref.~\cite{Beringer:1900zz}. }
\label{tab:c_masses_speedoflight}
\end{table}

\FloatBarrier
\section{\label{sec:2pt}Two-point functions and fit methods}
\FloatBarrier

\FloatBarrier
\subsection{Heavy baryon operators}
\FloatBarrier

This section describes how we combine the color and spin indices of the quark fields to form interpolating
operators for the baryon states of interest \cite{Bowler:1996ws}. Starting from three quark flavors $q$, $q'$, $q''$, we construct the following
basic types of baryon operators,
\begin{eqnarray}
O_5[q, q^\prime, q^{\prime\prime}]_\alpha &=& \epsilon_{abc}\:(C\gamma_5)_{\beta\gamma}\:\:q^a_\beta\:\:q^{\prime b}_\gamma\:\: (P_+ q^{\prime\prime})^c_\alpha, \\
O_5^\prime[q, q^\prime, q^{\prime\prime}]_\alpha &=& \frac{1}{\sqrt{2}} \epsilon_{abc}\:(C\gamma_5)_{\beta\gamma}
\Big[ q^a_\beta\:\:q^{\prime\prime b}_\gamma\:\: (P_+ q^{\prime})^c_\alpha + q^{\prime a}_\beta\:\:q^{\prime\prime b}_\gamma\:\: (P_+ q)^c_\alpha \Big], \\
O_j[q, q^\prime, q^{\prime\prime}]_\alpha &=& \epsilon_{abc}\:(C\gamma_j)_{\beta\gamma}\:\:q^a_\beta\:\:q^{\prime b}_\gamma\:\: (P_+ q^{\prime\prime})^c_\alpha,
\end{eqnarray}
where $a,b,c$ are color indices, $\alpha,\beta,\gamma$ are spinor indices, $C=\gamma_0\gamma_2$ is the charge conjugation matrix, and $P_+$ is the positive-parity projector
\begin{equation}
 P_+ = \sfrac12(1+\gamma_0).
\end{equation}
The operators $O_5$ and $O_5^\prime$ have positive parity and spin $1/2$. The operator $O_j$ (where $j=1,2,3$) has positive parity but couples to
both spin $1/2$ and spin $3/2$ in general.
Using the projectors \cite{Bowler:1996ws}
\begin{eqnarray}
 P^{(1/2)}_{jk}&=&\sfrac13\gamma_j\gamma_k, \\
 P^{(3/2)}_{jk}&=&\delta_{jk}-\sfrac13\gamma_j\gamma_k,
\end{eqnarray}
we construct operators $O_j^{(1/2)}$ and $O_j^{(3/2)}$ with definite spin:
\begin{eqnarray}
 O_j^{(1/2)}[q, q^\prime, q^{\prime\prime}]_\alpha &=& \left(P^{(1/2)}_{jk} O_k[q, q^\prime, q^{\prime\prime}]\right)_\alpha, \\
 O_j^{(3/2)}[q, q^\prime, q^{\prime\prime}]_\alpha &=& \left(P^{(3/2)}_{jk} O_k[q, q^\prime, q^{\prime\prime}]\right)_\alpha.
\end{eqnarray}
In Table \ref{tab:baryonops}, we list the names of the baryons we consider in this work, together with the interpolating
operators used to extract their energies. In the nonrelativistic Dirac gamma matrix basis, the four-spinor bottom-quark field, $b$, is given
in terms of the two-spinor NRQCD field, $\psi$, as
\begin{equation}
 b = \left( \begin{array}{c} \psi \\ 0 \end{array} \right).
\end{equation}
The charm quark field is denoted by $c$ in this section, and is identical to the field $Q$ appearing in Eq.~(\ref{eq:RHQ}).

The zero-momentum two-point functions are defined as
\begin{eqnarray}
 C_{jk\:\alpha\beta}^{(J)}(t) &=& \sum_{\mathbf{x}} \left\langle O_j^{(J)}[q, q^\prime, q^{\prime\prime}]_\alpha(\mathbf{x},t_{\rm src}+t)
 \:\:\:  \overline{O}_k^{(J)}[q, q^\prime, q^{\prime\prime}]_\beta (\mathbf{x}_{\rm src},t_{\rm src}) \right\rangle, \\
 C_{55\:\alpha\beta}(t) &=& \sum_{\mathbf{x}} \left\langle O_5^{(\prime)}[q, q^\prime, q^{\prime\prime}]_\alpha(\mathbf{x},t_{\rm src}+t)
 \:\:\:  \overline{O_5}^{(\prime)}[q, q^\prime, q^{\prime\prime}]_\beta (\mathbf{x}_{\rm src},t_{\rm src}) \right\rangle,
\end{eqnarray}
where we allow for different smearings of the quark fields at the source and sink (see Sec.~\ref{sec:twopt}).
For large $t$ (but $t$ small compared to the temporal extent of the lattice), the ground-state contribution dominates and these two-point functions approach the form
\begin{eqnarray}
 C_{jk\:\alpha\beta}^{(J)}(t) & \rightarrow & Z^{(J)}_{\rm snk}\: Z^{(J)}_{\rm src}\: e^{-E_J\:t}\:\left[P_+ P^{(J)}_{jk}\right]_{\alpha\beta}, \label{eq:twoptjk} \\
 C_{55\:\alpha\beta}(t) & \rightarrow & Z_{\rm snk}\: Z_{\rm src}\: e^{-E\:t}\:\left[P_+\right]_{\alpha\beta}. \label{eq:twopt5}
\end{eqnarray}
Before the fitting, we performed a weighted average over the non-zero $(j,k,\alpha\beta)$-components.

In most cases, the lowest-energy states with which the operators shown in Table \ref{tab:baryonops}
have a nonzero ``overlap'' are the desired baryons (for example, the mixing between $\Sigma_c$ and $\Lambda_c$ is forbidden
by isospin symmetry, which is exact in our calculation with $m_u=m_d$). The only exception 
occurs for the ``primed'' baryons such as the $\Xi_c'$. The interpolating operators listed for the primed
baryons also have a small amplitude to couple to the lighter non-primed states. For the singly-heavy
baryons, this mixing would vanish in the limit of infinite heavy-quark mass, in which
the angular momentum of the light degrees of freedom, $S_l$, becomes a conserved quantum number (the primed
baryons have $S_l=1$, while the unprimed baryons have $S_l=0$). To investigate the mixing at finite heavy-quark mass, we
also computed cross-correlation functions between the operators designed for the primed and unprimed baryons (such
as $\Xi^\prime_c$ and $\Xi_c$). This is discussed further in Sec.~\ref{sec:mixing}.

Finally, we note that some of the baryons we consider are unstable resonances in the real word (albeit with very narrow widths). For example,
the $\Sigma_c$ can decay through the strong interaction to $\Lambda_c \: \pi$, and the lightest state coupling to the $\Sigma_c$ interpolating
operator in infinite volume and with physical quark masses would actually be a $\Lambda_c\text{-}\pi$ $P$-wave state. However,
in our lattice calculation the $\Lambda_c\text{-}\pi$ state is shifted to higher energy due to the finite lattice size and the unphysically heavy pion masses.

\begin{table}
\begin{tabular}{lllllll}
\hline\hline
 \\[-2.5ex]
 Hadron     & \hspace{2ex} & $J^P$ & \hspace{2ex}  & Operator(s) \\
 \\[-2.5ex]
\hline
 \\[-2.5ex]
 $\Lambda_c$         &&  $\sfrac12^+$  &&  $O_5 [u,d,c]$ \\
 \\[-2.5ex]
 $\Sigma_c$          &&  $\sfrac12^+$  &&  $O^{(1/2)}_j[u,u,c]$ \\
 \\[-2.5ex]
 $\Sigma_c^*$        &&  $\sfrac32^+$  &&  $O^{(3/2)}_j[u,u,c]$ \\
 \\[-2.5ex]
 $\Xi_c$             &&  $\sfrac12^+$  &&  $O_5 [u,s,c]$  \\
 \\[-2.5ex]
 $\Xi_c'$            &&  $\sfrac12^+$  &&  $O^{(1/2)}_j[u,s,c]$, \hspace{2ex} $O_5^\prime [u,s,c]$  \\
 \\[-2.5ex]
 $\Xi_c^*$           &&  $\sfrac32^+$  &&  $O^{(3/2)}_j[u,s,c]$  \\
 \\[-2.5ex]
 $\Omega_c$          &&  $\sfrac12^+$  &&  $O^{(1/2)}_j[s,s,c]$  \\
 \\[-2.5ex]
 $\Omega_c^*$        &&  $\sfrac32^+$  &&  $O^{(3/2)}_j[s,s,c]$  \\
 \\[-2.5ex]
 $\Xi_{cc}$          &&  $\sfrac12^+$  &&  $O^{(1/2)}_j[c,c,u]$  \\
 \\[-2.5ex]
 $\Xi_{cc}^*$        &&  $\sfrac32^+$  &&  $O^{(3/2)}_j[c,c,u]$  \\
 \\[-2.5ex]
 $\Omega_{cc}$       &&  $\sfrac12^+$  &&  $O^{(1/2)}_j[c,c,s]$  \\
 \\[-2.5ex]
 $\Omega_{cc}^*$     &&  $\sfrac32^+$  &&  $O^{(3/2)}_j[c,c,s]$  \\
 \\[-2.5ex]
 $\Omega_{ccc}$      &&  $\sfrac32^+$  &&  $O^{(3/2)}_j[c,c,c]$  \\
 \\[-2.5ex]
 $\Lambda_b$         &&  $\sfrac12^+$  &&  $O_5 [u,d,b]$ \\
 \\[-2.5ex]
 $\Sigma_b$          &&  $\sfrac12^+$  &&  $O^{(1/2)}_j[u,u,b]$ \\
 \\[-2.5ex]
 $\Sigma_b^*$        &&  $\sfrac32^+$  &&  $O^{(3/2)}_j[u,u,b]$ \\
 \\[-2.5ex]
 $\Xi_b$             &&  $\sfrac12^+$  &&  $O_5 [u,s,b]$  \\
 \\[-2.5ex]
 $\Xi_b'$            &&  $\sfrac12^+$  &&  $O^{(1/2)}_j[u,s,b]$, \hspace{2ex}  $O_5^\prime [u,s,b]$  \\
 \\[-2.5ex]
 $\Xi_b^*$           &&  $\sfrac32^+$  &&  $O^{(3/2)}_j[u,s,b]$  \\
 \\[-2.5ex]
 $\Omega_b$          &&  $\sfrac12^+$  &&  $O^{(1/2)}_j[s,s,b]$  \\
 \\[-2.5ex]
 $\Omega_b^*$        &&  $\sfrac32^+$  &&  $O^{(3/2)}_j[s,s,b]$  \\
 \\[-2.5ex]
 $\Xi_{bb}$          &&  $\sfrac12^+$  &&  $O^{(1/2)}_j[b,b,u]$  \\
 \\[-2.5ex]
 $\Xi_{bb}^*$        &&  $\sfrac32^+$  &&  $O^{(3/2)}_j[b,b,u]$  \\
 \\[-2.5ex]
 $\Omega_{bb}$       &&  $\sfrac12^+$  &&  $O^{(1/2)}_j[b,b,s]$  \\
 \\[-2.5ex]
 $\Omega_{bb}^*$     &&  $\sfrac32^+$  &&  $O^{(3/2)}_j[b,b,s]$  \\
 \\[-2.5ex]
 $\Omega_{bbb}$      &&  $\sfrac32^+$  &&  $O^{(3/2)}_j[b,b,b]$  \\
 \\[-2.5ex]
 $\Xi_{cb}$          &&  $\sfrac12^+$  &&  $O_5 [u,c,b]$  \\
 \\[-2.5ex]
 $\Xi_{cb}'$         &&  $\sfrac12^+$  &&  $O^{(1/2)}_j[u,c,b]$, \hspace{2ex}  $O_5^\prime [u,c,b]$  \\
 \\[-2.5ex]
 $\Xi_{cb}^*$        &&  $\sfrac32^+$  &&  $O^{(3/2)}_j[u,c,b]$  \\
 \\[-2.5ex]
 $\Omega_{cb}$       &&  $\sfrac12^+$  &&  $O_5 [s,c,b]$  \\
 \\[-2.5ex]
 $\Omega_{cb}'$      &&  $\sfrac12^+$  &&  $O^{(1/2)}_j[s,c,b]$, \hspace{2ex}  $O_5^\prime [s,c,b]$  \\
 \\[-2.5ex]
 $\Omega_{cb}^*$     &&  $\sfrac32^+$  &&  $O^{(3/2)}_j[s,c,b]$  \\
 \\[-2.5ex]
 $\Omega_{ccb}$      &&  $\sfrac12^+$  &&  $O^{(1/2)}_j[c,c,b]$  \\
 \\[-2.5ex]
 $\Omega_{ccb}^*$    &&  $\sfrac32^+$  &&  $O^{(3/2)}_j[c,c,b]$  \\
 \\[-2.5ex]
 $\Omega_{cbb}$      &&  $\sfrac12^+$  &&  $O^{(1/2)}_j[b,b,c]$  \\
 \\[-2.5ex]
 $\Omega_{cbb}^*$    &&  $\sfrac32^+$  &&  $O^{(3/2)}_j[b,b,c]$  \\
 \\[-2.5ex]

\hline\hline
\end{tabular}
  \caption{\label{tab:baryonops}Heavy-baryon operators.}
\end{table}

\FloatBarrier
\subsection{\label{sec:mesonops}Heavy meson operators}
\FloatBarrier

We also computed the energies of the heavy-quarkonium states $\eta_c$, $\eta_b$, $J/\psi$, and $\Upsilon$ using two-point functions of the operators
\begin{eqnarray}
 O^{(M)}_5 &=& \bar{q} \gamma_5 q, \\
 O^{(M)}_j &=& \bar{q} \gamma_j q,
\end{eqnarray}
where $q=b,c$. These were already used for the tuning of the charm and bottom quark actions. In the later stages of the data analysis
we use the energy differences
\begin{equation}
 E_X^{(\rm sub)} = E_X - \frac{n_c}{2} \overline{E}_{c\bar{c}}  - \frac{n_b}{2} \overline{E}_{b\bar{b}},
\end{equation}
where $E_X$ is the energy of a baryon containing $n_c$ charm quarks and $n_b$ bottom quarks, and
$\overline{E}_{c\bar{c}}$ and $\overline{E}_{b\bar{b}}$ are the spin-averaged charmonium and bottomonium energies.
In these energy differences, the bulk of the dependence on the heavy-quark masses cancels and the uncertainty associated
with the conversion from lattice to physical units is reduced dramatically. Furthermore, for hadrons containing
$b$ quarks, using energy differences is necessary to cancel the overall unphysical NRQCD energy shift.

\FloatBarrier
\subsection{\label{sec:twopt}Two-point functions and fit methodology}
\FloatBarrier

From a given baryon or meson operator as discussed in the previous two sections, we obtained
multiple versions by applying Gaussian smearing to some or all of the quark fields. These
different operators couple to the same states but differ in their relative amplitudes
to couple to the ground and excited states and produce different amounts of statistical noise
in the correlation functions. We constructed the smeared quark fields, $\tilde{q}$, as
\begin{equation}
 \tilde{q} = \left(1 + \frac{r_S^2}{2 n_S}\Delta^{(2)}\right)^{n_S} q, \label{eq:smear_op}
\end{equation}
where the gauge-covariant three-dimensional lattice Laplace operator, $\Delta^{(2)}$, is defined as
\begin{equation}
 \Delta^{(2)} q(\mathbf{x}, t) = -\frac{1}{a^2} \sum_{j=1}^3\left(  U_j(\mathbf{x}, t) q(\mathbf{x}+a\mathbf{\hat{j}},t) - 2q(\mathbf{x}, t)
 + U_{-j}(\mathbf{x}, t) q(\mathbf{x}-a\mathbf{\hat{j}},t) \right).  \label{eq:Laplace}
\end{equation}
We used different smearing parameters for the light and strange, charm, and bottom quark fields, as detailed
in Table \ref{tab:smearing}. Since this work reuses domain-wall light and strange quark propagators computed by us
in earlier work \cite{Detmold:2011bp, Detmold:2012ge, Detmold:2012vy}, the smearing parameters for light and strange quarks
could not be changed here. For the charm and bottom quarks, we used different values of $a r_S$ for the two different lattice spacings in order
to keep the smearing width in physical units, $r_S$, fixed. For the charm quarks, we used ``stout''-smeared gauge links \cite{Morningstar:2003gk} in
Eq.~(\ref{eq:Laplace}), with ten iterations and staple weight $\rho=0.08$ in the spatial directions.

\begin{table}
  \begin{tabular}{lllllll}
    \hline\hline
    Data set & \hspace{2ex} &  $a r_S,\: n_S$ (light/strange)  & \hspace{2ex} & $a r_S,\: n_S$ (charm) & \hspace{2ex} & $a r_S,\: n_S$ (bottom) \\
    \hline
    \texttt{C104}, \texttt{C14}, \texttt{C24}, \texttt{C54}, \texttt{C53} &&  $3.08,\: 30$  && $2.12,\: 70$  &&  $1.41,\: 10$ \\
    \texttt{F23}, \texttt{F43}, \texttt{F63}                              &&  $3.08,\: 30$  && $2.83,\: 70$  &&  $1.89,\: 10$ \\
    \hline\hline
  \end{tabular}
  \caption{\label{tab:smearing}Parameters used for the smearing of the quark fields in the baryon and meson interpolating operators. }
\end{table}

For the triply-heavy baryons, we either applied the smearing to all three quarks or to none of the quarks. This leads to
two-by-two matrices of two-point functions. For example, in the case of a $ccb$ baryon, we have (schematically)
\begin{equation}
 C_{2\times2} = \left(  \begin{array}{cc}   \langle O[c, c, b] \:  \overline{O}[c, c, b] \rangle &
                              \langle O[\tilde{c}, \tilde{c}, \tilde{b}] \:  \overline{O}[c, c, b] \rangle  \\
                              \langle O[c, c, b] \:  \overline{O}[\tilde{c}, \tilde{c}, \tilde{b}] \rangle &
                              \langle O[\tilde{c}, \tilde{c}, \tilde{b}] \:  \overline{O}[\tilde{c}, \tilde{c}, \tilde{b}] \rangle 
         \end{array} \right).
\end{equation}
The domain-wall propagators for the up, down, and strange quarks all had smeared sources. At the sink, we either smeared
all domain-wall quarks, or kept them local. Thus, for the baryons containing both heavy and light valence quarks, we constructed $(2\times 4)$-matrices of correlation
functions; for example, for a baryon with $usb$ valence quarks,
\begin{equation}
 C_{2\times4} = \left(  \begin{array}{cccc}  \langle O[\tilde{u}, \tilde{s}, b] \:  \overline{O}[\tilde{u}, \tilde{s}, b] \rangle &
                              \langle O[\tilde{u}, \tilde{s}, \tilde{b}] \:  \overline{O}[\tilde{u}, \tilde{s}, b] \rangle &
                              \langle O[u, s, b] \:  \overline{O}[\tilde{u}, \tilde{s}, b] \rangle &
                              \langle O[u, s, \tilde{b}] \:  \overline{O}[\tilde{u}, \tilde{s}, b] \rangle \\
                              \langle O[\tilde{u}, \tilde{s}, b] \:  \overline{O}[\tilde{u}, \tilde{s}, \tilde{b}] \rangle &
                              \langle O[\tilde{u}, \tilde{s}, \tilde{b}] \:  \overline{O}[\tilde{u}, \tilde{s}, \tilde{b}] \rangle &
                              \langle O[u, s, b] \:  \overline{O}[\tilde{u}, \tilde{s}, \tilde{b}] \rangle &
                              \langle O[u, s, \tilde{b}] \:  \overline{O}[\tilde{u}, \tilde{s}, \tilde{b}] \rangle
         \end{array} \right).
\end{equation}
To extract the energies from exponential fits of the correlation functions, we used both single-correlator fits and matrix fits,
as well as different procedures for choosing the time ranges to include in the fit. For the single-correlator fits, we selected only
the correlator with all quarks smeared at source and sink, using, for example for a $usb$ baryon,
\begin{equation}
 \langle O[\tilde{u}, \tilde{s}, \tilde{b}] \:  \overline{O}[\tilde{u}, \tilde{s}, \tilde{b}] \rangle \: \underset{{\rm large}\:\:t}{\longrightarrow} \: A^2\:e^{-E\,t},
\end{equation}
with fit parameters $A$ and $E$. The $(2\times 2)$-matrix fits were performed using
\begin{equation}
 C_{2\times2}(t)\:\: \underset{{\rm large}\:\:t}{\longrightarrow}\:\:
 \left(  \begin{array}{cc  }  A_1  A_1 \,  e^{-E t} &
                              A_2  A_1 \,  e^{-E t} \\
                              A_1  A_2 \,  e^{-E t} &
                              A_2  A_2 \,  e^{-E t} 
         \end{array} \right), \label{eq:Omegaccbfit}
\end{equation}
with parameters $A_1$, $A_2$, and $E$, while the $(2\times 4)$-matrix fits had the form
\begin{equation}
 C_{2\times4}(t)\:\: \underset{{\rm large}\:\:t}{\longrightarrow}\:\:
 \left(  \begin{array}{cccc}  A_1  A_1 \,  e^{-E t} &
                              A_2  A_1 \,  e^{-E t} &
                              A_3  A_1 \,  e^{-E t} &
                              A_4  A_1 \,  e^{-E t} \\
                              A_1  A_2 \,  e^{-E t} &
                              A_2  A_2 \,  e^{-E t} &
                              A_3  A_2 \,  e^{-E t} &
                              A_4  A_2 \,  e^{-E t}
         \end{array} \right), \label{eq:Xibfit}
\end{equation}
with parameters $A_1$, $A_2$, $A_3$, $A_4$, and $E$.
The starting times $t_{\rm min}$ after which the data points are included in the fit must be chosen such that
contributions from excited states have decayed sufficiently and have become smaller than the statistical uncertainties.
While contributions from excited states decay exponentially with $t$, the statistical uncertainties grow
exponentially with $t$ \cite{Lepage:1991ui}. The individual component correlators in a matrix fit have
different amounts of excited-state contamination as well as different amounts of statistical noise. Therefore, the optimal choices
of $t_{\rm min}$ may be different for the different components, and we choose them independently in order
to get the highest possible precision for the matrix fit. We also choose $t_{\rm max}$ independently for
each component. The choice of $t_{\rm max}$ is limited in the positive direction by two requirements: avoiding contamination from backward-propagating/thermal states,
and avoiding too many degrees of freedom in the fit (having too many degrees of freedom relative to the number of data samples
leads to a poorly estimated, or even singular, covariance matrix in the definition of the $\chi^2$ function).

\begin{figure}
 \includegraphics[width=0.6\linewidth]{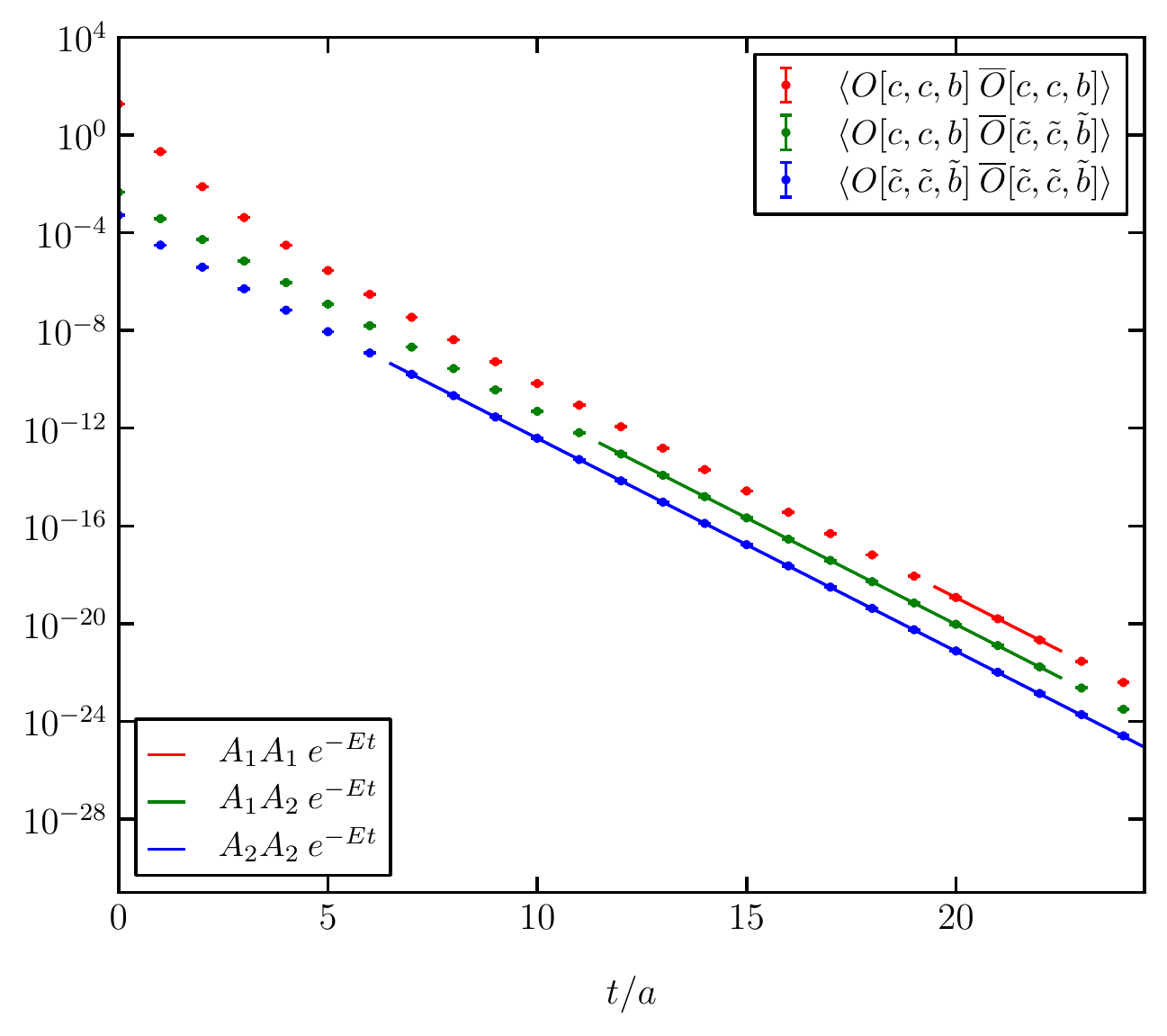}
 \caption{\label{fig:Omegaccbfit}Matrix fit of the $\Omega_{ccb}$ two-point functions using Eq.~(\ref{eq:Omegaccbfit}). The data shown here are from the \texttt{C54} set.
 The correlator $\langle O[\tilde{c}, \tilde{c}, \tilde{b}] \:  \overline{O}[c, c, b] \rangle$,
 which equals  $\langle O[c, c, b] \:  \overline{O}[\tilde{c}, \tilde{c}, \tilde{b}] \rangle$ in the limit of infinite statistics,
 is not shown for clarity.}
\end{figure}

\begin{figure}
 \includegraphics[width=0.6\linewidth]{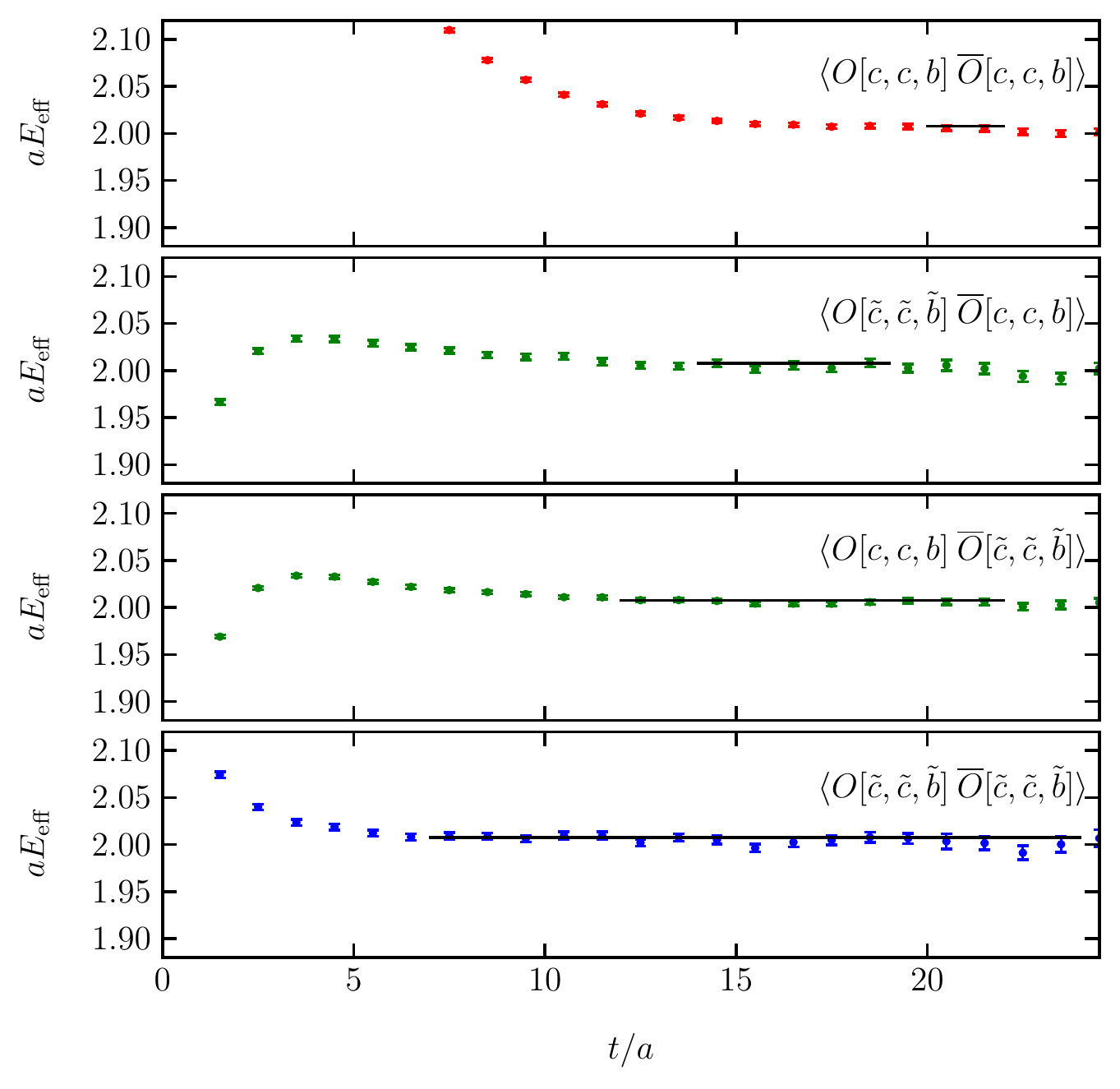}
 \caption{\label{fig:Omegaccbfitem}Effective-energy plot for the $2\times2$ matrix of $\Omega_{ccb}$ two-point functions from the \texttt{C54} set.
 The effective energy for a correlator $C(t)$ is computed as $\:aE_{\rm eff}(t+\frac{a}{2})=\ln\left[C(t)/C(t+a)\right]$. The lines indicate the time ranges and the energy
 obtained from the fit shown in Fig.~\protect\ref{fig:Omegaccbfit}.}
\end{figure}

The benefits of allowing individual fit ranges for the correlators within a matrix fit are illustrated for the case
of the $\Omega_{ccb}$ baryon in Figs.~\ref{fig:Omegaccbfit} and \ref{fig:Omegaccbfitem}.
The smeared-source, smeared-sink correlator $\langle O[\tilde{c}, \tilde{c}, \tilde{b}] \:  \overline{O}[\tilde{c}, \tilde{c}, \tilde{b}] \rangle$ is noisy, but $t_{\rm min}$
can be chosen very small. In contrast, the local-source, local-sink correlator $\langle O[c, c, b] \:  \overline{O}[c, c, b] \rangle$ is statistically
most precise, but $t_{\rm min}$ has to be chosen very large to avoid excited-state contamination. Performing the coupled matrix fit with individual time
ranges allows us to extract the best possible result for the energy, using the best regions of all correlators.

Given the large number of different correlators and data sets used in this work, optimizing all fit ranges by hand
would be impractical and prone to bias. We therefore implemented several procedures for automatically choosing the fit
ranges according to criteria including the quality-of-fit and the size of the statistical uncertainty of the extracted
energy. We used four different procedures:
\begin{itemize}
 \item Method 1: This method was applied to perform $(2\times4)$-matrix fits for the heavy-light baryons and $(2\times2)$-matrix fits for the heavy quarkonia and triply
        heavy baryons. Initial guesses for $(t_{\rm min}, t_{\rm max})$ for each component correlator were obtained by first performing individual fits of the
        form $A\: e^{-E t}$ of each component correlator, requiring $\chi^2/{\rm d.o.f.}\lesssim 1$ while preferring fits with smaller uncertainty. Simultaneous matrix
        fits were then performed with these initial fit ranges. These initial matrix fits typically had $\chi^2/{\rm d.o.f.}>1$. This is expected because
        the coupled fit achieves a smaller statistical uncertainty, requiring that excited states be negligible to a higher level of precision, and because
        the numbers of parameters in Eqs.~(\ref{eq:Omegaccbfit}) and (\ref{eq:Xibfit}) are not increased proportionally to the larger number of the degrees of freedom.
        We then applied a Monte-Carlo search for improved fit ranges of the matrix fit to achieve $\chi^2/{\rm d.o.f.}\lesssim 1$. The algorithm used for this
        repeatedly attempts small (multidimensional) random shifts to the fit region, accepting a shift only if $\chi^2/{\rm d.o.f.}$ decreases.
 \item Method 2: This method is a modified version of Method 1, where we constrained the Monte-Carlo search for the fit domain by requiring that
       each component correlator contributes at least three time slices to the fit, i.e., $t_{\rm max}-t_{\rm min} \geq 3a$.
 \item Method 3: This deterministic method performed five-dimensional scans of the fit ranges of $(2\times2)$ matrix fits in small intervals around the initial
       ranges (the initial ranges were chosen as in Method 1). For the heavy-light baryons, we only used those correlators in which all light quarks at the source and
       sink were smeared in order to obtain $(2\times2)$ matrices.
       Here, five-dimensional scan means that we independently varied the values of $t_{\rm min}$ for all four component
       correlators, but varied the values of $t_{\rm max}$ only by a common shift for all four component correlators relative to the initial ranges,
       to keep the computational cost within bounds. The scans were constrained by the requirement that each component correlator contributes at least 5 time slices (for the coarse lattices)
       or 7 time slices (for fine lattice) to the fit. Of all the matrix fits performed with this scanning procedure, only those with
       $\chi^2/{\rm d.o.f.}\leq1$ and $Q\geq 0.5$ were kept, and then the fit with the smallest uncertainty for the energy was chosen.
 \item Method 4: This deterministic method performed fits only to the single correlator in which all quarks are smeared at source and sink (this correlator
       is expected to have the least excited-state contamination). Two-dimensional scans of $t_{\rm min}$ and $t_{\max}$ were performed in a wide range.
       As in Method 3, the scans were constrained by the requirement that each component correlator contributes at least 5 time slices (for the coarse lattices)
       or 7 time slices (for fine lattice) to the fit. Of all fits, those with $\chi^2/{\rm d.o.f.}\leq1$ and $Q\geq 0.5$ were kept,
       and then the fit with the smallest uncertainty for the energy was chosen.
\end{itemize}
When applying each procedure, we enforced common fit ranges for hyperfine partners such as the $\Sigma_b$ and $\Sigma_b^*$,
to ensure the optimal cancellation of statistical uncertainties and excited-state contamination in the small hyperfine splittings.

\begin{figure}
 \includegraphics[height=4cm]{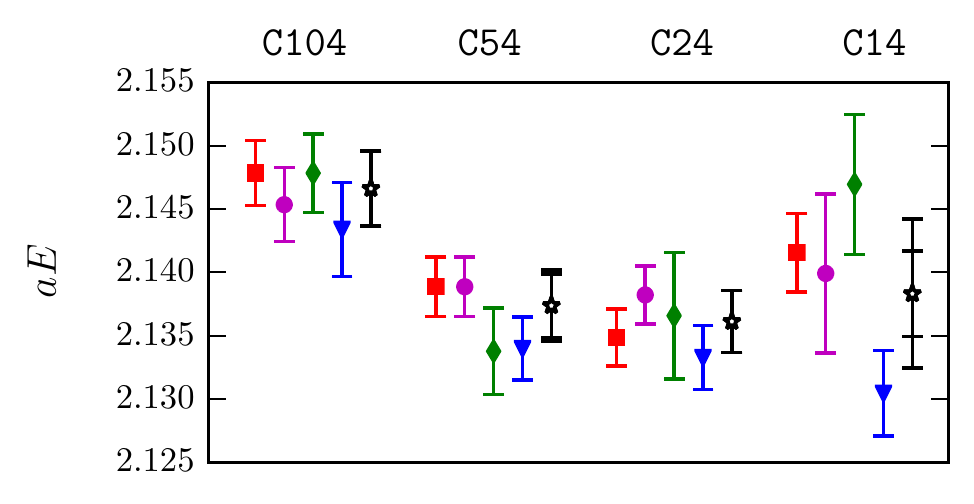} \hfill \includegraphics[height=4cm]{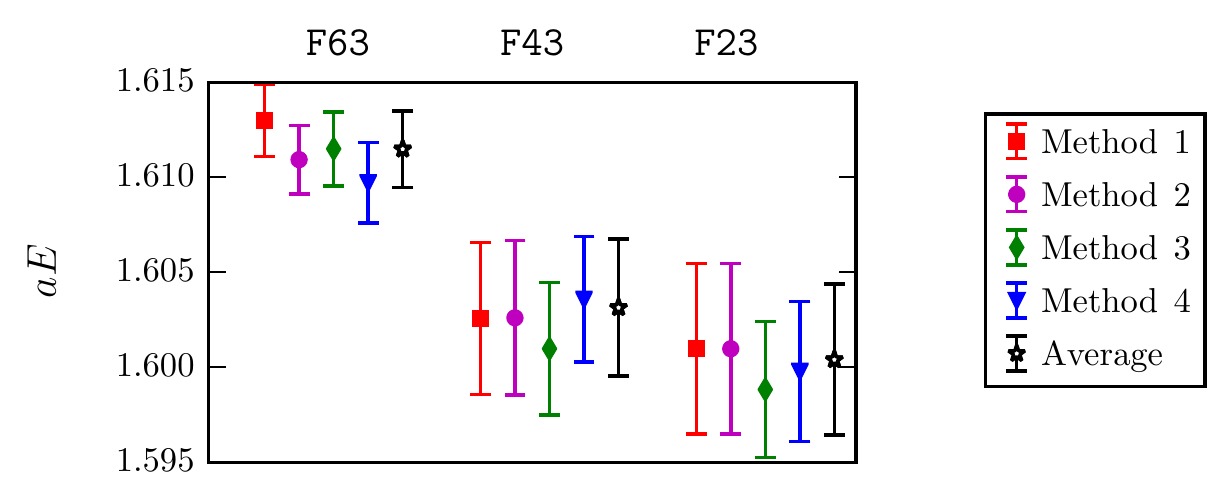}
 \caption{\label{fig:Xccstarenergies}$\Xi_{cc}^*$ energies obtained using the four different fit methods for each data set as explained in the main text.
 Also shown are the method-averaged energies (correlations are taken into account).
 For the method-averaged energies, the outer error bars include a scale factor in the cases where the average has
 $\chi^2/{\rm d.o.f.}>1$ (here, for the \texttt{C14} and \texttt{C54} data sets).}
\end{figure}

To illustrate how the results from methods 1 through 4 compare with each other, we show the $\Xi_{cc}^*$ energies in Fig.~\ref{fig:Xccstarenergies}.
The different methods generally give quite consistent results, and we use the correlated weighted average for the 
further analysis. The correlations between the energies from the different methods are taken into account using statistical bootstrap; we perform
the weighted averages for each bootstrap sample to obtain a new bootstrap distribution for the average energy. The statistical
uncertainty of the average energy is then obtained from the width of this distribution. In some cases, the energies obtained using
the different fit methods are not consistent with each other (as can be seen for the \texttt{C14} data set in Fig.~\ref{fig:Xccstarenergies}),
and we inflate the uncertainty of the average using a scale factor.
To this end, we compute the value of $\chi^2$ for a constant fit to the four energies. If $\chi^2/(N-1)>1$ (where $N=4$ is the number of data points),
we inflate the uncertainty of the weighted average by a factor of \cite{Beringer:1900zz}
\begin{equation}
 S=\sqrt{\chi^2/(N-1)}.
\end{equation}
The averaged baryon and quarkonium energies from all data sets are given in Tables \ref{tab:baryonenergies} and \ref{tab:quarkoniumenergies}, respectively.

\begin{table}
\begin{tabular}{lcccccccccccccccc}
\hline\hline
State             & & $\mathtt{C104}$ & & $\mathtt{C14}$ & & $\mathtt{C24}$ & & $\mathtt{C54}$ & & $\mathtt{C53}$ & & $\mathtt{F23}$ & & $\mathtt{F43}$ & & $\mathtt{F63}$  \\
\hline
 $\Lambda_c$      &&  1.4068(24)  &&  1.3542(51)  &&  1.3628(59)  &&  1.3748(38)  &&  $\hdots$    &&  1.007(19)   &&  1.020(14)   &&  1.0344(29)    \\
 \\[-3ex]
 $\Sigma_c$       &&  1.4920(46)  &&  1.4549(77)  &&  1.4634(51)  &&  1.4653(43)  &&  $\hdots$    &&  1.0825(72)  &&  1.0929(51)  &&  1.1008(41)    \\
 \\[-3ex]
 $\Sigma_c^*$     &&  1.5452(52)  &&  1.4813(94)  &&  1.5137(54)  &&  1.5115(46)  &&  $\hdots$    &&  1.1087(96)  &&  1.1231(57)  &&  1.1372(51)    \\
 \\[-3ex]
 $\Xi_c$          &&  1.4715(19)  &&  1.4481(29)  &&  1.4469(39)  &&  1.4557(32)  &&  1.4308(37)  &&  1.0838(92)  &&  1.0877(86)  &&  1.0958(19)    \\
 \\[-3ex]
 $\Xi_c'$         &&  1.5318(29)  &&  1.5169(36)  &&  1.5157(36)  &&  1.5231(30)  &&  1.5060(40)  &&  1.1302(36)  &&  1.1334(32)  &&  1.1440(22)    \\
 \\[-3ex]
 $\Xi_c^*$        &&  1.5802(36)  &&  1.5593(45)  &&  1.5605(42)  &&  1.5661(34)  &&  1.5446(46)  &&  1.1626(46)  &&  1.1664(42)  &&  1.1798(27)    \\
 \\[-3ex]
 $\Omega_c$       &&  1.5797(24)  &&  $\hdots$    &&  $\hdots$    &&  1.5790(21)  &&  1.5452(29)  &&  $\hdots$    &&  1.1763(23)  &&  1.1856(17)    \\
 \\[-3ex]
 $\Omega_c^*$     &&  1.6256(27)  &&  $\hdots$    &&  $\hdots$    &&  1.6206(24)  &&  1.5858(36)  &&  $\hdots$    &&  1.2105(31)  &&  1.2179(21)    \\
 \\[-3ex]
 $\Xi_{cc}$       &&  2.0916(24)  &&  2.0826(24)  &&  2.0835(22)  &&  2.0863(21)  &&  $\hdots$    &&  1.5630(27)  &&  1.5659(25)  &&  1.5738(16)    \\
 \\[-3ex]
 $\Xi_{cc}^*$     &&  2.1466(29)  &&  2.1383(59)  &&  2.1361(25)  &&  2.1374(28)  &&  $\hdots$    &&  1.6004(40)  &&  1.6031(36)  &&  1.6115(20)    \\
 \\[-3ex]
 $\Omega_{cc}$    &&  2.1407(15)  &&  $\hdots$    &&  $\hdots$    &&  2.1388(16)  &&  2.1231(20)  &&  $\hdots$    &&  1.6109(16)  &&  1.6158(14)    \\
 \\[-3ex]
 $\Omega_{cc}^*$  &&  2.1907(16)  &&  $\hdots$    &&  $\hdots$    &&  2.1858(30)  &&  2.1720(22)  &&  $\hdots$    &&  1.6460(22)  &&  1.6508(16)    \\
 \\[-3ex]
 $\Omega_{ccc}$   &&  2.7352(14)  &&  $\hdots$    &&  $\hdots$    &&  2.7315(14)  &&  $\hdots$    &&  $\hdots$    &&  2.0654(17)  &&  2.06756(99)   \\
 \\[-3ex]
 $\Lambda_b$      &&  0.7559(48)  &&  0.7134(54)  &&  0.7113(59)  &&  0.7216(60)  &&  $\hdots$    &&  0.552(19)   &&  0.562(13)   &&  0.5672(66)    \\
 \\[-3ex]
 $\Sigma_b$       &&  0.8581(85)  &&  0.849(12)   &&  0.8385(63)  &&  0.8389(57)  &&  $\hdots$    &&  0.620(11)   &&  0.639(11)   &&  0.6411(83)    \\
 \\[-3ex]
 $\Sigma_b^*$     &&  0.875(13)   &&  0.866(15)   &&  0.8508(61)  &&  0.8483(64)  &&  $\hdots$    &&  0.636(12)   &&  0.654(11)   &&  0.6525(82)    \\
 \\[-3ex]
 $\Xi_b$          &&  0.8158(31)  &&  0.7978(42)  &&  0.7984(84)  &&  0.8020(47)  &&  0.7833(68)  &&  0.6085(66)  &&  0.610(10)   &&  0.6203(40)    \\
 \\[-3ex]
 $\Xi_b'$         &&  0.8962(82)  &&  0.8861(50)  &&  0.8849(54)  &&  0.8879(41)  &&  0.8781(68)  &&  0.6718(60)  &&  0.6760(54)  &&  0.6787(45)    \\
 \\[-3ex]
 $\Xi_b^*$        &&  0.909(11)   &&  0.9019(57)  &&  0.8971(72)  &&  0.9028(40)  &&  0.8857(69)  &&  0.6825(61)  &&  0.6894(55)  &&  0.6944(47)    \\
 \\[-3ex]
 $\Omega_b$       &&  0.9382(44)  &&  $\hdots$    &&  $\hdots$    &&  0.9406(29)  &&  0.9136(43)  &&  $\hdots$    &&  0.7229(62)  &&  0.7182(30)    \\
 \\[-3ex]
 $\Omega_b^*$     &&  0.9535(46)  &&  $\hdots$    &&  $\hdots$    &&  0.9567(29)  &&  0.9269(45)  &&  $\hdots$    &&  0.7347(52)  &&  0.7321(31)    \\
 \\[-3ex]
 $\Xi_{bb}$       &&  0.7242(49)  &&  0.7187(30)  &&  0.7145(39)  &&  0.7173(31)  &&  $\hdots$    &&  0.5721(54)  &&  0.5785(49)  &&  0.5877(71)    \\
 \\[-3ex]
 $\Xi_{bb}^*$     &&  0.7486(63)  &&  0.7410(33)  &&  0.7351(38)  &&  0.7381(32)  &&  $\hdots$    &&  0.5825(66)  &&  0.589(10)   &&  0.6032(59)    \\
 \\[-3ex]
 $\Omega_{bb}$    &&  0.7590(22)  &&  $\hdots$    &&  $\hdots$    &&  0.7586(19)  &&  0.7450(27)  &&  $\hdots$    &&  0.6161(27)  &&  0.6183(18)    \\
 \\[-3ex]
 $\Omega_{bb}^*$  &&  0.7820(28)  &&  $\hdots$    &&  $\hdots$    &&  0.7792(20)  &&  0.7654(30)  &&  $\hdots$    &&  0.6291(32)  &&  0.6330(19)    \\
 \\[-3ex]
 $\Omega_{bbb}$   &&  0.5335(14)  &&  $\hdots$    &&  $\hdots$    &&  0.5311(21)  &&  $\hdots$    &&  $\hdots$    &&  0.4735(18)  &&  0.4717(13)    \\
 \\[-3ex]
 $\Xi_{cb}$       &&  1.4174(52)  &&  1.4102(42)  &&  1.4165(40)  &&  1.4176(33)  &&  $\hdots$    &&  1.0869(87)  &&  1.0833(94)  &&  1.0898(30)    \\
 \\[-3ex]
 $\Xi_{cb}'$      &&  1.4486(49)  &&  1.4369(46)  &&  1.4423(42)  &&  1.4431(32)  &&  $\hdots$    &&  1.104(11)   &&  1.101(11)   &&  1.1088(28)    \\
 \\[-3ex]
 $\Xi_{cb}^*$     &&  1.4630(67)  &&  1.4548(48)  &&  1.4584(48)  &&  1.4592(38)  &&  $\hdots$    &&  1.1172(96)  &&  1.1092(97)  &&  1.1214(57)    \\
 \\[-3ex]
 $\Omega_{cb}$    &&  1.4603(35)  &&  $\hdots$    &&  $\hdots$    &&  1.4647(23)  &&  1.4468(38)  &&  $\hdots$    &&  1.1183(31)  &&  1.1240(24)    \\
 \\[-3ex]
 $\Omega_{cb}'$   &&  1.4816(26)  &&  $\hdots$    &&  $\hdots$    &&  1.4866(21)  &&  1.4673(36)  &&  $\hdots$    &&  1.1349(31)  &&  1.1397(24)    \\
 \\[-3ex]
 $\Omega_{cb}^*$  &&  1.4983(30)  &&  $\hdots$    &&  $\hdots$    &&  1.5027(38)  &&  1.4841(40)  &&  $\hdots$    &&  1.1438(37)  &&  1.1520(37)    \\
 \\[-3ex]
 $\Omega_{ccb}$   &&  2.0071(14)  &&  $\hdots$    &&  $\hdots$    &&  2.0079(15)  &&  $\hdots$    &&  $\hdots$    &&  1.5406(21)  &&  1.5413(12)    \\
 \\[-3ex]
 $\Omega_{ccb}^*$ &&  2.0247(15)  &&  $\hdots$    &&  $\hdots$    &&  2.0253(16)  &&  $\hdots$    &&  $\hdots$    &&  1.5512(25)  &&  1.5539(12)    \\
 \\[-3ex]
 $\Omega_{cbb}$   &&  1.2689(13)  &&  $\hdots$    &&  $\hdots$    &&  1.2678(13)  &&  $\hdots$    &&  $\hdots$    &&  1.0076(18)  &&  1.0066(12)    \\
 \\[-3ex]
 $\Omega_{cbb}^*$ &&  1.2888(14)  &&  $\hdots$    &&  $\hdots$    &&  1.2873(15)  &&  $\hdots$    &&  $\hdots$    &&  1.0210(21)  &&  1.0206(14)    \\
 \\[-3ex]
\hline\hline
\end{tabular}
\caption{\label{tab:baryonenergies}Baryon energies (in lattice units) extracted from the eight different data sets (see Table \protect\ref{tab:params}).
The results given here are averages over the different fit methods; the uncertainties include a scale factor as explained in the main text. For the partially quenched data
sets ($\mathtt{C14}$, $\mathtt{C24}$, $\mathtt{C53}$, $\mathtt{F23}$), results are given only for those baryons containing a light or strange valence quark affected by the partial quenching.
Note that for baryons containing $b$ quarks, the energies are shifted from their physical values because of the use of NRQCD. These shifts cancel in appropriate energy differences.}
\end{table}

\FloatBarrier

\begin{table}
\begin{tabular}{lcccccccc}
\hline\hline
State        & & $\mathtt{C104}$ & & $\mathtt{C54}$ & & $\mathtt{F43}$ & & $\mathtt{F63}$  \\
\hline
 $\eta_c$    & & 1.69288(24)     & & 1.69110(25)    & & 1.28293(35)    & & 1.28340(20)     \\
 \\[-3ex]
 $J/\psi$    & & 1.75571(38)     & & 1.75263(40)    & & 1.32983(53)    & & 1.33006(32)     \\
 \\[-3ex]
 $\eta_b$    & & 0.24928(22)     & & 0.24838(27)    & & 0.23607(34)    & & 0.23566(24)     \\
 \\[-3ex]
 $\Upsilon$  & & 0.28528(29)     & & 0.28413(35)    & & 0.26429(45)    & & 0.26374(31)     \\
 \\[-3ex]
\hline\hline
\end{tabular}
\caption{\label{tab:quarkoniumenergies}Quarkonium energies (in lattice units) extracted from the four data sets that correspond to independent gauge-field ensembles (see Table \protect\ref{tab:params}).
The results given here are averages over the different fit methods; the uncertainties include a scale factor as explained in the main text. Note that the bottomonium energies are shifted from their physical values because of the use of NRQCD.}
\end{table}

\FloatBarrier

\subsection{\label{sec:mixing}Mixing effects}
\FloatBarrier

Before moving on to the chiral and continuum extrapolations in Sec.~\ref{sec:extrap}, we briefly return here to the issue of the mixing
between the ``primed'' and ``unprimed'' baryons with $J^P=\frac12^+$. Of the baryons considered in this work, this affects the
$\Xi_c^\prime$ and $\Xi_c$, the $\Xi_b^\prime$ and $\Xi_b$, the $\Xi_{cb}^\prime$ and $\Xi_{cb}$, and the $\Omega_{cb}^\prime$ and $\Omega_{cb}$.
In each case, the interpolating operators we use for the ``primed'' and ``unprimed'' baryons (see Table \ref{tab:baryonops}) do not differ in any of the exactly conserved
quantum numbers. Thus, the two-point functions of both the ``primed'' and ``unprimed'' operators asymptotically approach the same ground state, which
is the ``unprimed'' baryon, while the ``primed'' baryon only appears as an excited state in both two-point functions.

\begin{figure}
\includegraphics[width=0.48\linewidth]{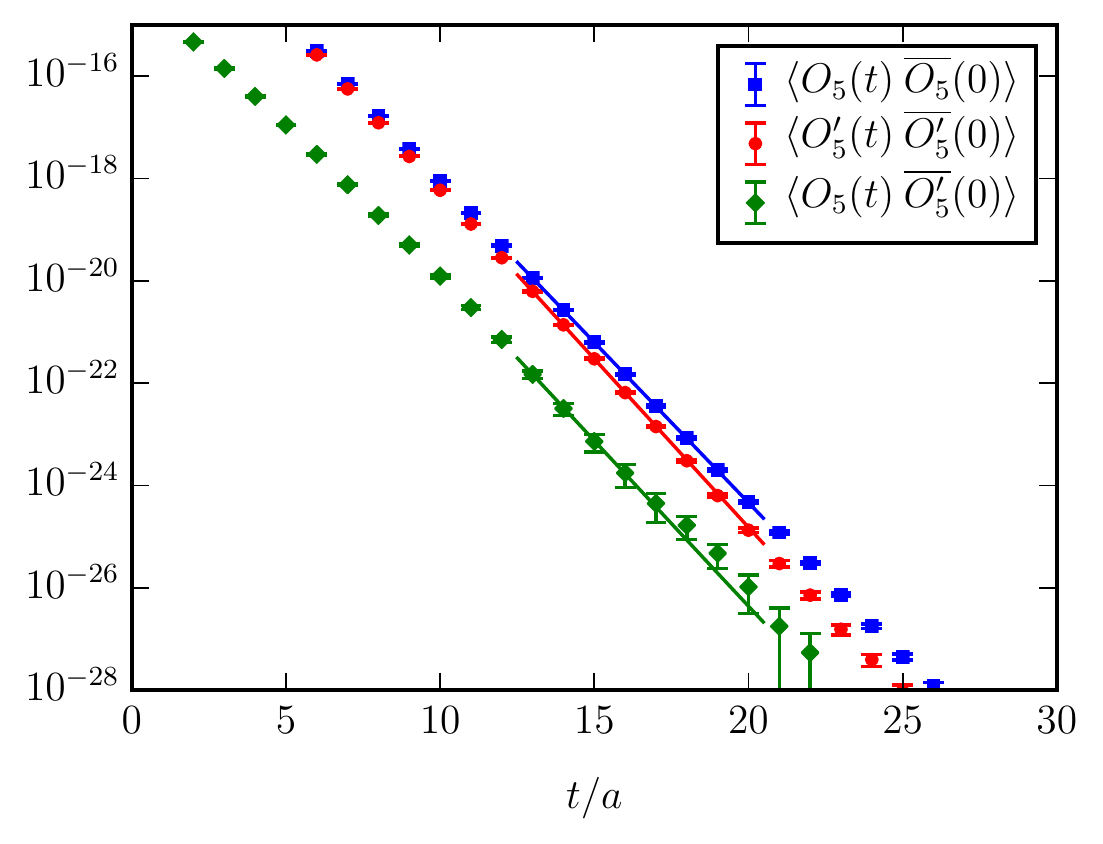}
\hfill \includegraphics[width=0.50\linewidth]{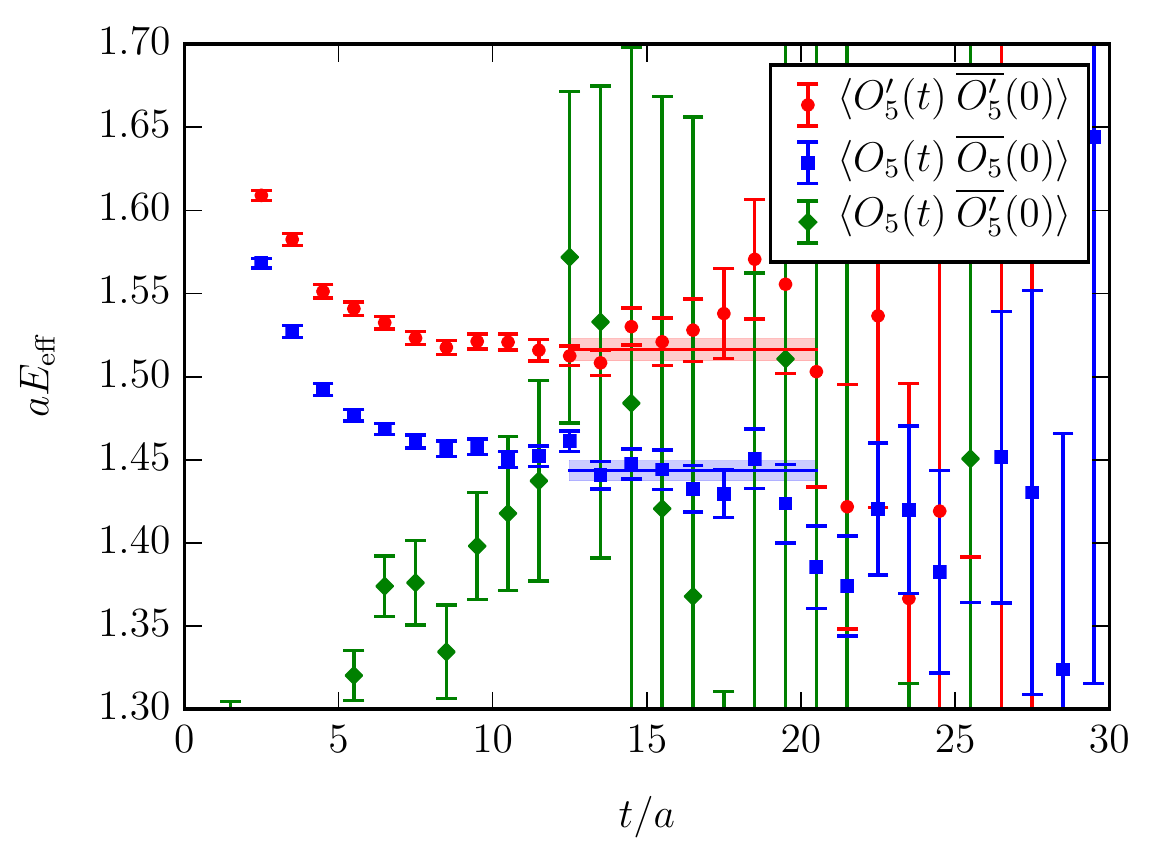}
\caption{\label{fig:XiXiprimedmixing}Coupled two-exponential fit to the correlators
$\langle O_5[\tilde{u}, \tilde{s}, \tilde{c}](t) \:\:\overline{O_5}[\tilde{u}, \tilde{s}, \tilde{c}](0) \rangle$,
$\langle O_5[\tilde{u}, \tilde{s}, \tilde{c}](t) \:\:\overline{O_5^\prime}[\tilde{u}, \tilde{s}, \tilde{c}](0) \rangle$,
and $\langle O_5^\prime[\tilde{u}, \tilde{s}, \tilde{c}](t) \:\:\overline{O_5^\prime}[\tilde{u}, \tilde{s}, \tilde{c}](0) \rangle$
 using Eqs.~(\ref{eq:Xic}), (\ref{eq:XicXicprime}), and (\ref{eq:Xicprime}). The data shown here are from the \texttt{C54} set.}
\end{figure}

\begin{figure}
 \includegraphics[width=0.48\linewidth]{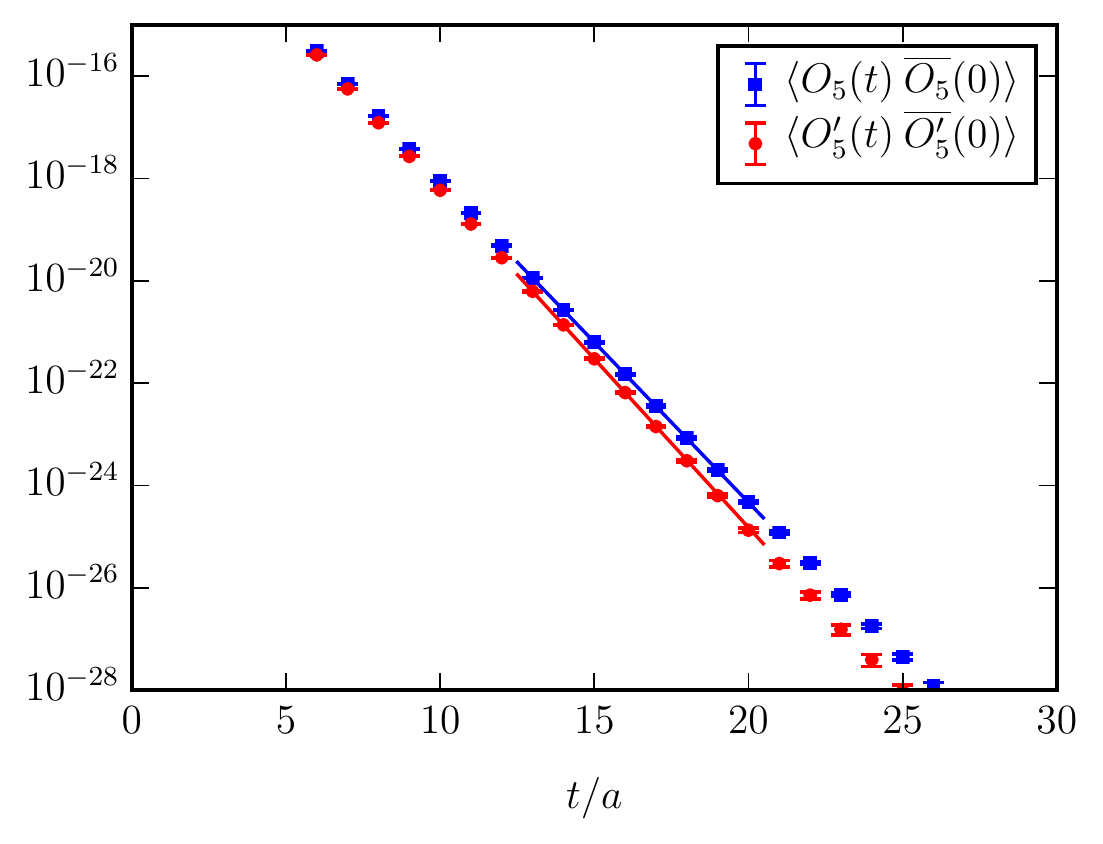}
 \hfill \includegraphics[width=0.50\linewidth]{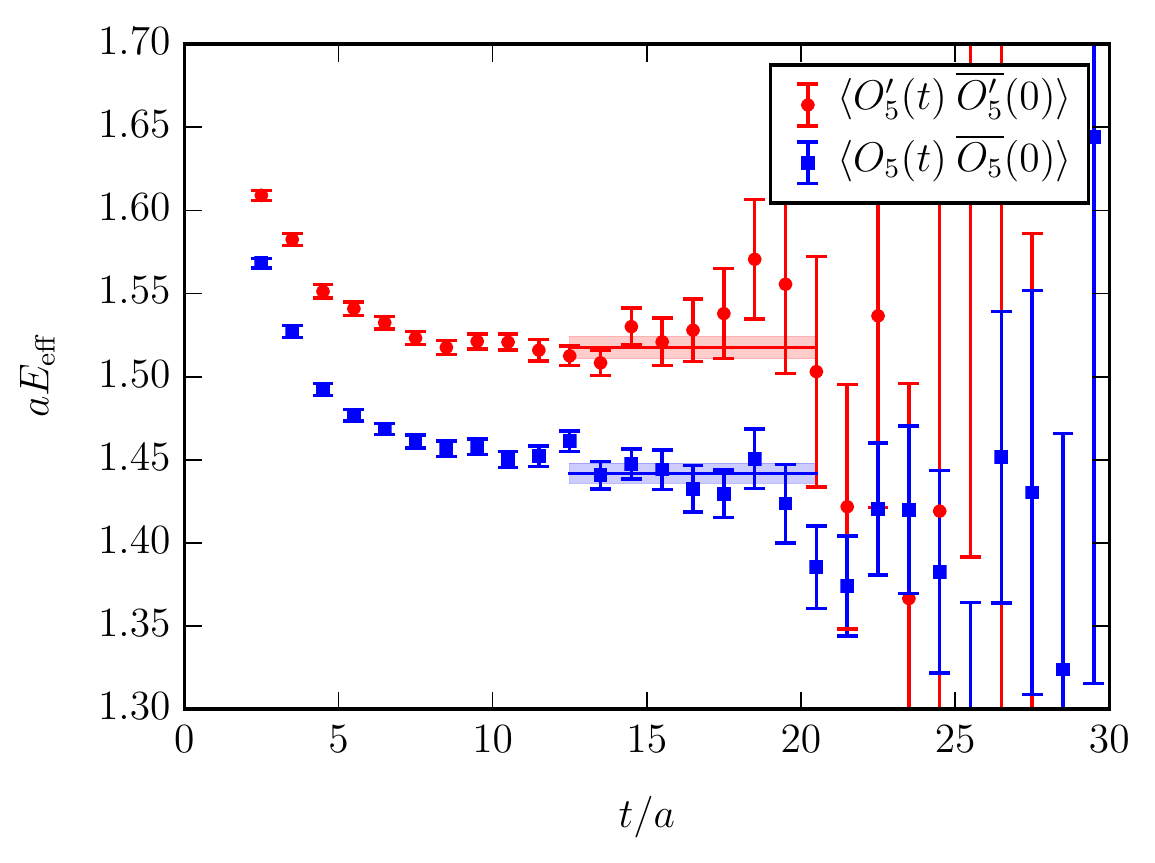}
\caption{\label{fig:XiXiprimednomixing}Independent single-exponential fits to the correlators
$\langle O_5[\tilde{u}, \tilde{s}, \tilde{c}](t) \:\:\overline{O_5}[\tilde{u}, \tilde{s}, \tilde{c}](0) \rangle$
and $\langle O^\prime_5[\tilde{u}, \tilde{s}, \tilde{c}](t) \:\:\overline{O^\prime_5}[\tilde{u}, \tilde{s}, \tilde{c}](0) \rangle$ according to
Eqs.~(\ref{eq:Xicnomixing}) and (\ref{eq:Xicprimenomixing}). The data shown here are from the \texttt{C54} set.}
\end{figure}

To be more concrete, let us consider the case of the $\Xi_c^\prime$ and $\Xi_c$, and let us consider only
the interpolating operators $O_5^\prime[\tilde{u}, \tilde{s}, \tilde{c}]$ and $O_5[\tilde{u}, \tilde{s}, \tilde{c}]$ in which all quarks are smeared
(the the following denoted more briefly as just $O_5^\prime$ and $O_5$). The spectral decomposition of the two-point correlators of $O_5$ and $O_5^\prime$ is given by
\begin{eqnarray}
  \langle O_5(t) \:\:\overline{O_5}(0) \rangle  &\:\:\underset{{\rm large}\:\:t}{\longrightarrow}\:\:&
  \langle 0 | O_5 | \Xi_c \rangle \langle \Xi_c | \overline{O_5} | 0 \rangle \,e^{-E_{\Xi_c} t}
  + \langle 0 | O_5 | \Xi^\prime_c \rangle \langle \Xi^\prime_c | \overline{O_5} | 0 \rangle \,e^{-E_{\Xi^\prime_c} t}\, ,  \label{eq:Xic} \\
  \langle O_5(t) \:\:\overline{O^\prime_5}(0) \rangle  &\:\:\underset{{\rm large}\:\:t}{\longrightarrow}\:\:&
  \langle 0 | O_5 | \Xi_c \rangle \langle \Xi_c | \overline{O^\prime_5} | 0 \rangle \,e^{-E_{\Xi_c} t}
  + \langle 0 | O_5 | \Xi^\prime_c \rangle \langle \Xi^\prime_c | \overline{O^\prime_5} | 0 \rangle \,e^{-E_{\Xi^\prime_c} t} \, , \label{eq:XicXicprime} \\
  \langle O^\prime_5(t) \:\:\overline{O^\prime_5}(0) \rangle  &\:\:\underset{{\rm large}\:\:t}{\longrightarrow}\:\:&
  \langle 0 | O^\prime_5 | \Xi_c \rangle \langle \Xi_c | \overline{O^\prime_5} | 0 \rangle \,e^{-E_{\Xi_c} t}
  + \langle 0 | O^\prime_5 | \Xi^\prime_c \rangle \langle \Xi^\prime_c | \overline{O^\prime_5} | 0 \rangle \,e^{-E_{\Xi^\prime_c} t} \, , \label{eq:Xicprime}
\end{eqnarray}
where only the contributions from the ground state and the first excited state are shown (the contributions
from higher excited states decay exponentially faster with $t$). Numerical result for these three correlators from the \texttt{C54} data set, as well as a coupled
two-exponential fit of the form given by Eqs.~(\ref{eq:Xic}), (\ref{eq:XicXicprime}), and (\ref{eq:Xicprime}), are shown in the left panel of Fig.~\ref{fig:XiXiprimedmixing}.
The fit range is $13\leq t/a \leq 20$, and the resulting energies are
\begin{equation}
 a E_{\Xi_c} =1.4435(61), \hspace{4ex} a E_{\Xi_c}^\prime= 1.5163(64). \label{eq:XiXiprimed2expfit}
\end{equation}
These energies are indicated with the horizontal bands in the effective-energy plot on the right-hand side of Fig.~\ref{fig:XiXiprimedmixing}.
We also performed naive, independent single-exponential fits of just the ``diagonal'' correlators in the same time range, using the form
\begin{eqnarray}
  \langle O_5(t) \:\:\overline{O_5}(0) \rangle  &\:\:\underset{{\rm large}\:\:t}{\longrightarrow}\:\:&
  \langle 0 | O_5 | \Xi_c \rangle \langle \Xi_c | \overline{O_5} | 0 \rangle \,e^{-E_{\Xi_c} t} \hspace{2ex}({\rm no\:\:mixing})\, ,  \label{eq:Xicnomixing} \\
  \langle O_5^\prime(t) \:\:\overline{O^\prime_5}(0) \rangle  &\:\:\underset{{\rm large}\:\:t}{\longrightarrow}\:\:&
  \langle 0 | O_5^\prime | \Xi^\prime_c \rangle \langle \Xi^\prime_c | \overline{O^\prime_5} | 0 \rangle \,e^{-E_{\Xi^\prime_c} t} \hspace{2ex}({\rm no\:\:mixing}) \, ,    \label{eq:Xicprimenomixing}
\end{eqnarray}
which neglects the overlap of the operator $\overline{O^\prime_5}$ with the ground state. This fit is shown in Fig.~\ref{fig:XiXiprimednomixing} and gives the energies
\begin{equation}
 a E_{\Xi_c} = 1.4419(61), \hspace{4ex} a E_{\Xi^\prime_c} = 1.5176(67).
\end{equation}
This result is in fact perfectly consistent with the full two-exponential fit result (\ref{eq:XiXiprimed2expfit}). The reason is that the ``wrong state''
overlap matrix elements $\langle 0 | O^\prime_5 | \Xi_c \rangle$ and $\langle 0 | O_5 | \Xi^\prime_c \rangle$ are highly suppressed relative to the ``right state''
matrix elements $\langle 0 | O^\prime_5 | \Xi^\prime_c \rangle$ and $\langle 0 | O_5 | \Xi_c \rangle$. The coupled two-exponential fit using Eqs.~(\ref{eq:Xic}), (\ref{eq:XicXicprime}),
and (\ref{eq:Xicprime}) gives
\begin{equation}
 \frac{\langle 0 | O^\prime_5 | \Xi_c \rangle}{\langle 0 | O^\prime_5 | \Xi^\prime_c \rangle}  = 0.003(17),
 \hspace{4ex} \frac{\langle 0 | O_5 | \Xi^\prime_c \rangle}{\langle 0 | O_5 | \Xi_c \rangle} = 0.020(43).
\end{equation}
Because of this suppression, the effective-energy plot of the
two-point function $\langle O_5^\prime(t) \:\:\overline{O^\prime_5}(0) \rangle$ shows a clean plateau at the $\Xi^\prime_c$ energy at intermediate $t$,
with no obvious sign of the ground-state contribution before the signal disappears into noise (see the right-hand sides of Figs.~\ref{fig:XiXiprimedmixing} and \ref{fig:XiXiprimednomixing}).
The physical reason for the smallness of the mixing is a double suppression through the approximate heavy-quark spin symmetry and the approximate $SU(3)$ flavor symmetry.
In the heavy-quark limit $m_c\to\infty$, the spin of the light degrees of freedom, $S_l$, becomes exactly conserved, and the $\Xi_c$, $\Xi_c^\prime$ have $S_l=0$ and $S_l=1$,
respectively. Furthermore, in the limit of degenerate $u$, $d$, and $s$-quark masses, the $\Xi_c$ belongs to the a $\overline{\mathbf{3}}$ (anti-fundamental) irreducible representation
of the $SU(3)$ flavor symmetry, while the $\Xi_c^\prime$ is part of a $\mathbf{6}$ (sextet) irreducible representation.
We also performed analogous comparisons of coupled two-exponential fits and naive single exponential fits for the other affected baryons, and obtained consistent results for the energies
from both fit methods in all cases.

\FloatBarrier
\section{\label{sec:extrap}Chiral and continuum extrapolations}
\FloatBarrier

Having extracted the baryon energies for multiple values values of
the light and strange quark masses and for two different lattice spacings, the last stage of the analysis is to
extrapolate these results to the physical values of the quark masses and the continuum limit. For the heavy-light baryons,
we also remove the small effects of the finite volume.
We perform the extrapolations not directly for the baryon energies $E_X$, but rather for the ``subtracted'' energies
\begin{equation}
 E_X^{(\rm sub)} = E_X - \frac{n_c}{2} \overline{E}_{c\bar{c}}  - \frac{n_b}{2} \overline{E}_{b\bar{b}} \, , \label{eq:Esub}
\end{equation}
where $n_c$, $n_b$ are the numbers of charm and bottom quarks in the baryon, and $\overline{E}_{c\bar{c}}$, $\overline{E}_{b\bar{b}}$
are the spin-averaged charmonium and bottomonium energies, defined as
\begin{equation}
\overline{E}_{c\bar{c}} = \frac34 E_{J/\psi} + \frac 14 E_{\eta_c}, \hspace{4ex}\overline{E}_{b\bar{b}} = \frac34 E_{\Upsilon} + \frac 14 E_{\eta_b}.
\end{equation}
After extrapolating $E_X^{(\rm sub)}$ to the physical point, the full baryon energies can then be obtained simply by adding the
experimental values of $\frac{n_c}{2} \overline{E}_{c\bar{c}} + \frac{n_b}{2} \overline{E}_{b\bar{b}}$, which are known with high precision \cite{Beringer:1900zz},
to the results. The main reasons for using $E_X^{(\rm sub)}$ rather than $E_X$ are the following: the NRQCD energy shift cancels (this is relevant
only for baryons containing $b$ quarks), the leading dependence on the heavy-quark masses cancels, and the contribution of the uncertainty in the lattice spacing
is reduced.

We started by computing $[aE_X^{(\rm sub)}]_i$, where $i$ labels the data set, for the bootstrap ensembles of the method-averaged energies
(obtained as described in Sec.~\ref{sec:twopt}). For the chiral/continuum extrapolation fits, we need the values $[E_X^{(\rm sub)}]_i$ in physical units,
and we also need the covariances ${\rm Cov}\left([E_X^{(\rm sub)}]_i, [E_Y^{(\rm sub)}]_j\right)$ between the different baryon energies (these are nonzero only
if the data sets $i$ and $j$ correspond to the same ensemble of gauge configurations; e.g. $i=\mathtt{F23}$, $j=\mathtt{F43}$). To convert to physical units, we
used the inverse lattice spacings $[a^{-1}]_i$ determined in Ref.~\cite{Meinel:2010pv} from the bottomonium $2S-1S$ energy splittings. The covariances were then computed
as follows:
\begin{eqnarray}
 \nonumber {\rm Cov}\left([E_X^{(\rm sub)}]_i,\, [E_Y^{(\rm sub)}]_j\right)
 &=& [a^{-1}]^2 \, S([aE_X^{(\rm sub)}]_i)\, S([aE_Y^{(\rm sub)}]_j) \:\, {\rm Cov}\left([aE_X^{(\rm sub)}]_i,\, [aE_Y^{(\rm sub)}]_j\right) \\
 && + \left( \delta [a^{-1}] \right)^2 \,[aE_X^{(\rm sub)}]_i \, [aE_Y^{(\rm sub)}]_j \, ,
\end{eqnarray}
where $S([aE_X^{(\rm sub)}]_i)\geq1$, $S([aE_Y^{(\rm sub)}]_j)\geq1$ are the scale factors associated with the method-averages (see Sec.~\ref{sec:twopt}),
${\rm Cov}\left([aE_X^{(\rm sub)}]_i,\, [aE_Y^{(\rm sub)}]_j\right)$ is the covariance of the bootstrap ensembles
of the method-averaged energies in lattice units, and $\delta[a^{-1}]$ is the uncertainty of the lattice spacing (above, we assume
that $i$ and $j$ correspond to the same ensemble of gauge fields, so that $a^{-1}_i=a^{-1}_j=a^{-1}$). The uncertainties of
the lattice spacings are around 1.5\% (see Table \ref{tab:params}), while the uncertainties of $[aE_X^{(\rm sub)}]_i$
range from approximately 0.3\% to 3\%.

To predict the dependence of the heavy-light baryon masses on the light-quark masses in a model-independent
way, we use heavy-hadron chiral perturbation theory \cite{Wise:1992hn, Burdman:1992gh, Yan:1992gz, Cho:1992cf, Hu:2005gf},
the low-energy effective field theory of heavy hadrons and pions that combines heavy-quark symmetry and chiral symmetry.
Because we utilized partial quenching with $am_{u,d}^{(\mathrm{val})}< am_{u,d}^{(\mathrm{sea})}$ for some of our data sets
(to reach lower pion masses without having to generate new ensembles of gauge fields), we need to use
partially quenched \cite{Bernard:1993sv, Sharpe:2000bc, Bernard:2013kwa} heavy hadron chiral perturbation theory
to fit the dependence on both $m_{u,d}^{(\mathrm{sea})}$ and $am_{u,d}^{(\mathrm{val})}$. Next-to-leading
order expressions for the masses of singly and doubly heavy baryons in partially quenched
heavy-hadron perturbation theory were derived in Refs.~\cite{Tiburzi:2004kd} and \cite{Mehen:2006}, respectively.
We use the two-flavor $SU(4|2)$ theory, which is expected to converge faster than the $SU(6|3)$ theory. For the strange
baryons, we start from the $SU(6|3)$ theory but integrate out mesons containing valence or sea strange quarks
to obtain $SU(4|2)$ expressions for the baryon masses the different strangeness sectors. We also allow for
analytic dependence on $am_{s}^{(\mathrm{val})}$.
For the singly heavy baryons, we generalized the expressions given in \cite{Tiburzi:2004kd} to include
the leading $1/m_Q$ corrections, which introduce nonzero hyperfine splittings. We also include the leading
finite-volume and lattice spacing effects in our fits. The following sections describe the fits in detail
for the different types of heavy baryons. The final results for the baryon masses at the physical pion mass
and in the continuum limit can be found in Sec.~\ref{sec:finalresults}, which also contains
a discussion of the systematic uncertainties.

\FloatBarrier
\subsection{Singly heavy baryons}
\FloatBarrier

In the following we denote the heavy quark by $Q=c,b$. The fits in the charm and bottom sectors are done independently,
but the expressions used are the same. We group the baryon states according to their strangeness, $S$,
and perform coupled fits of the lattice data within each group. For the $S=0$ states
$\{\Lambda_Q, \Sigma_Q, \Sigma_Q^*\}$, the fit functions have the form
\begin{eqnarray}
 E^{(\rm sub)}_{\Lambda_Q} &=& E^{(\rm sub, 0)} + d_{\pi}^{(\rm vv)} \frac{[m_\pi^{(\rm vv)}]^2}{4\pi f}
 + d_{\pi}^{(\rm ss)} \frac{[m_\pi^{(\rm ss)}]^2}{4\pi f} + \mathcal{M}_{\Lambda_Q} + d_a\:a^2 \Lambda^3 \, , \\
 E^{(\rm sub)}_{\Sigma_Q} &=& E^{(\rm sub, 0)} + \Delta^{(0)} + c_{\pi}^{(\rm vv)} \frac{[m_\pi^{(\rm vv)}]^2}{4\pi f}
 + c_{\pi}^{(\rm ss)} \frac{[m_\pi^{(\rm ss)}]^2}{4\pi f} + \mathcal{M}_{\Sigma_Q} + c_a\:a^2 \Lambda^3 \, , \\
 E^{(\rm sub)}_{\Sigma^*_Q} &=& E^{(\rm sub, 0)} + \Delta^{(0)} + \Delta_*^{(0)}  + c_{\pi}^{(\rm vv)} \frac{[m_\pi^{(\rm vv)}]^2}{4\pi f}
 + c_{\pi}^{(\rm ss)} \frac{[m_\pi^{(\rm ss)}]^2}{4\pi f} + \mathcal{M}_{\Sigma^*_Q} + c_a\:a^2 \Lambda^3 \, ,
\end{eqnarray}
where $\mathcal{M}_{\Lambda_Q}$, $\mathcal{M}_{\Sigma_Q}$, and $\mathcal{M}_{\Sigma^*_Q}$ are the nonanalytic loop corrections \cite{Tiburzi:2004kd},
generalized here to include a nonzero hyperfine splitting $\Delta_*$ between the $\Sigma^*_Q$ and $\Sigma_Q$ baryons:
\begin{eqnarray}
 \mathcal{M}_{\Lambda_Q} &=& -\frac{g_3^2}{12\pi^2 f^2}\Bigg[  2 \mathcal{F}(m_\pi^{(\rm vs)},\Delta+\Delta_*,\mu )+ \mathcal{F}(m_\pi^{(\rm vs)},\Delta,\mu )+
   \mathcal{F}(m_\pi^{(\rm vv)},\Delta+\Delta_*,\mu )+\frac{1}{2}\mathcal{F}(m_\pi^{(\rm vv)},\Delta,\mu )\Bigg]\, , \label{eq:ELambdaQ} \\
 \mathcal{M}_{\Sigma_Q} &=& \frac{g_2^2}{12\pi^2 f^2}\Bigg[ \frac23  \mathcal{F}(m_\pi^{(\rm vs)},\Delta_*,\mu )+ \frac13 \mathcal{F}(m_\pi^{(\rm vs)},0,\mu ) \Bigg]
 + \frac{g_3^2}{12\pi^2 f^2}\Bigg[ -  \mathcal{F}(m_\pi^{(\rm vs)},-\Delta,\mu )+ \frac12 \mathcal{F}(m_\pi^{(\rm vv)},-\Delta,\mu )\Bigg] \, , \label{eq:ESigmaQ} \\
\nonumber \mathcal{M}_{\Sigma^*_Q} &=& \frac{g_2^2}{12\pi^2 f^2}\Bigg[  \frac{1}{6}\mathcal{F}(m_\pi^{(\rm vs)},-\Delta_*,\mu )+\frac{5}{6} \mathcal{F}(m_\pi^{(\rm vs)},0,\mu ) \Bigg] \\
&& +\frac{g_3^2}{12\pi^2 f^2}\Bigg[-\mathcal{F}(m_\pi^{(\rm vs)},-\Delta-\Delta_*,\mu )+\frac12 \mathcal{F}(m_\pi^{(\rm vv)},-\Delta-\Delta_*,\mu
   )\Bigg] \, \label{eq:ESigmaQstar} .
\end{eqnarray}
In the unitary case $m_\pi^{(\rm vv)}=m_\pi^{(\rm vs)}$, these expressions reduce to the expressions obtained previously in Ref.~\cite{Briceno:2012wt}.
The chiral function $\mathcal{F}$ includes finite-volume corrections and is defined in Appendix \ref{sec:F}. We did not treat the $\Sigma_Q-\Lambda_Q$ and $\Sigma^*_Q-\Sigma_Q$
splittings $\Delta$ and $\Delta_*$ used for the evaluation of the chiral functions as fit parameters. Instead, we used the results
of linear extrapolations to the chiral limit of the splittings determined for each data set (neglecting lattice-spacing dependence). The
values used are given in Table \ref{tab:DeltaDeltastarsinglyheavy} (for the very small $\Sigma_b^*-\Sigma_b$ hyperfine splitting, we used the average splitting
instead of the extrapolated splitting). The scheme ambiguity for choosing $\Delta$ and $\Delta^*$ in the evaluations of $\mathcal{F}$ only
affects the baryon masses at next-to-next-to-leading order, and is included in our estimates of the systematic uncertainties in Sec.~\ref{sec:finalresults}.
\begin{table}
\begin{tabular}{lcccccccc}
\hline\hline
                            & & $\{\Lambda_c, \Sigma_c, \Sigma^*_c\}$ & & $\{\Lambda_b, \Sigma_b, \Sigma^*_b\}$ & & $\{\Xi_c, \Xi^\prime_c, \Xi^*_c\}$ & & $\{\Xi_b, \Xi^\prime_b, \Xi^*_b\}$  \\
\hline
$\Delta$ (MeV)              & & 199(18)                               & & 253(20)                               & & 139(11)                            & & 155(16)                             \\
$\Delta_*$ (MeV)            & & 68(10)                                & & 22.7(4.8)                             & & 70.0(8.2)                          & & 28.8(3.1)                           \\
\hline\hline
\end{tabular}
\caption{\label{tab:DeltaDeltastarsinglyheavy}Values of $\Delta$ and $\Delta_*$ (in MeV) used in the evaluation of the chiral loop integrals for the singly heavy baryons.}
\end{table}
The chiral loop corrections also depend on the valence-valence pion masses $m_\pi^{(\rm vv)}$, which can be found in Table \ref{tab:params}, and on the valence-sea 
pion masses $m_\pi^{(\rm vs)}$, which we set equal to
\begin{equation}
 m_\pi^{(\rm vs)} = \sqrt{\frac{[m_\pi^{(\rm vv)}]^2 + [m_\pi^{(\rm ss)}]^2}{2}}. \label{eq:mpivs}
\end{equation}
The sea-sea pion masses $m_\pi^{(\rm ss)}$ in Eq.~(\ref{eq:mpivs}) can also be read off from Table \ref{tab:params} by taking the valence-valence pion masses
at $am_{u,d}^{(\mathrm{val})} = am_{u,d}^{(\mathrm{sea})}$. We chose the renormalization scale to be $\mu=4\pi f$, where $f$ is the pion decay constant,
\begin{equation}
 f = 132\:\:{\rm MeV}.
\end{equation}
The free parameters of the fit are $E^{(\rm sub, 0)}$, $\Delta^{(0)}$, $\Delta_*^{(0)}$, $d_{\pi}^{(\rm vv)}$, $d_{\pi}^{(\rm ss)}$, $d_a$,
$c_{\pi}^{(\rm vv)}$, $c_{\pi}^{(\rm ss)}$, and $c_a$. The ``$d$'' parameters describe the analytic quark-mass and lattice-spacing
dependence of the isosinget baryon $\Lambda_Q$, while the ``$c$'' parameters describe these dependencies for both isotriplet baryons $\Sigma_Q$ and $\Sigma_Q^*$,
as predicted by the chiral Lagrangian \cite{Tiburzi:2004kd}. The leading lattice-spacing dependence is quadratic because we used a chirally symmetric
domain-wall action for the light quarks and $\mathcal{O}(a)$-improved heavy-quark actions for the charm and bottom quarks (gluon discretization errors
also start at order $a^2$). To make the ``$c$'' and
``$d$'' parameters dimensionless, we introduced appropriate powers of $4\pi f$ and
\begin{equation}
 \Lambda = 500\:\:{\rm MeV}.
\end{equation}
Note that our inclusion of nonzero hyperfine splittings in the chiral loop corrections in principle requires higher-order analytic counterterms to cancel
the renormalization-scale dependence exactly. However, we find the renormalization-scale dependence in the absence of these terms to be sufficiently weak
(the changes in the extrapolated energies when replacing $\mu \mapsto 2\mu$ are well below the statistical uncertainties). In our analysis
of systematic uncertainties (see Sec.~\ref{sec:finalresults}) we consider the effect of including these higher-order counterterms with Bayesian constraints.

The axial couplings $g_2$ and $g_3$ in Eqs.~(\ref{eq:ELambdaQ}), (\ref{eq:ESigmaQ}), and (\ref{eq:ESigmaQstar}) are also fit parameters, but we constrained 
them with Gaussian priors to remain in the vicinity of the static-limit values calculated previously in lattice QCD \cite{Detmold:2011bp, Detmold:2012ge}.
To this end, we added the term
\begin{equation}
 \frac{\big[g_2 - g_2^{(0)}\big]^2}{\sigma_{g_2}^2} + \frac{\big[g_3 - g_3^{(0)}\big]^2}{\sigma_{g_3}^2}
\end{equation}
to the $\chi^2$ function of the fit. Here, $g_2^{(0)}=0.84$ and $g_3^{(0)}=0.71$ are the central values obtained in Refs.~\cite{Detmold:2011bp, Detmold:2012ge}.
The widths $\sigma_{g_2}$ and $\sigma_{g_3}$ were set by adding in quadrature to the uncertainties from \cite{Detmold:2011bp, Detmold:2012ge} an additional 10\% width (for $Q=b$)
or 30\% width (for $Q=c$) to account for $1/m_Q$ corrections.

\begin{figure}
\hspace{25pt}\includegraphics[height=18pt]{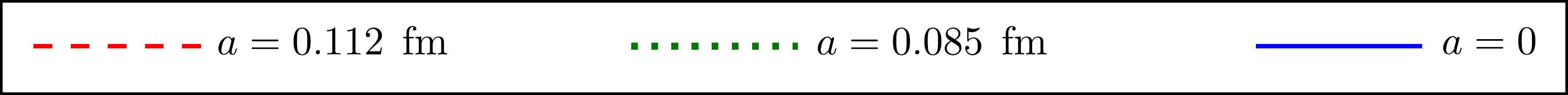}

\includegraphics[width=0.495\linewidth]{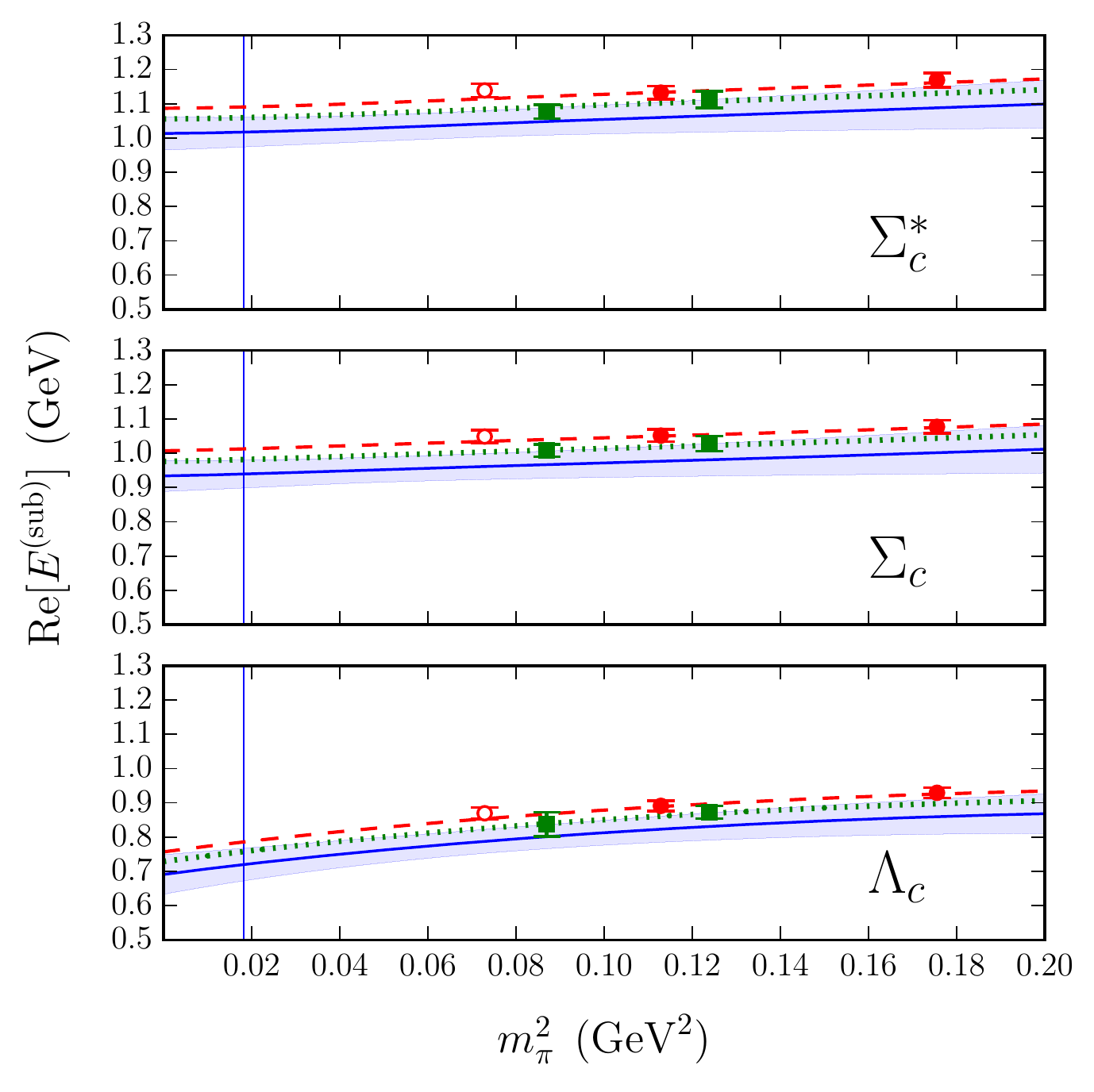} \hfill \includegraphics[width=0.495\linewidth]{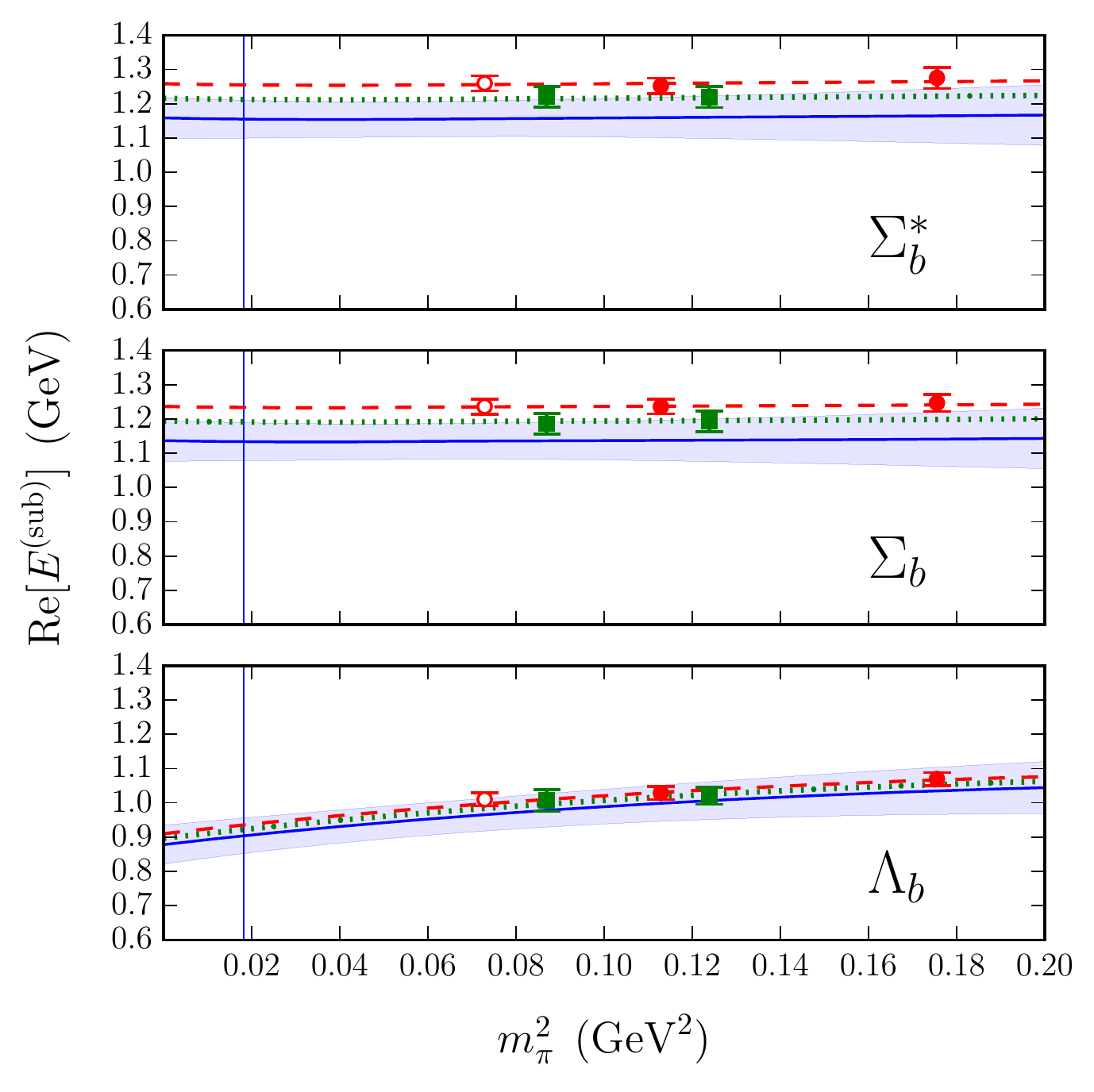}
\caption{\label{fig:fitLambda}Chiral and continuum extrapolations for the $\{\Lambda_Q, \Sigma_Q, \Sigma^*_Q\}$ baryons. The curves show the fit functions
in infinite volume at $m_\pi^{(\mathrm{vv})}=m_\pi^{(\mathrm{vs})}=m_\pi$, for the two different lattice spacing where we have data, and in the continuum limit.
For the continuum curves, the shaded bands indicate the $1\sigma$ uncertainty. The lattice data have been shifted to
infinite volume (see Table \protect\ref{tab:volumeshiftssinglyheavy} for the values of the shifts); data points at the coarse lattice spacing are plotted with circles,
and data points at the fine lattice spacing are plotted with squares. The partially quenched data points, which have
$m_\pi^{(\mathrm{vv})}<m_\pi^{(\mathrm{vs})}$, are included in the plot with open symbols at $m_\pi=m_\pi^{(\mathrm{vv})}$,
even though the fit functions actually have slightly different values for these points.
The data sets with the lowest two pion masses (\texttt{C14} and \texttt{F23}) are excluded here because our treatment of
finite-volume effects in HH$\chi$PT breaks down below the $\Sigma_Q^{(*)}\to\Lambda_Q\:\pi$ strong decay thresholds.
The vertical lines indicate the physical value of the pion mass.}
\end{figure}

\begin{figure}
\includegraphics[width=0.495\linewidth]{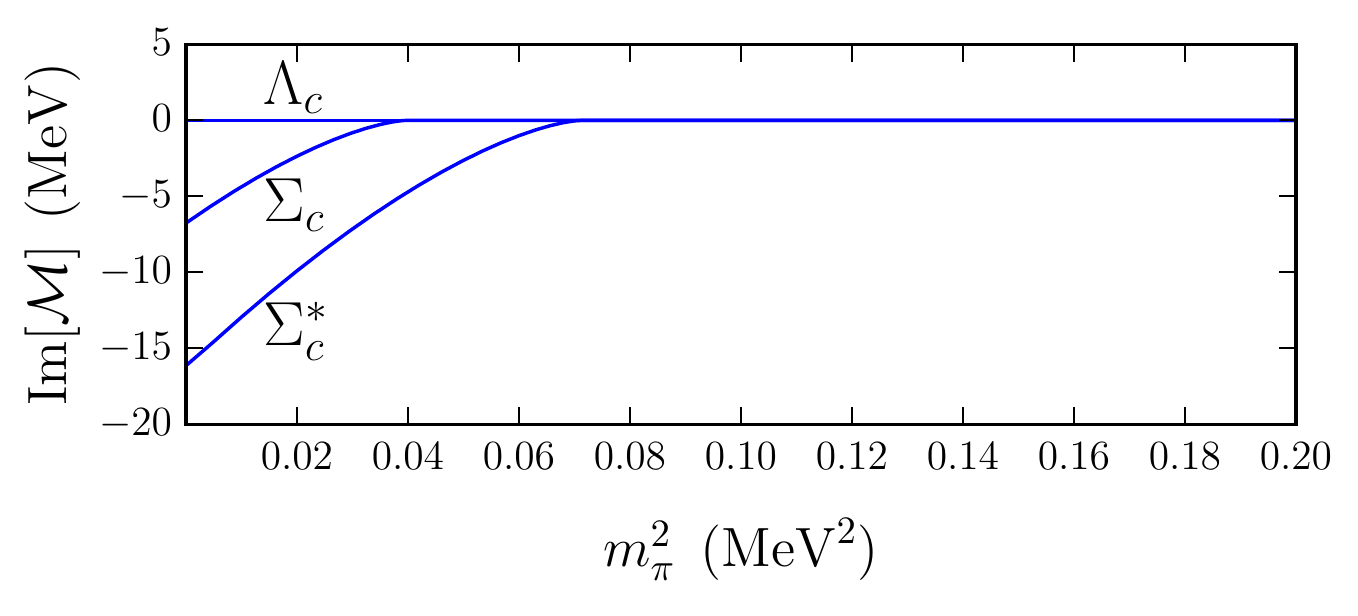} \hfill \includegraphics[width=0.495\linewidth]{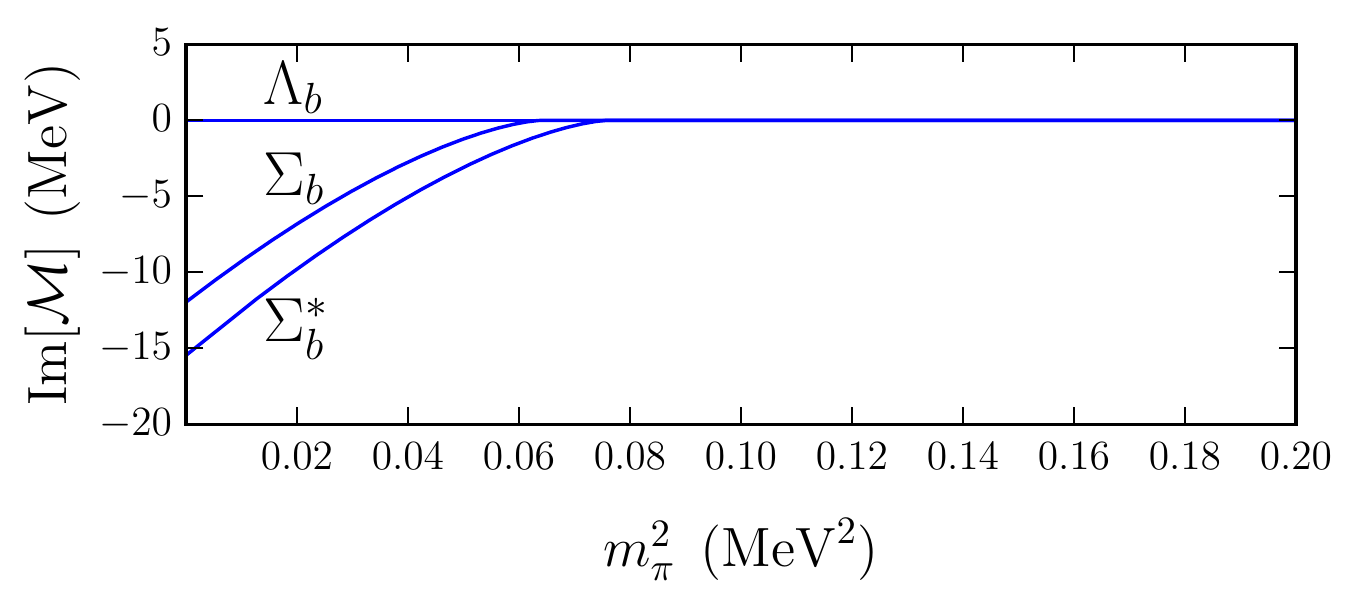}
\caption{\label{fig:fitLambdaimag}Imaginary parts of the $\{\Lambda_Q, \Sigma_Q, \Sigma^*_Q\}$ baryon energies in infinite volume, obtained from HH$\chi$PT. The imaginary
parts depend strongly on $\Delta$ and $\Delta+\Delta_*$; the values used here are given Table \ref{tab:DeltaDeltastarsinglyheavy}.}
\end{figure}

The resulting fit parameters for the chiral and continuum extrapolations of the $\{\Lambda_c, \Sigma_c, \Sigma_c^*\}$ and $\{\Lambda_b, \Sigma_b, \Sigma_b^*\}$ energies
are given in Table \ref{tab:fitLambda}; the covariance matrix of the fit parameters was obtained as 2 times the inverse of the Hessian of $\chi^2$.
The fits are illustrated in Fig.~\ref{fig:fitLambda}. Note that the chiral loop corrections $\mathcal{M}_{\Sigma_Q}$ and $\mathcal{M}_{\Sigma_Q^*}$
develop nonzero imaginary parts for $m_\pi < \Delta$ and $m_\pi < \Delta + \Delta_*$, respectively, as shown in Fig.~\ref{fig:fitLambdaimag}.
This is because at these pion masses, the strong decays $\Sigma_Q\to\Lambda_Q\:\pi$ and $\Sigma_Q^{*}\to\Lambda_Q\:\pi$ are kinematically allowed (in infinite volume),
and one can cut the loop diagram. The imaginary parts are related to the widths of these decays,
\begin{equation}
 \Gamma[\Sigma_Q^{(*)} \to \Lambda_Q\: \pi] = -2\: \mathrm{Im}[ \mathcal{M}_{\Sigma_Q^{(*)}} ] = \frac{g_3^2}{6\pi f^2} |\mathbf{p}_\pi|^3,
\end{equation}
where $|\mathbf{p}_\pi|$ is the magnitude of the pion momentum in the $\Sigma_Q^{(*)}$ rest frame. Note that, while the real parts of the
$\Sigma_Q$ and $\Sigma_Q^{(*)}$ energies depend only weakly on $\Delta$ and $\Delta+\Delta_*$, the imaginary parts
depend strongly on $\Delta$ and $\Delta+\Delta_*$ (therefore, to precisely calculate the decay widths one should use the experimental
values of these splittings \cite{Detmold:2011bp, Detmold:2012ge}).

This discussion of the $\Sigma_Q^{(*)} \to \Lambda_Q\: \pi$ decays is appropriate only in the chiral effective theory in infinite volume.
The lattice calculation itself yields eigenvalues of the QCD Hamiltonian in a finite volume, which are of course real-valued. In principle,
$\Lambda_Q\text{-}\pi$ scattering phase shifts (and hence the $\Sigma_Q^{(*)}$ resonance parameters) can be extracted from the finite-volume energy
energy spectrum using the L\"uscher method \cite{Luscher:1986pf, Luscher:1990ux}, but this is beyond the scope of this work.
Because of the momentum quantization in a finite box with periodic boundary conditions, the $\Lambda_Q\text{-}\pi$ $P$-wave states are expected to
have higher energy than the $\Sigma_Q^{(*)}$ states for all of our data sets. Nevertheless, we exclude the data sets with $m_\pi^{(\rm vv)}<\Delta + \Delta_*$
from the chiral extrapolation fits, because our treatment of finite-volume effects in the $\Sigma_Q^{(*)}$ energies using HH$\chi$PT (see Appendix \ref{sec:F})
breaks down below the strong-decay thresholds.

\begin{figure}
\hspace{25pt}\includegraphics[height=18pt]{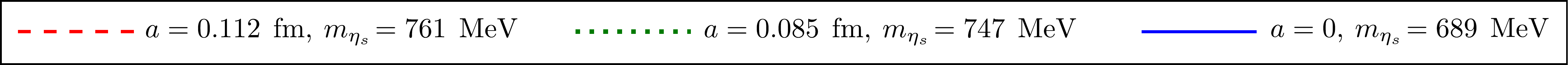}

\includegraphics[width=0.495\linewidth]{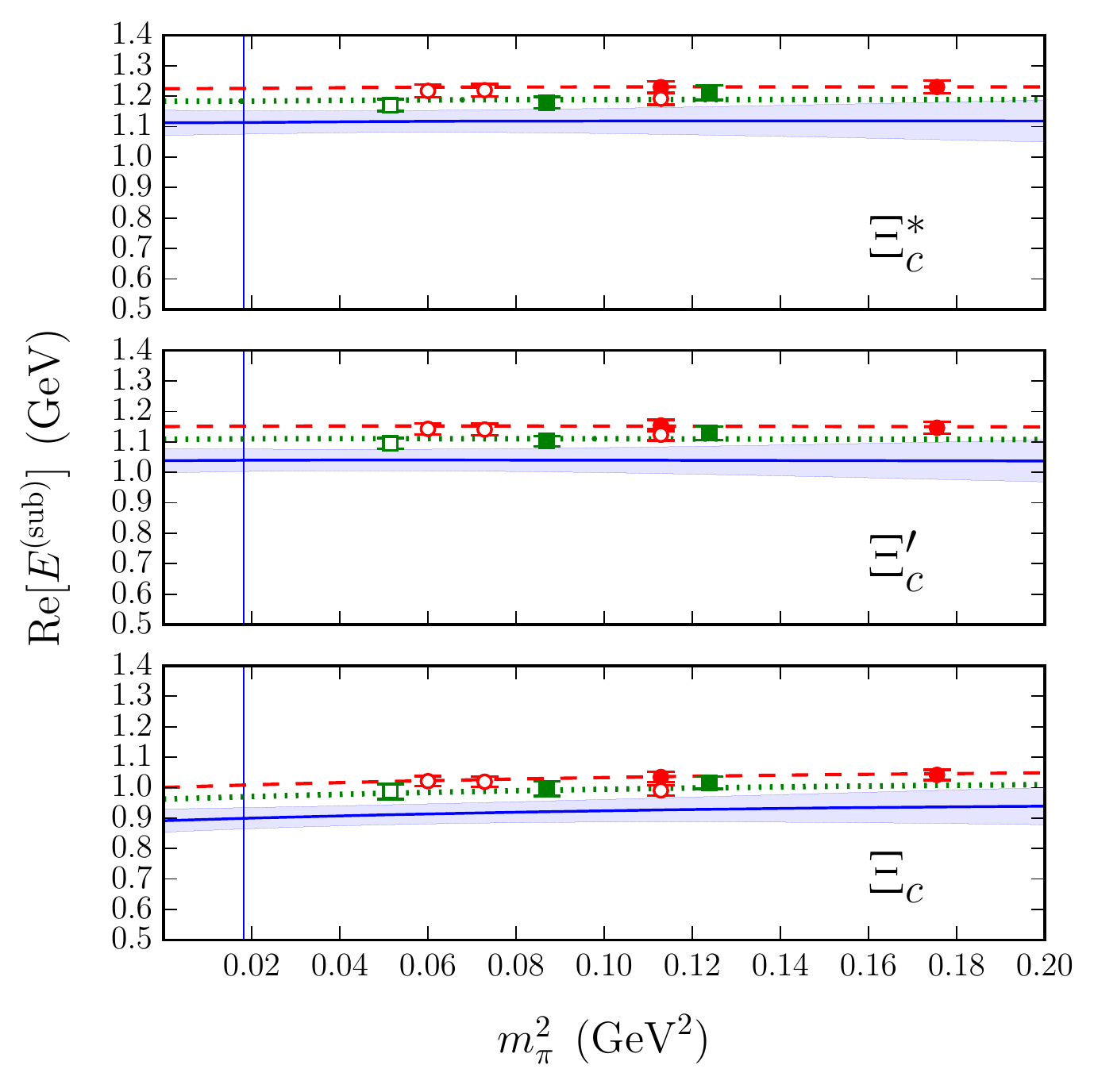} \hfill \includegraphics[width=0.495\linewidth]{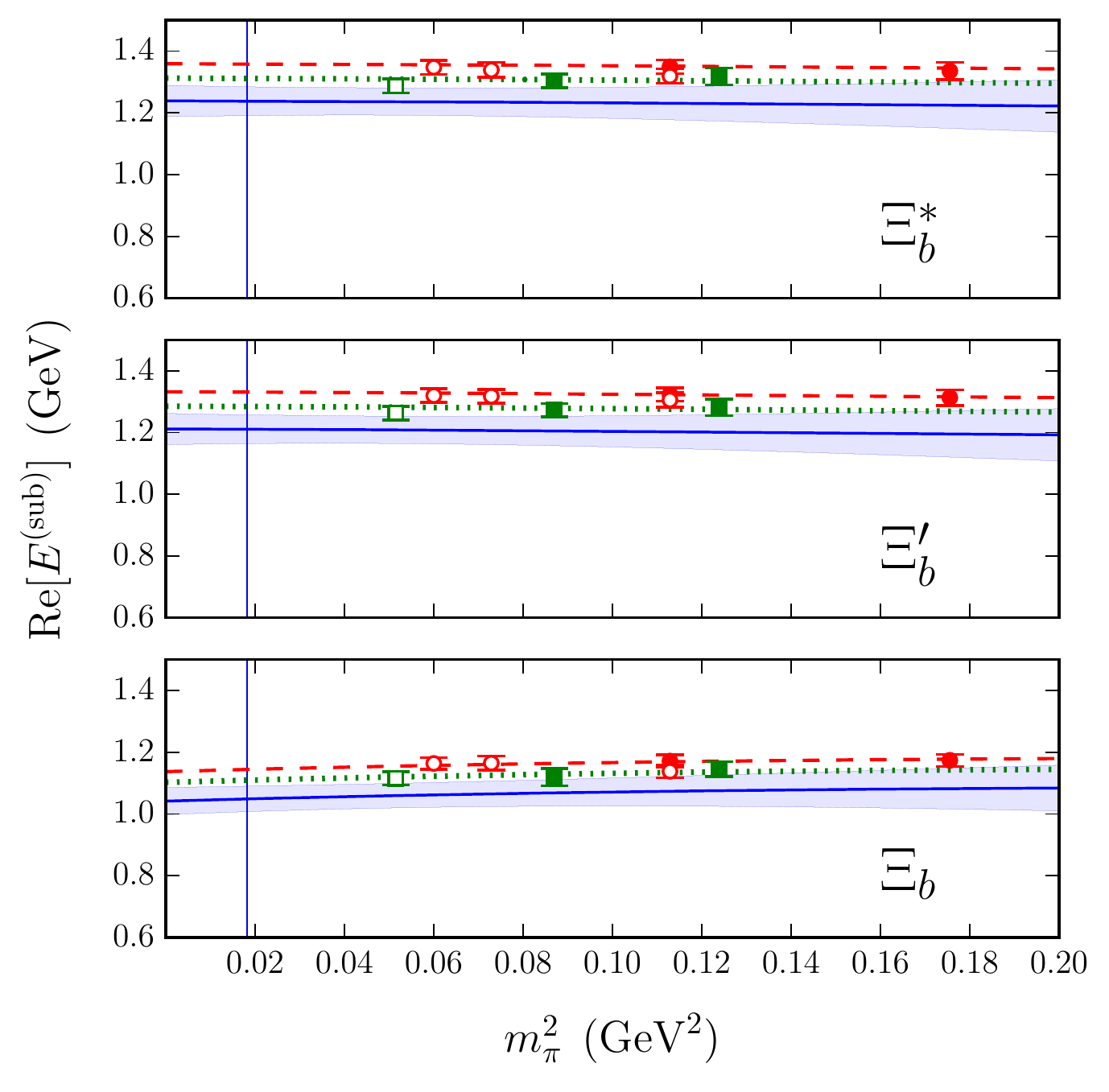}
\caption{\label{fig:fitXi} Chiral and continuum extrapolations for the $\{\Xi_Q, \Xi^\prime_Q, \Xi^*_Q\}$ baryons. The details are as explained in the caption of Fig.~\protect\ref{fig:fitLambda}, except that
now the curves also correspond to different values of $m_{\eta_s}^{(\rm vv)}$ as shown in the legend, and no data points are excluded here. Two of the data points
at the coarse lattice spacing have equal pion masses; these points are from the \texttt{C54} and \texttt{C53} data sets, which have different valence strange-quark masses.}
\end{figure}

For the $S=-1$ states $\{\Xi_Q, \Xi_Q^\prime, \Xi_Q^*\}$, the loop corrections from $SU(4|2)$ chiral perturbation theory read
\begin{eqnarray}
 \mathcal{M}_{\Xi_Q} &=& - \frac{g_3^2}{12\pi^2 f^2}\Bigg[  \mathcal{F}(m_\pi^{(\rm vs)},\Delta+\Delta_*,\mu )+\frac12\mathcal{F}(m_\pi^{(\rm vs)},\Delta,\mu )-\frac14
   \mathcal{F}(m_\pi^{(\rm vv)},\Delta+\Delta_*,\mu )-\frac18\mathcal{F}(m_\pi^{(\rm vv)},\Delta,\mu )\Bigg], \\
\nonumber \mathcal{M}_{\Xi^\prime_Q} &=& \frac{g_2^2}{12\pi^2 f^2}\Bigg[ \frac{1}{3}  \mathcal{F}(m_\pi^{(\rm vs)},\Delta_*,\mu )
+\frac{1}{6} \mathcal{F}(m_\pi^{(\rm vs)},0,\mu )-\frac{1}{12} 
   \mathcal{F}(m_\pi^{(\rm vv)},\Delta_*,\mu )- \frac{1}{24}\mathcal{F}(m_\pi^{(\rm vv)},0,\mu )\Bigg] \\
   && + \frac{g_3^2}{12\pi^2 f^2}\Bigg[-\frac{1}{2}  \mathcal{F}(m_\pi^{(\rm vs)},-\Delta,\mu )+\frac{1}{8}
    \mathcal{F}(m_\pi^{(\rm vv)},-\Delta,\mu ) \Bigg], \\
\nonumber \mathcal{M}_{\Xi^*_Q} &=& \frac{g_2^2}{12\pi^2 f^2}\Bigg[ \frac{1}{12}  \mathcal{F}(m_\pi^{(\rm vs)},-\Delta_*,\mu )+\frac{5}{12} \mathcal{F}(m_\pi^{(\rm vs)},0,\mu )-
   \frac{1}{48}\mathcal{F}(m_\pi^{(\rm vv)},-\Delta_*,\mu )-\frac{5}{48}  \mathcal{F}(m_\pi^{(\rm vv)},0,\mu )\Bigg]\\
  && + \frac{g_3^2}{12\pi^2 f^2}\Bigg[-\frac{1}{2}  \mathcal{F}(m_\pi^{(\rm vs)},-\Delta-\Delta_*,\mu )+\frac{1}{8}  \mathcal{F}(m_\pi^{(\rm vv)},-\Delta-\Delta_*,\mu ) \Bigg].
\end{eqnarray}
In this case, we also perform an interpolation of the valence strange-quark mass to its physical value, so that the fit functions become
\begin{eqnarray}
 E^{(\rm sub)}_{\Xi_Q} &=& E^{(\rm sub, 0)} + d_{\pi}^{(\rm vv)} \frac{[m_\pi^{(\rm vv)}]^2}{4\pi f}
 + d_{{\eta_s}}^{(\rm vv)} \frac{[m_{\eta_s}^{(\rm vv)}]^2-[m_{\eta_s}^{(\rm phys)}]^2}{4\pi f}
 + d_{\pi}^{(\rm ss)} \frac{[m_\pi^{(\rm ss)}]^2}{4\pi f} + \mathcal{M}_{\Xi_Q} + d_a\:a^2 \Lambda^3 , \\
 E^{(\rm sub)}_{\Xi^\prime_Q} &=& E^{(\rm sub, 0)} + \Delta^{(0)} + c_{\pi}^{(\rm vv)} \frac{[m_\pi^{(\rm vv)}]^2}{4\pi f}
 + c_{{\eta_s}}^{(\rm vv)} \frac{[m_{\eta_s}^{(\rm vv)}]^2-[m_{\eta_s}^{(\rm phys)}]^2}{4\pi f}
 + c_{\pi}^{(\rm ss)} \frac{[m_\pi^{(\rm ss)}]^2}{4\pi f} + \mathcal{M}_{\Xi^\prime_Q} + c_a\:a^2 \Lambda^3  , \\
 E^{(\rm sub)}_{\Xi^*_Q} &=& E^{(\rm sub, 0)} + \Delta^{(0)} + \Delta_*^{(0)} + c_{\pi}^{(\rm vv)} \frac{[m_\pi^{(\rm vv)}]^2}{4\pi f}
 + c_{{\eta_s}}^{(\rm vv)} \frac{[m_{\eta_s}^{(\rm vv)}]^2-[m_{\eta_s}^{(\rm phys)}]^2}{4\pi f}
 + c_{\pi}^{(\rm ss)} \frac{[m_\pi^{(\rm ss)}]^2}{4\pi f} + \mathcal{M}_{\Xi^*_Q} + c_a\:a^2 \Lambda^3 , \hspace{1ex}
\end{eqnarray}
with the two additional parameters $d_{{\eta_s}}^{(\rm vv)}$ and $c_{{\eta_s}}^{(\rm vv)}$. As already discussed in Sec.~\ref{sec:lightquarks}, we use the
square of the ``$\eta_s$'' pseudoscalar meson mass as a proxy for the strange-quark mass. The $\eta_s$ meson is defined by treating the $s$ and $\bar{s}$
as different, but degenerate flavors, so that the meson becomes stable and no disconnected quark contractions arise in the lattice calculation of the two-point function.
At the physical value of the strange-quark mass, one has $m_{\eta_s}^{(\rm phys)}=689.3(1.2)\:\:{\rm MeV}$ \cite{Dowdall:2011wh}.
The fit parameters obtained for the $\{\Xi_Q, \Xi_Q^\prime, \Xi_Q^*\}$ systems are given in the last two columns of Table \ref{tab:fitLambda}, and plots
of the fits are shown in Fig.~\ref{fig:fitXi}. In this case, all data sets were included in the fit, because all of them satisfy $m_\pi^{(\rm vv)} > \Delta+\Delta_*$
(see Table \ref{tab:DeltaDeltastarsinglyheavy} for the values of $\Delta$ and $\Delta_*$).

\begin{table}
\begin{tabular}{lrrrrrrrr}
\hline\hline
                            & & $\{\Lambda_c, \Sigma_c, \Sigma^*_c\}$ & & $\{\Lambda_b, \Sigma_b, \Sigma^*_b\}$ & & $\{\Xi_c, \Xi^\prime_c, \Xi^*_c\}$ & & $\{\Xi_b, \Xi^\prime_b, \Xi^*_b\}$  \\
\hline
 $E^{(\rm sub,0)}$ (MeV)    \hspace{1ex} & &  691(58)    \hspace{1ex} & &  878(57)    \hspace{1ex} & & 891(38)     \hspace{1ex} & & 1042(44)  \hspace{1ex}   \\
 \\[-3ex]
 $\Delta^{(0)}$ (MeV)       \hspace{1ex} & &  243(45)    \hspace{1ex} & &  259(51)    \hspace{1ex} & & 147(18)     \hspace{1ex} & & 170(32)   \hspace{1ex}   \\
 \\[-3ex]
 $\Delta_*^{(0)}$ (MeV)     \hspace{1ex} & &  79.3(8.7)  \hspace{1ex} & &  21.8(5.2)  \hspace{1ex} & & 74.2(5.4)   \hspace{1ex} & & 27.2(3.3) \hspace{1ex}   \\
 \\[-3ex]
 $d_{\pi}^{(\rm vv)}$       \hspace{1ex} & &  2.9(1.3)   \hspace{1ex} & &  2.56(81)   \hspace{1ex} & & 0.81(21)    \hspace{1ex} & & 0.55(20)  \hspace{1ex}   \\
 \\[-3ex]
 $d_{\eta_s}^{(\rm vv)}$    \hspace{1ex} & &  $\hdots$   \hspace{1ex} & & $\hdots$    \hspace{1ex} & & 0.566(69)   \hspace{1ex} & & 0.41(12)  \hspace{1ex}   \\
 \\[-3ex]
 $d_{\pi}^{(\rm ss)}$       \hspace{1ex} & &  1.5(1.1)   \hspace{1ex} & &  1.35(95)   \hspace{1ex} & & 0.34(72)    \hspace{1ex} & & 0.50(75)  \hspace{1ex}   \\
 \\[-3ex]
 $d_a$                      \hspace{1ex} & &  1.6(1.2)   \hspace{1ex} & &  0.8(1.6)   \hspace{1ex} & & 1.8(1.2)    \hspace{1ex} & & 1.7(1.5)  \hspace{1ex}   \\
 \\[-3ex]
 $c_{\pi}^{(\rm vv)}$       \hspace{1ex} & & $-0.39(39)$ \hspace{1ex} & & $-0.47(35)$ \hspace{1ex} & & 0.36(12)    \hspace{1ex} & & 0.31(14)  \hspace{1ex}   \\
 \\[-3ex]
 $c_{\eta_s}^{(\rm vv)}$    \hspace{1ex} & & $\hdots$    \hspace{1ex} & &  $\hdots$   \hspace{1ex} & & 0.417(58)   \hspace{1ex} & & 0.32(10)  \hspace{1ex}   \\
 \\[-3ex]
 $c_{\pi}^{(\rm ss)}$       \hspace{1ex} & & 0.36(81)    \hspace{1ex} & & $-0.26(92)$ \hspace{1ex} & & $-0.52(68)$ \hspace{1ex} & & $-0.65(82)$  \hspace{1ex}   \\
 \\[-3ex]
 $c_a$                      \hspace{1ex} & &  1.8(1.4)   \hspace{1ex} & & 2.5(1.8)    \hspace{1ex} & & 2.1(1.4)    \hspace{1ex} & & 2.5(1.7)   \hspace{1ex}  \\
 \\[-3ex]
 $g_2$                      \hspace{1ex} & &  0.82(32)   \hspace{1ex} & & 0.84(22)    \hspace{1ex} & & 0.82(32)    \hspace{1ex} & & 0.83(22)  \hspace{1ex}   \\
 \\[-3ex]
 $g_3$                      \hspace{1ex} & &  0.75(24)   \hspace{1ex} & & 0.70(15)    \hspace{1ex} & & 0.75(24)    \hspace{1ex} & & 0.72(15)  \hspace{1ex}   \\
\hline\hline
\end{tabular}
\caption{\label{tab:fitLambda}Chiral and continuum extrapolation fit parameters for the singly heavy baryons containing $u/d$ valence quarks.}
\end{table}

\begin{table}
\begin{tabular}{lrrrrrrrrrrrrrrrr}
\hline\hline
State             & & $\mathtt{C104}$ & & $\mathtt{C14}$ & & $\mathtt{C24}$ & & $\mathtt{C54}$ & & $\mathtt{C53}$ & & $\mathtt{F23}$ & & $\mathtt{F43}$ & & $\mathtt{F63}$  \\
\hline
 $\Lambda_c$      & &   0.6           & &  $\hdots$      & & 1.7            & &  1.2           & & $\hdots$       & & $\hdots$       & & 1.7            & & 1.0             \\
 \\[-3ex]
 $\Sigma_c$       & &   0.1           & &  $\hdots$      & & $-0.1$         & &  0.3           & & $\hdots$       & & $\hdots$       & & 0.5            & & 0.2             \\
 \\[-3ex]
 $\Sigma_c^*$     & &   0.2           & &  $\hdots$      & & $-1.5$         & &  0.9           & & $\hdots$       & & $\hdots$       & & 2.1            & & 0.7             \\
 \\[-3ex]
 $\Xi_c$          & &   0.2           & &  0.4           & & 0.4            & &  0.3           & & 0.3            & & 0.5            & & 0.5            & & 0.3             \\
 \\[-3ex]
 $\Xi_c'$         & &   0.1           & &  0.2           & & 0.2            & &  0.2           & & 0.2            & & 0.4            & & 0.3            & & 0.2             \\
 \\[-3ex]
 $\Xi_c^*$        & &   0.1           & &  0.4           & & 0.5            & &  0.4           & & 0.4            & & 0.7            & & 0.7            & & 0.3             \\
 \\[-3ex]
 $\Lambda_b$      & &   0.5           & &  $\hdots$      & & 1.4            & &  1.0           & & $\hdots$       & & $\hdots$       & & 1.3            & & 0.8             \\
 \\[-3ex]
 $\Sigma_b$       & &   0.1           & &  $\hdots$      & & $-0.6$         & &  0.5           & & $\hdots$       & & $\hdots$       & & 1.2            & & 0.4             \\
 \\[-3ex]
 $\Sigma_b^*$     & &   0.2           & &  $\hdots$      & & $-2.3$         & &  0.9           & & $\hdots$       & & $\hdots$       & & 2.2            & & 0.6             \\
 \\[-3ex]
 $\Xi_b$          & &   0.2           & &  0.4           & & 0.4            & &  0.3           & & 0.3            & & 0.5            & & 0.4            & & 0.3             \\
 \\[-3ex]
 $\Xi_b'$         & &   0.1           & &  0.2           & & 0.2            & &  0.2           & & 0.3            & & 0.3            & & 0.3            & & 0.2             \\
 \\[-3ex]
 $\Xi_b^*$        & &   0.1           & &  0.3           & & 0.3            & &  0.2           & & 0.2            & & 0.5            & & 0.4            & & 0.2             \\
 \\[-3ex]
\hline\hline
\end{tabular}
\caption{\label{tab:volumeshiftssinglyheavy}Finite-volume energy shifts $E(L)-E(\infty)$ (in MeV) for the singly heavy baryons containing $u/d$ valence quarks.}
\end{table}

\FloatBarrier

\begin{figure}
\hspace{25pt}\includegraphics[height=18pt]{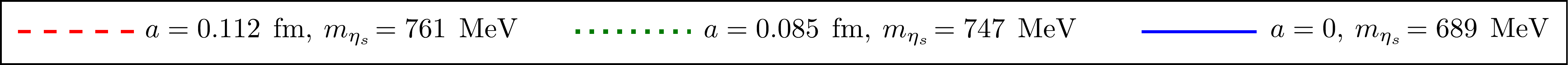}

\includegraphics[width=0.495\linewidth]{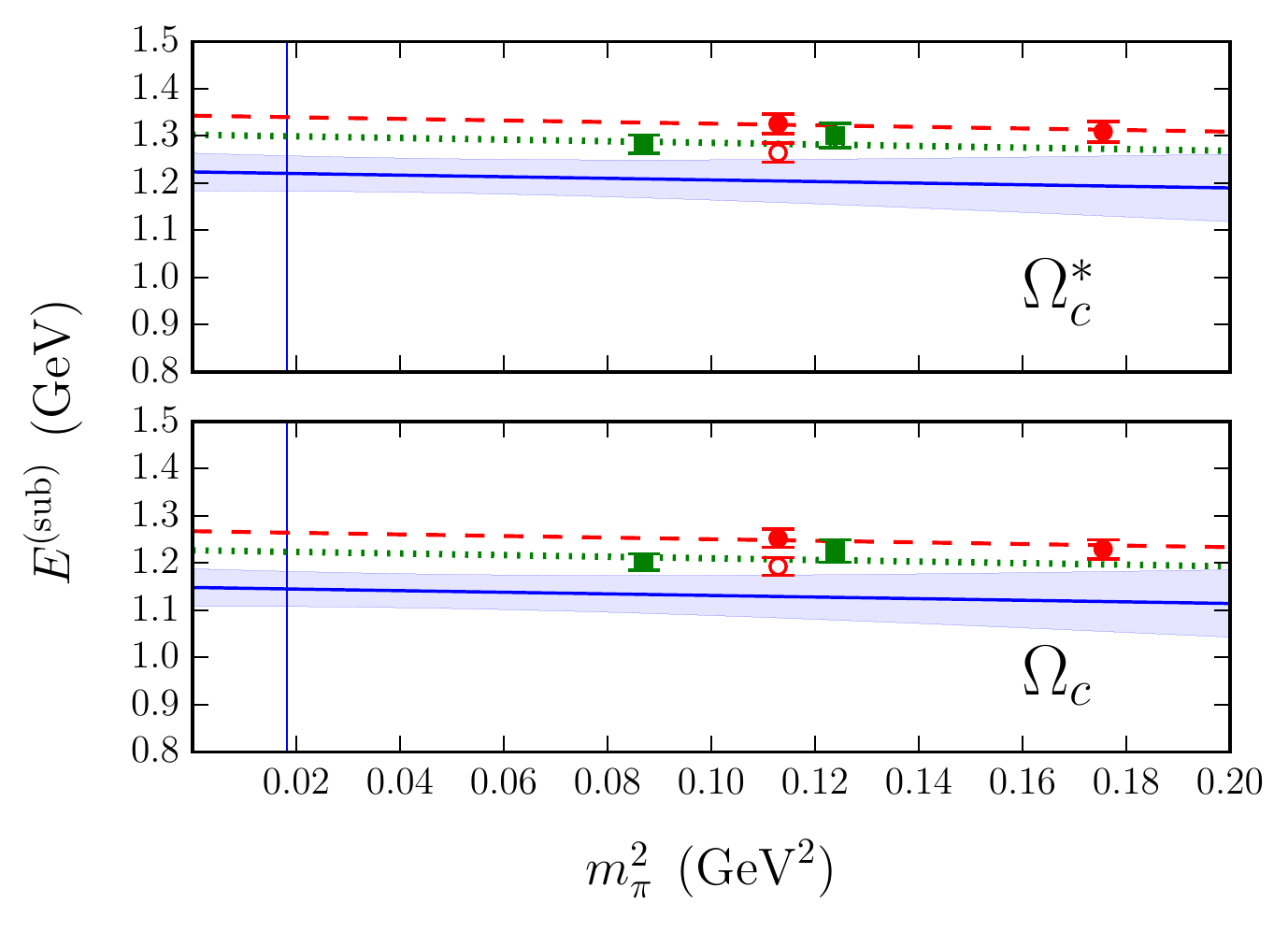} \hfill  \includegraphics[width=0.495\linewidth]{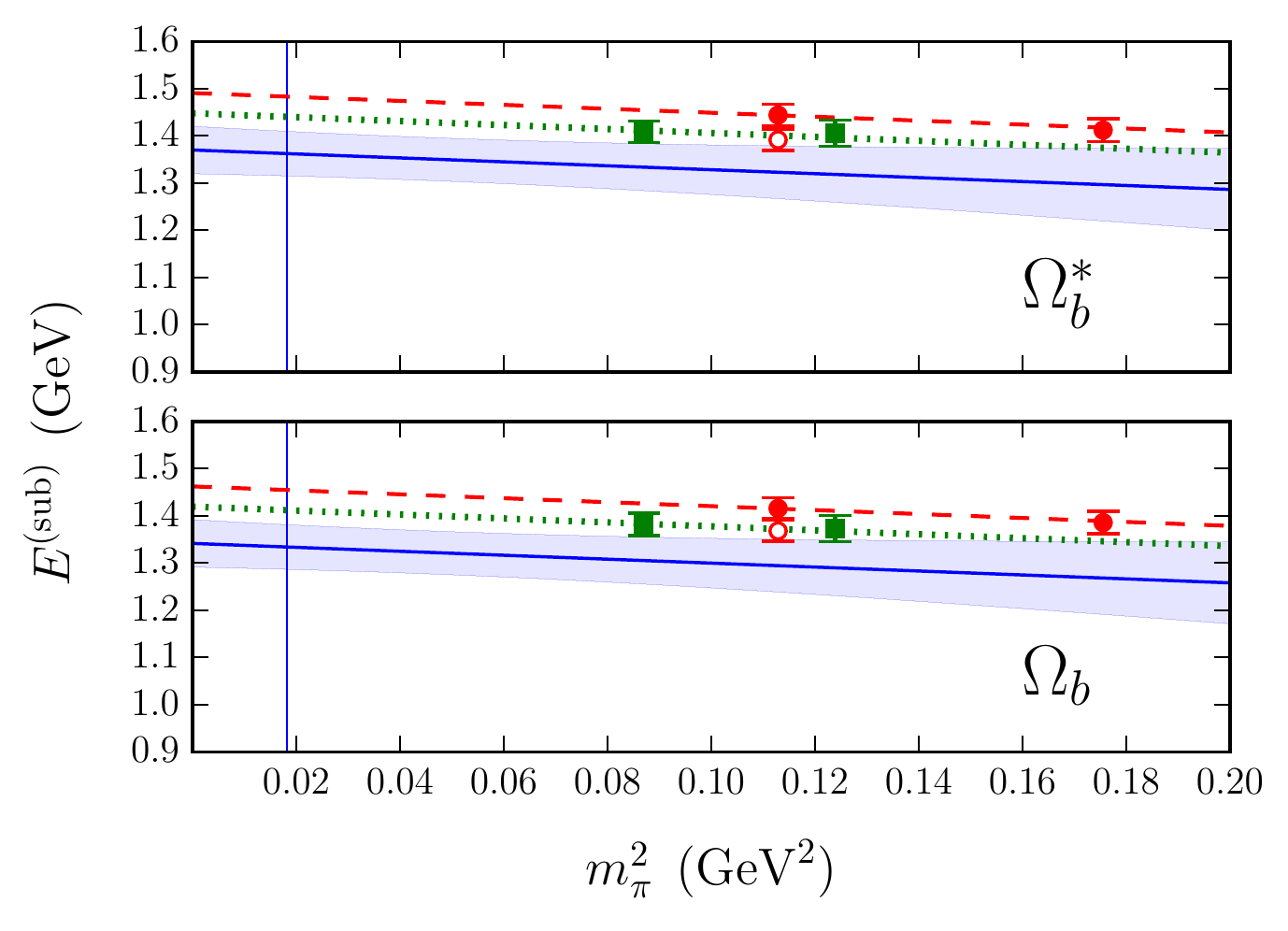}
\caption{\label{fig:Omegafit}Chiral and continuum extrapolations for the $\{\Omega_Q, \Omega^*_Q\}$ baryons. The curves show the fit functions
for the two different lattice spacing where we have data, and in the continuum limit, evaluated at appropriate values of $m_{\eta_s}^{(\rm vv)}$ as shown in the legend.
For the continuum curves, the shaded bands indicate the $1\sigma$ uncertainty. Data points at the coarse lattice spacing are plotted with circles,
and data points at the fine lattice spacing are plotted with squares. The open circles are from the \texttt{C53} data set with a lower-than-physical
valence strange-quark mass.}
\end{figure}

The $S=-2$ baryons $\{\Omega_Q, \Omega_Q^*\}$ do not contain light valence quarks and therefore do not receive any loop corrections at next-to-leading-order in $SU(4|2)$ HH$\chi$PT.
In this case, we still allow for a linear dependence on the light sea-quark mass, and, as before, interpolate linearly in the valence strange-quark mass. Thus,
the fit functions are
\begin{eqnarray}
 E^{(\rm sub)}_{\Omega_{Q}} &=& E^{(\rm sub, 0)} + c_{{\eta_s}}^{(\rm vv)} \frac{[m_{\eta_s}^{(\rm vv)}]^2-[m_{\eta_s}^{(\rm phys)}]^2}{4\pi f}
 + c_{\pi}^{(\rm ss)} \frac{[m_\pi^{(\rm ss)}]^2}{4\pi f} + c_a\:a^2 \Lambda^3, \\
 E^{(\rm sub)}_{\Omega^*_{Q}} &=& E^{(\rm sub, 0)} + \Delta_*^{(0)} + c_{{\eta_s}}^{(\rm vv)} \frac{[m_{\eta_s}^{(\rm vv)}]^2-[m_{\eta_s}^{(\rm phys)}]^2}{4\pi f}
 + c_{\pi}^{(\rm ss)} \frac{[m_\pi^{(\rm ss)}]^2}{4\pi f} + c_a\:a^2 \Lambda^3,
\end{eqnarray}
with parameters $E^{(\rm sub, 0)}$, $\Delta_*^{(0)}$, $c_{{\eta_s}}^{(\rm vv)}$, $c_{\pi}^{(\rm ss)}$, and $c_a$. The fit results are given in Table \ref{tab:Omegafit} and
are plotted in Fig.~\ref{fig:Omegafit}.

\begin{table}
\begin{tabular}{lrrrr}
\hline\hline
                            & & $\{\Omega_c, \Omega^*_c\}$ & & $\{\Omega_b, \Omega^*_b\}$  \\
\hline
 $E^{(\rm sub,0)}$ (MeV)    & & 1148(40)      & &   1342(50)      \\
 \\[-3ex]
 $\Delta_*^{(0)}$ (MeV)     & & 75.3(1.9)     & &   28.4(2.2)      \\
 \\[-3ex]
 $c_{\eta_s}^{(\rm vv)}$    & & 0.722(49)     & &   0.604(72)      \\
 \\[-3ex]
 $c_{\pi}^{(\rm ss)}$       & & $-0.28(65)$   & &  $-0.69(79)$       \\
 \\[-3ex]
 $c_a$                      & & 1.8(1.4)      & &  2.1(1.8)       \\
\hline\hline
\end{tabular}
\caption{\label{tab:Omegafit}Chiral and continuum extrapolation fit parameters for the $\Omega_Q$ and $\Omega^*_Q$ baryons.}
\end{table}

\FloatBarrier
\subsection{Doubly heavy baryons}
\FloatBarrier

Heavy quark-diquark symmetry relates the properties of doubly heavy baryons and heavy-light mesons \cite{Savage:1990di}; consequently,
both can be included in HH$\chi$PT in a single supermultiplet field, and their interaction strength with pions is given by the same
axial coupling, $g_1$ (in the heavy-quark limit) \cite{Hu:2005gf}. The masses of baryons with two heavy quarks of equal flavor
have been calculated to next-to-leading-order in partially quenched $SU(6|3)$ HH$\chi$PT in Ref.~\cite{Mehen:2006}. Here, we modify
these expressions for the $SU(4|2)$ case, and also extend them to the case of different-flavor heavy quarks.
In the case of equal-flavor heavy quarks, the Pauli exclusion principle implies that the two heavy quarks must form a spin-1 diquark
in the ground state ($S$-wave), which can then combine with the light degrees of freedom to form a hyperfine doublet with $J^P=\frac12^+,\frac32^+$;
these baryons are denoted as $\{ \Xi_{QQ}, \Xi_{QQ}^* \}$ for strangeness 0 and $\{\Omega_{QQ}, \Omega_{QQ}^* \}$ for strangeness $-1$ (here, $Q=c$ or $Q=b$).
In the case of two different heavy-quark flavors $Q=c$, $Q^\prime=b$, the two heavy quarks can form an $S$-wave diquark with either spin 1 or spin 0,
leading to three different states with $J^P=\{\frac12^+, \frac12^+, \frac32^+ \}$ in each strangeness sector: $\{ \Xi_{QQ^\prime}, \Xi_{QQ^\prime}^\prime, \Xi_{QQ^\prime}^* \}$ and
$\{ \Omega_{QQ^\prime}, \Omega_{QQ^\prime}^\prime, \Omega_{QQ^\prime}^* \}$, where the latter two of the three states contain a spin-1 heavy diquark (in the heavy-quark limit).

Let us consider the equal-heavy-flavor case with strangeness 0 first. We performed fits to the lattice data for the $\{ \Xi_{QQ}, \Xi_{QQ}^* \}$ doublets using the functions
\begin{eqnarray}
 E^{(\rm sub)}_{\Xi_{Q Q}} &=& E^{(\rm sub, 0)} + c_{\pi}^{(\rm vv)} \frac{[m_\pi^{(\rm vv)}]^2}{4\pi f} + c_{\pi}^{(\rm ss)} \frac{[m_\pi^{(\rm ss)}]^2}{4\pi f} + \mathcal{M}_{\Xi_{Q Q}} + c_a\:a^2 \Lambda^3, \\
 E^{(\rm sub)}_{\Xi^*_{Q Q}} &=& E^{(\rm sub, 0)} + \Delta_*^{(0)} + c_{\pi}^{(\rm vv)} \frac{[m_\pi^{(\rm vv)}]^2}{4\pi f} + c_{\pi}^{(\rm ss)} \frac{[m_\pi^{(\rm ss)}]^2}{4\pi f} + \mathcal{M}_{\Xi^*_{Q Q}} + c_a\:a^2 \Lambda^3,
\end{eqnarray}
where $\mathcal{M}_{\Xi_{Q Q}}$ and $\mathcal{M}_{\Xi^*_{Q Q}}$ are the nonanalytic loop corrections,
\begin{eqnarray}
 \mathcal{M}_{\Xi_{Q Q}} &=& -\frac{g_1^2}{16\pi^2 f^2}\Bigg[  \frac{32}{9} \mathcal{F}(m_\pi^{(\rm vs)},\Delta_*,\mu )+\frac{4}{9} \mathcal{F}(m_\pi^{(\rm vs)},0,\mu )
 -\frac{8}{9} \mathcal{F}(m_\pi^{(\rm vv)},\Delta_*,\mu )-\frac{1}{9}\mathcal{F}(m_\pi^{(\rm vv)},0,\mu ) \Bigg],\\
 \mathcal{M}_{\Xi^*_{Q Q}} &=& -\frac{g_1^2}{16\pi^2 f^2}\Bigg[  \frac{16}{9} \mathcal{F}(m_\pi^{(\rm vs)},-\Delta_*,\mu )+\frac{20}{9} \mathcal{F}(m_\pi^{(\rm vs)},0,\mu )
 -\frac{4}{9} \mathcal{F}(m_\pi^{(\rm vv)},-\Delta_*,\mu )-\frac{5}{9}   \mathcal{F}(m_\pi^{(\rm vv)},0,\mu ) \Bigg].
\end{eqnarray}
The unconstrained fit parameters are $E^{(\rm sub, 0)}$, $\Delta_*^{(0)}$, $c_{\pi}^{(\rm vv)}$, $c_{\pi}^{(\rm ss)}$, and $c_a$. In our scheme for the function $\mathcal{F}$
(see Appendix \ref{sec:F}), the parameter $\Delta_*^{(0)}$ is equal to the $\Xi^*_{Q Q}-\Xi_{Q Q}$ hyperfine splitting in the chiral limit. It is related to the heavy-meson hyperfine splitting parameter
$\Delta_H^{(0)}$ by $\Delta_*^{(0)} = \frac 34 \Delta_H^{(0)}$ \cite{Mehen:2006}.
We constrained the axial coupling $g_1$ by adding the term
\begin{equation}
 \frac{\big[g_1 - g_1^{(0)}\big]^2}{\sigma_{g_1}^2}
\end{equation}
to the $\chi^2$ function of the fit. Here, $g_1^{(0)}=0.449$ is the central value of the static-limit axial coupling calculated using lattice QCD in Refs.~\cite{Detmold:2011bp, Detmold:2012ge},
and the width $\sigma_{g_1}$ was set by adding in quadrature to the uncertainty from \cite{Detmold:2011bp, Detmold:2012ge} an additional 20\% width (for $Q=b$)
or 60\% width (for $Q=c$) to account for $1/m_Q$ corrections (chosen twice as large as for singly heavy baryons because of the additional breaking of heavy quark-diquark symmetry).

As in the case of the singly-heavy baryons,
we distinguish the splitting $\Delta_*$ used in the evaluation of the chiral loop corrections from the fit parameter $\Delta_*^{(0)}$.  We determined $\Delta_*$ 
prior to the main fit by linearly extrapolating the lattice results for this splitting to the chiral limit (for $Q=c$) or by taking the average over all data sets
(in the case $Q=b$, where the splitting is smaller and has a larger relative statistical uncertainty). The values of $\Delta_*$ are given in Table \ref{tab:Deltastardoublyheavy}.

\begin{figure}
\hspace{25pt}\includegraphics[height=18pt]{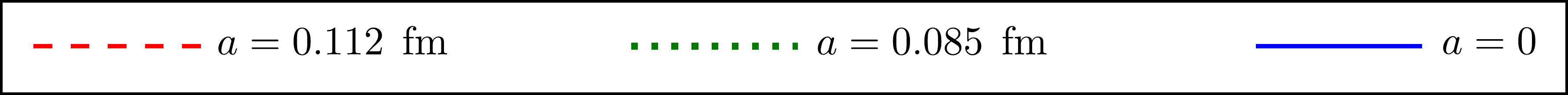}

\includegraphics[width=0.495\linewidth]{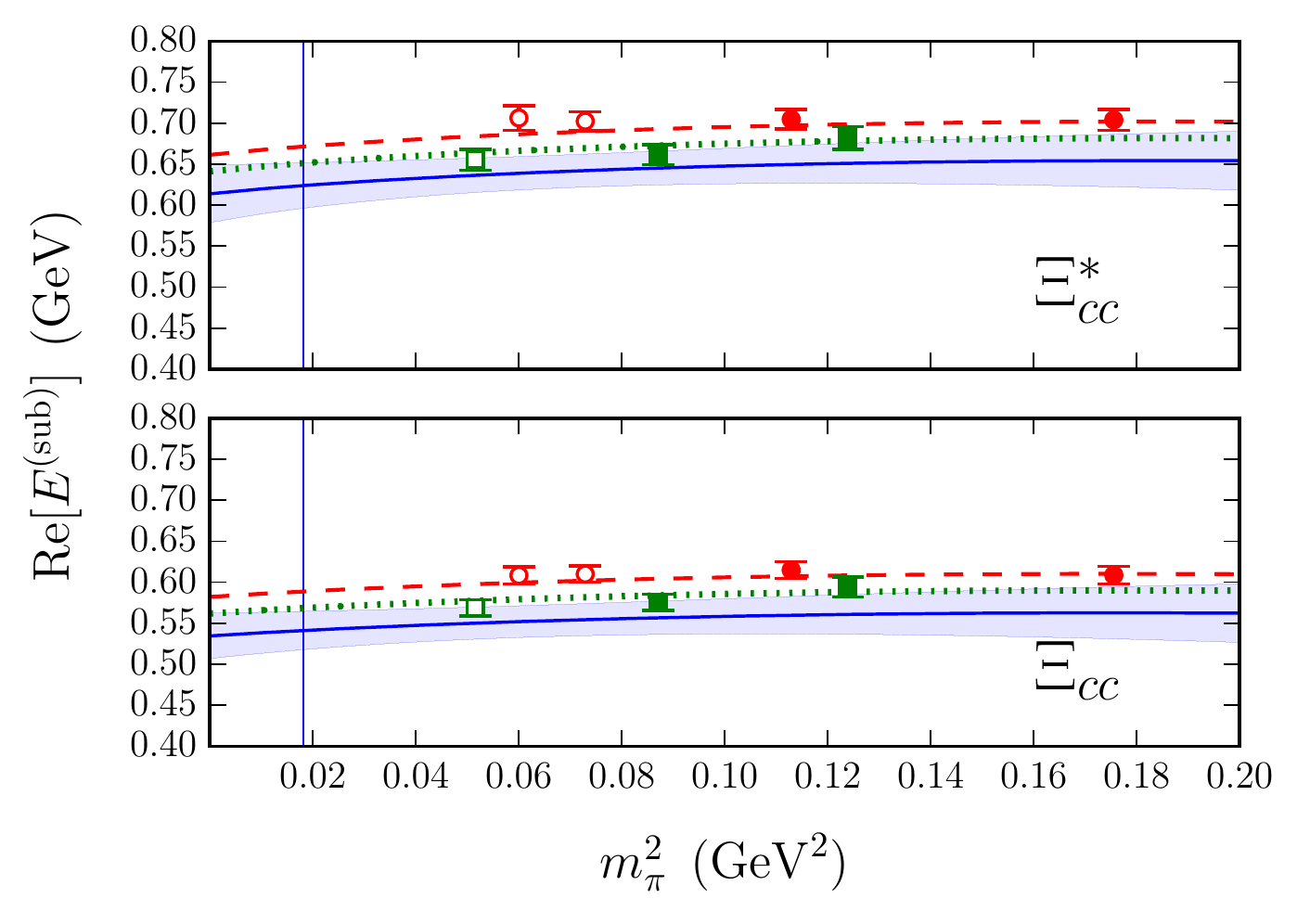}  \hfill \includegraphics[width=0.495\linewidth]{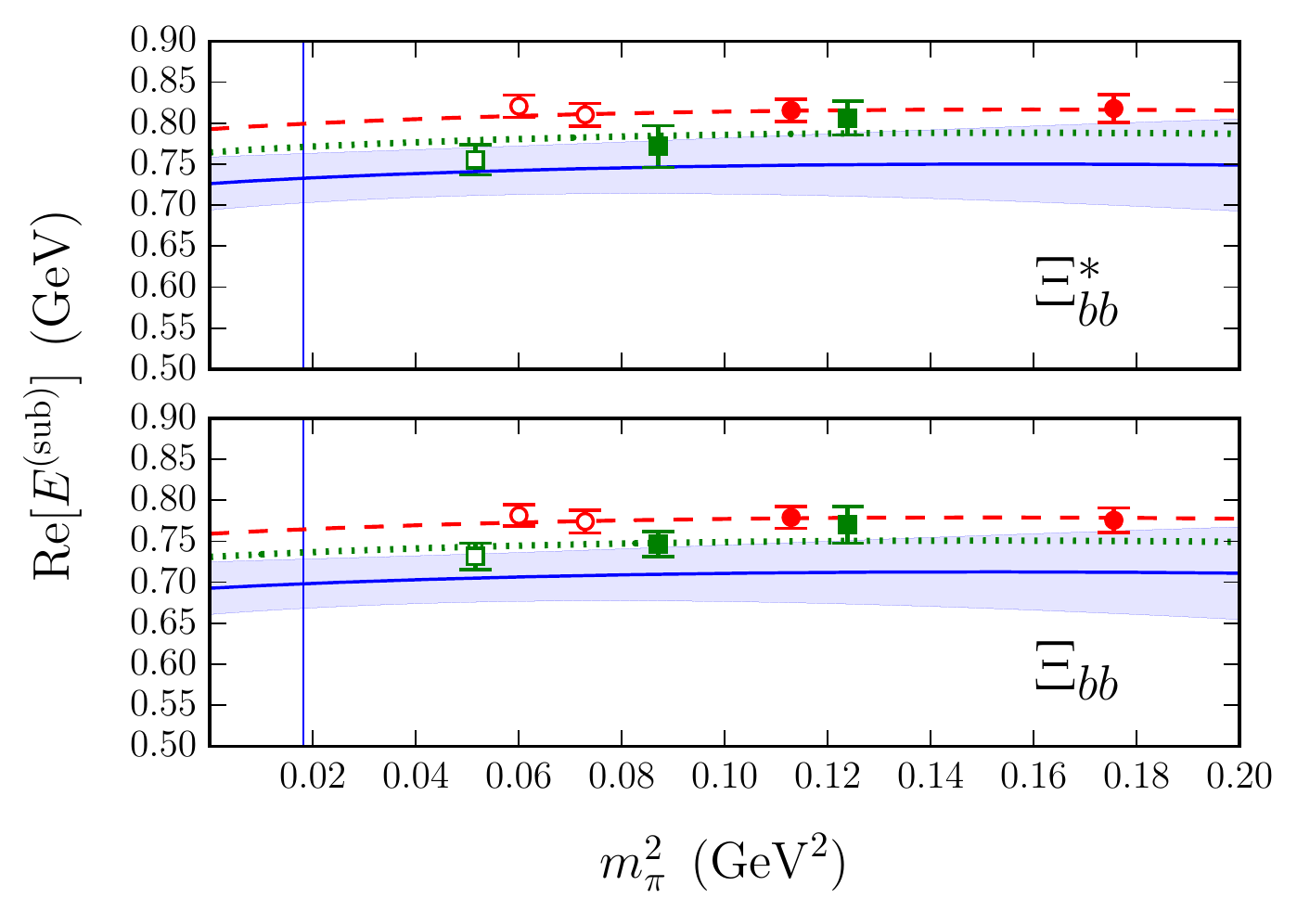}
\caption{\label{fig:doublyfeavyfit}Chiral and continuum extrapolations for the $\{\Xi_{QQ}, \Xi^*_{QQ}\}$ baryons. The details of the plots are as explained in the caption of
Fig.~\protect\ref{fig:fitLambda}, except that no data sets are excluded here.}
\end{figure}

The resulting parameters from the main fits to the $\{ \Xi_{QQ}, \Xi_{QQ}^* \}$ data are given in Table \ref{tab:doublyfeavyfit}, and plots of the fits are shown in Fig.~\ref{fig:doublyfeavyfit}.
The plots show the fit functions evaluated in infinite volume; the data points have also been shifted to infinite volume (see Table \ref{tab:volumeshiftsdoublyheavy} for the values of the shifts).
The final results for the baryon masses can be found in Sec.~\ref{sec:finalresults}.

\begin{table}
\begin{tabular}{lcccccccc}
\hline\hline
                            & & $\{\Xi_{cc}, \Xi^*_{cc}\}$ & & $\{\Xi_{bb}, \Xi^*_{bb}\}$ & & $\{\Xi_{cb}, \Xi^\prime_{cb}, \Xi^*_{cb}\}$  \\
\hline
$\Delta_*$ (MeV)            & & 91.9(5.4)                  & & 37.2(2.3)                  & & 28.7(3.0)                                    \\
\hline\hline
\end{tabular}
\caption{\label{tab:Deltastardoublyheavy}Values of $\Delta_*$ (in MeV) used in the evaluation of the chiral loop integrals for the doubly heavy baryons.}
\end{table}

\begin{table}
\begin{tabular}{lrrrrrr}
\hline\hline
                            & & $\{\Xi_{cc}, \Xi^*_{cc}\}$ & & $\{\Xi_{bb}, \Xi^*_{bb}\}$ & & $\{\Xi_{cb}, \Xi^\prime_{cb}, \Xi^*_{cb}\}$          \\
\hline
 $E^{(\rm sub,0)}$ (MeV)    & & 534(28)                    & & 693(32)                    & & 688(36)        \hspace{2ex}                          \\
 \\[-3ex]
 $\Delta^{(0)}$ (MeV)       & & $\hdots\hspace{4ex}$       & & $\hdots\hspace{4ex}$       & & 10(21)         \hspace{2ex}                          \\
 \\[-3ex]
 $\Delta_*^{(0)}$ (MeV)     & & 79(10)                     & & 33.5(2.8)                  & & 25.9(3.6)      \hspace{2ex}                          \\
 \\[-3ex]
 $d_{\pi}^{(\rm vv)}$       & & $\hdots\hspace{4ex}$       & & $\hdots\hspace{4ex}$       & & 0.21(16)       \hspace{2ex}                          \\
 \\[-3ex]
 $d_{\pi}^{(\rm ss)}$       & & $\hdots\hspace{4ex}$       & & $\hdots\hspace{4ex}$       & & $-0.39(52)$    \hspace{2ex}                          \\
 \\[-3ex]
 $d_a$                      & & $\hdots\hspace{4ex}$       & & $\hdots\hspace{4ex}$       & & 1.18(97)       \hspace{2ex}                          \\
 \\[-3ex]
 $c_{\pi}^{(\rm vv)}$       & & 0.47(24)                   & & 0.39(13)                   & & 0.30(23)       \hspace{2ex}                          \\
 \\[-3ex]
 $c_{\pi}^{(\rm ss)}$       & & 0.47(68)                   & & 0.33(55)                   & & 0.29(65)       \hspace{2ex}                          \\
 \\[-3ex]
 $c_a$                      & & 1.18(72)                   & & 1.7(1.1)                   & & 1.4(1.0)       \hspace{2ex}                          \\
 \\[-3ex]
 $g_1$                      & & 0.51(23)                   & & 0.465(10)                  & & 0.44(18)       \hspace{2ex}                          \\
\hline\hline
\end{tabular}
\caption{\label{tab:doublyfeavyfit}Chiral and continuum extrapolation fit parameters for the doubly heavy baryons containing $u/d$ valence quarks.}
\end{table}

\begin{table}
\begin{tabular}{lcccccccccccccc}
\hline\hline
State             & & $\mathtt{C104}$ & & $\mathtt{C14}$ & & $\mathtt{C24}$ & & $\mathtt{C54}$  & & $\mathtt{F23}$ & & $\mathtt{F43}$ & & $\mathtt{F63}$  \\
\hline
 $\Xi_{cc}$       & & 0.2             & & 0.5            & & 0.5            & & 0.4             & & 0.7            & & 0.6            & & 0.4             \\
 \\[-3ex]
 $\Xi_{cc}^*$     & & 0.3             & & 0.9            & & 0.8            & & 0.7             & & 1.2            & & 1.0            & & 0.6             \\
 \\[-3ex]
 $\Xi_{bb}$       & & 0.2             & & 0.5            & & 0.5            & & 0.4             & & 0.7            & & 0.6            & & 0.4             \\
 \\[-3ex]
 $\Xi_{bb}^*$     & & 0.2             & & 0.6            & & 0.6            & & 0.5             & & 0.9            & & 0.7            & & 0.4             \\
 \\[-3ex]
 $\Xi_{cb}$       & & 0.0             & & 0.1            & & 0.1            & & 0.1             & & 0.1            & & 0.1            & & 0.0             \\
 \\[-3ex]
 $\Xi_{cb}'$      & & 0.2             & & 0.5            & & 0.5            & & 0.4             & & 0.6            & & 0.6            & & 0.3             \\
 \\[-3ex]
 $\Xi_{cb}^*$     & & 0.2             & & 0.6            & & 0.5            & & 0.5             & & 0.8            & & 0.7            & & 0.4             \\
 \\[-3ex]
\hline\hline
\end{tabular}
\caption{\label{tab:volumeshiftsdoublyheavy}Finite-volume energy shifts $E(L)-E(\infty)$ (in MeV) for the doubly heavy baryons containing $u/d$ valence quarks.}
\end{table}

\begin{figure}
\hspace{25pt}\includegraphics[height=18pt]{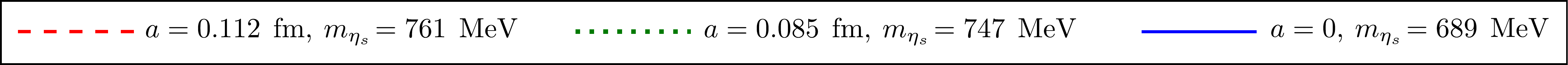}

\includegraphics[width=0.495\linewidth]{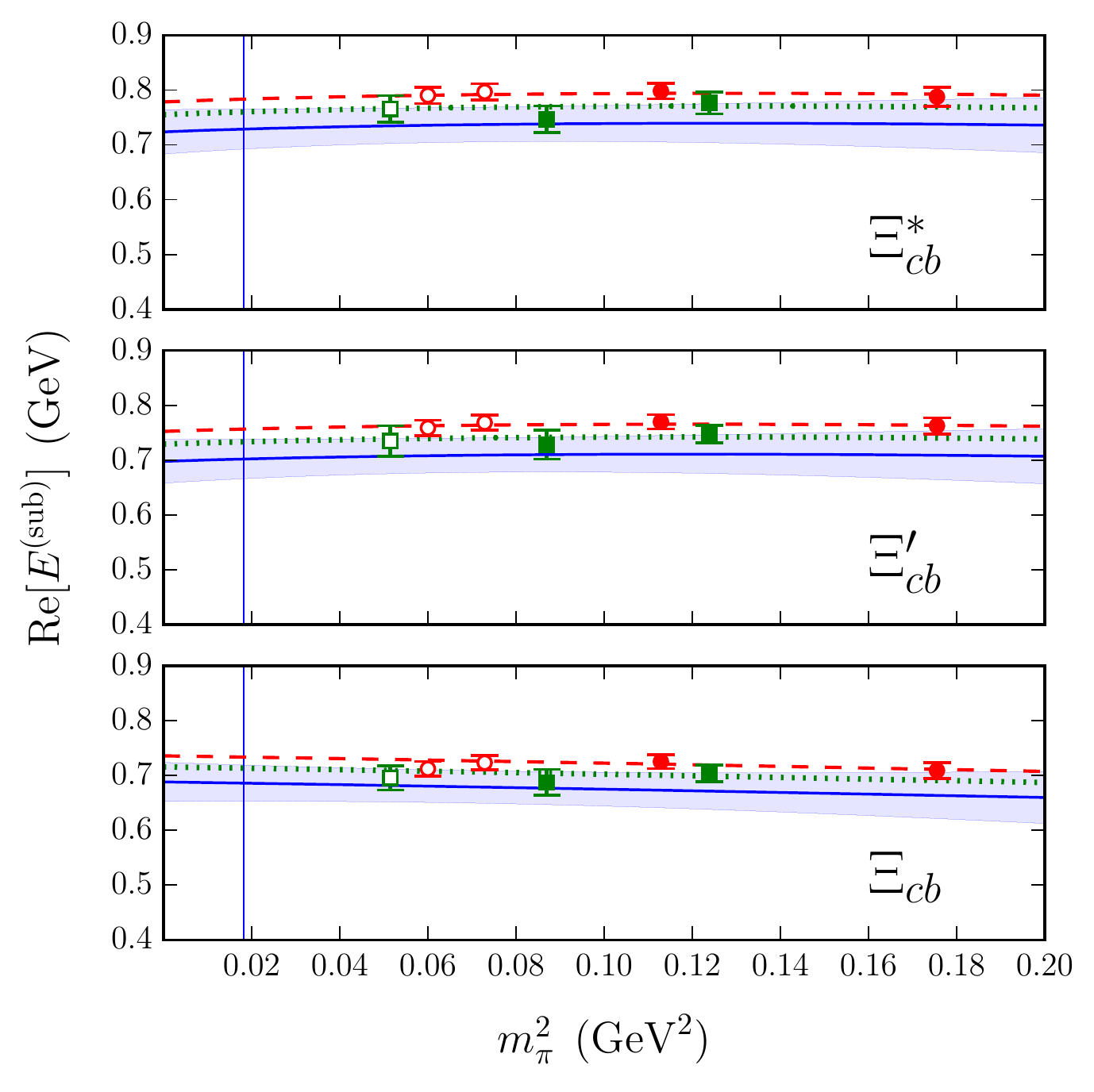}  \hfill \includegraphics[width=0.495\linewidth]{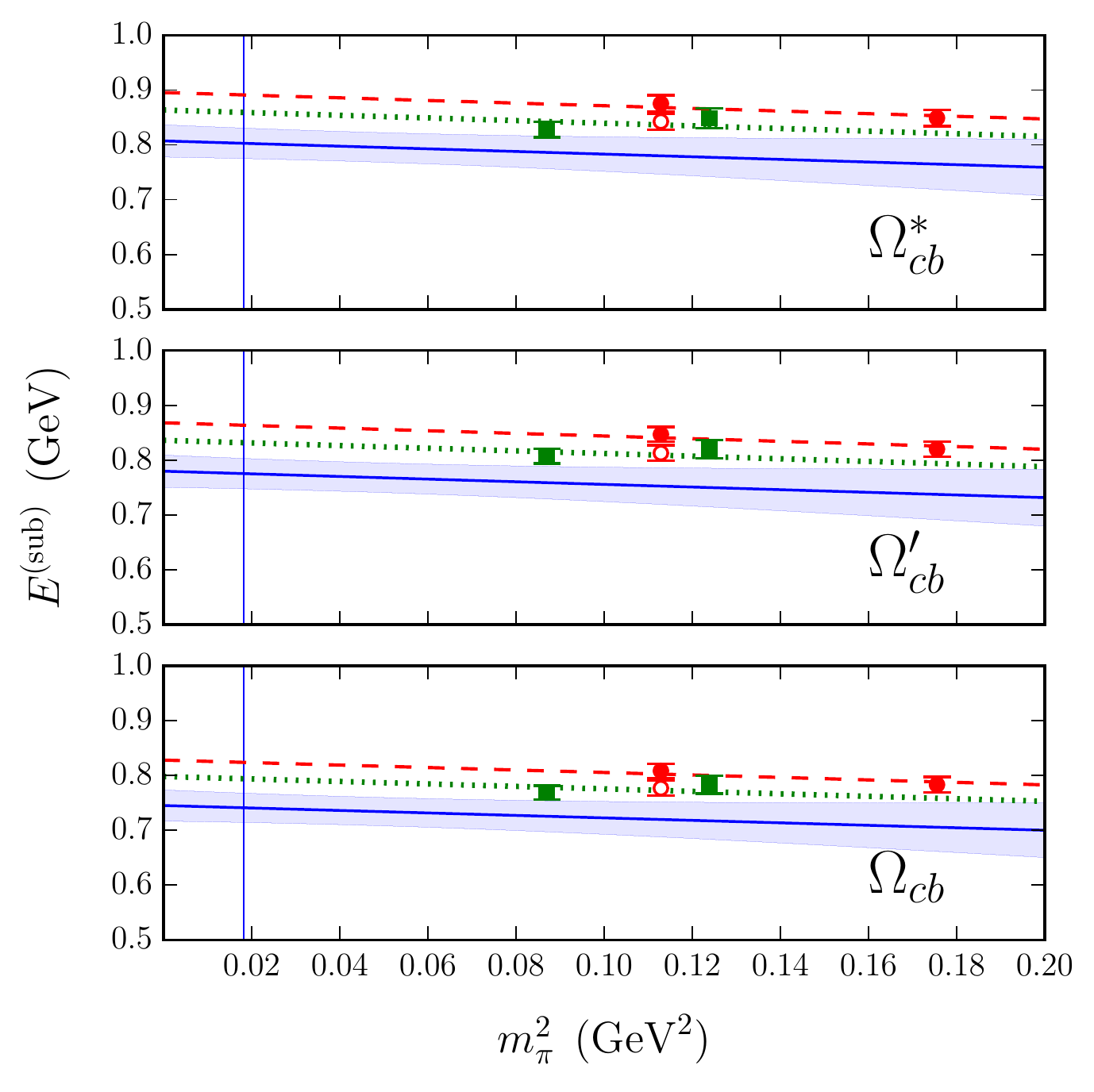}
\caption{\label{fig:doublyheavyCB}Left panel: chiral and continuum extrapolations for the $\{\Xi_{cb}, \Xi^\prime_{cb}, \Xi^*_{cb}\}$ baryons.
The details of the plots are as explained in the caption of Fig.~\protect\ref{fig:fitLambda}, except that no data sets are excluded here.
Right panel: chiral and continuum extrapolations for the $\{\Omega_{cb}, \Omega^\prime_{cb}, \Omega^*_{cb}\}$ baryons.
See the caption of Fig.~\protect\ref{fig:Omegafit} for explanations.}
\end{figure}

The chiral loop corrections for the non-strange baryons with different-flavor heavy quarks read
\begin{eqnarray}
 \mathcal{M}_{\Xi_{Q Q^\prime}} &=& -\frac{g_1^2}{16\pi^2 f^2}\Bigg[  \frac{4}{9} \mathcal{F}(m_\pi^{(\rm vs)},0,\mu )-\frac{1}{9}\mathcal{F}(m_\pi^{(\rm vv)},0,\mu ) \Bigg], \\
 \mathcal{M}_{\Xi^\prime_{Q Q^\prime}} &=& -\frac{g_1^2}{16\pi^2 f^2}\Bigg[  \frac{32}{9} \mathcal{F}(m_\pi^{(\rm vs)},\Delta_*,\mu )+\frac{4}{9} \mathcal{F}(m_\pi^{(\rm vs)},0,\mu )
 -\frac{8}{9} \mathcal{F}(m_\pi^{(\rm vv)},\Delta_*,\mu )-\frac{1}{9}\mathcal{F}(m_\pi^{(\rm vv)},0,\mu ) \Bigg],\\
 \mathcal{M}_{\Xi^*_{Q Q^\prime}} &=& -\frac{g_1^2}{16\pi^2 f^2}\Bigg[  \frac{16}{9} \mathcal{F}(m_\pi^{(\rm vs)},-\Delta_*,\mu )+\frac{20}{9} \mathcal{F}(m_\pi^{(\rm vs)},0,\mu )
 -\frac{4}{9} \mathcal{F}(m_\pi^{(\rm vv)},-\Delta_*,\mu )-\frac{5}{9}   \mathcal{F}(m_\pi^{(\rm vv)},0,\mu ) \Bigg].
\end{eqnarray}
Note that the chiral loop corrections for the $\{\Xi^\prime_{Q Q^\prime}, \Xi^*_{Q Q^\prime}\}$ hyperfine doublet are equal
to those for the $\{ \Xi_{QQ}, \Xi_{QQ}^* \}$ hyperfine doublet, while for the $\Xi_{Q Q^\prime}$ (which contains a spin-0 heavy diquark), the terms with $\Delta_*$ are missing.
This is because pion emission/absorption cannot change the spin of the heavy diquark, and hence the intermediate baryon in the self-energy
diagram for the $\Xi_{Q Q^\prime}$ also has to be a $\Xi_{Q Q^\prime}$. Because of this structure, the $\Xi_{Q Q^\prime}$ also requires
independent analytic counterterms, and we fit the lattice results for the $\{ \Xi_{QQ^\prime}, \Xi_{QQ^\prime}^\prime, \Xi_{QQ^\prime}^* \}$ energies using the functions
\begin{eqnarray}
 E^{(\rm sub)}_{\Xi_{Q Q^\prime}} &=& E^{(\rm sub, 0)} + d_{\pi}^{(\rm vv)} \frac{[m_\pi^{(\rm vv)}]^2}{4\pi f}
 + d_{\pi}^{(\rm ss)} \frac{[m_\pi^{(\rm ss)}]^2}{4\pi f} + \mathcal{M}_{\Xi_{Q Q^\prime}} + d_a\:a^2 \Lambda^3, \\
 E^{(\rm sub)}_{\Xi^\prime_{Q Q^\prime}} &=& E^{(\rm sub, 0)} + \Delta^{(0)} + c_{\pi}^{(\rm vv)} \frac{[m_\pi^{(\rm vv)}]^2}{4\pi f}
 + c_{\pi}^{(\rm ss)} \frac{[m_\pi^{(\rm ss)}]^2}{4\pi f} + \mathcal{M}_{\Xi^\prime_{Q Q^\prime}} + c_a\:a^2 \Lambda^3, \\
 E^{(\rm sub)}_{\Xi^*_{Q Q^\prime}} &=& E^{(\rm sub, 0)} + \Delta^{(0)} + \Delta_*^{(0)} + c_{\pi}^{(\rm vv)} \frac{[m_\pi^{(\rm vv)}]^2}{4\pi f}
 + c_{\pi}^{(\rm ss)} \frac{[m_\pi^{(\rm ss)}]^2}{4\pi f} + \mathcal{M}_{\Xi^*_{Q Q^\prime}} + c_a\:a^2 \Lambda^3,
\end{eqnarray}
with free fit parameters $E^{(\rm sub, 0)}$, $\Delta^{(0)}$, $\Delta_*^{(0)}$, $d_{\pi}^{(\rm vv)}$,
$d_{\pi}^{(\rm ss)}$, $d_a$, $c_{\pi}^{(\rm vv)}$, $c_{\pi}^{(\rm ss)}$, and $c_a$.
In this case, we included an extra 40\% width in $\sigma_{g_1}$ to account for unknown $1/m_Q$ effects in the axial coupling,
halfway between our choices for the $cc$ and $bb$ baryons. 
The results for all fit parameters are given in the last column of Table \ref{tab:doublyfeavyfit}, and the fits are visualized
in the left panel of Fig.~\ref{fig:doublyheavyCB}.

\begin{figure}
\hspace{25pt}\includegraphics[height=18pt]{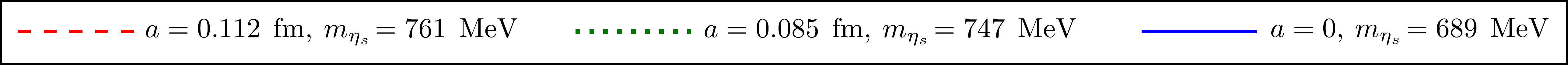}

\includegraphics[width=0.495\linewidth]{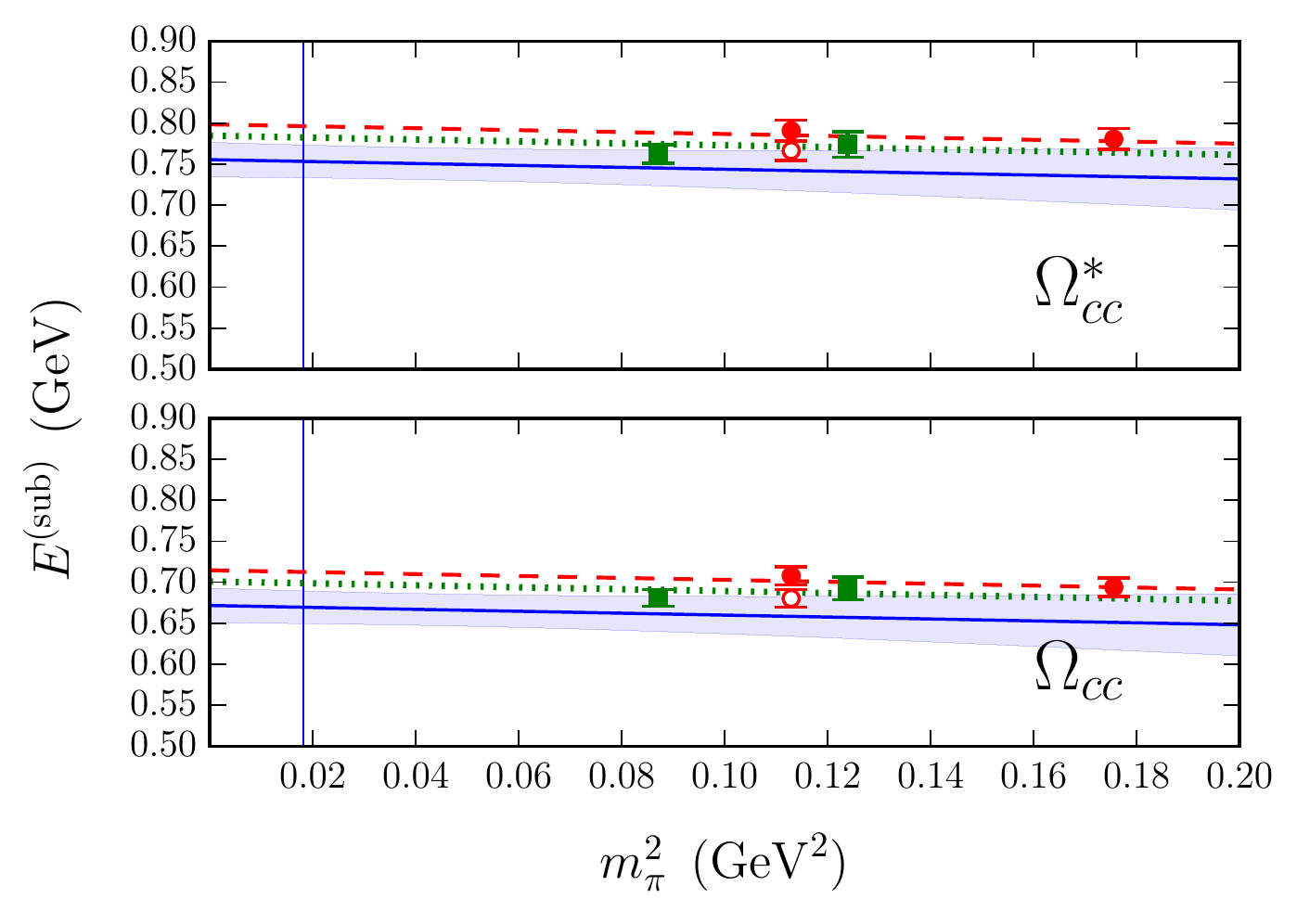}  \hfill \includegraphics[width=0.495\linewidth]{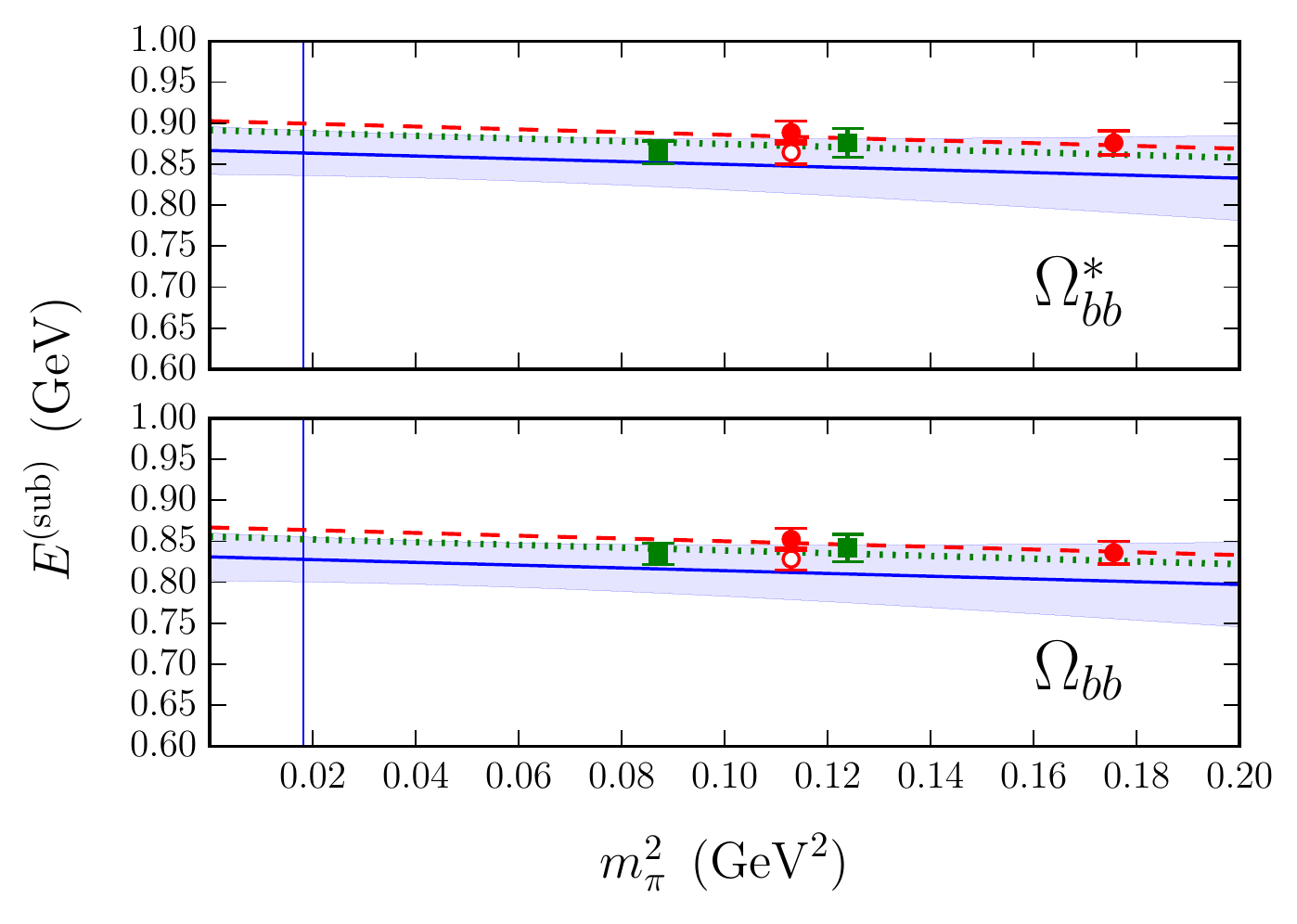}
\caption{\label{fig:doublyheavyOmega}Chiral and continuum extrapolations for the $\{\Omega_{QQ}, \Omega^*_{QQ}\}$ baryons.
See the caption of Fig.~\protect\ref{fig:Omegafit} for explanations.}
\end{figure}

The $S=-1$ doubly heavy baryons do not receive chiral loop corrections at next-to-leading-order in $SU(4|2)$ HH$\chi$PT. As in the case
of the $S=-2$ singly heavy baryons, we interpolate the energies linearly in the valence strange quark mass, and also allow for a linear dependence
on the light-sea-quark mass. We fit the lattice data for the $\{\Omega_{QQ}, \Omega_{QQ}^* \}$ and $\{ \Omega_{QQ^\prime}, \Omega_{QQ^\prime}^\prime, \Omega_{QQ^\prime}^* \}$
systems using the functions
\begin{eqnarray}
 E^{(\rm sub)}_{\Omega_{Q Q}} &=& E^{(\rm sub, 0)} + c_{{\eta_s}}^{(\rm vv)} \frac{[m_{\eta_s}^{(\rm vv)}]^2-[m_{\eta_s}^{(\rm phys)}]^2}{4\pi f}
 + c_{\pi}^{(\rm ss)} \frac{[m_\pi^{(\rm ss)}]^2}{4\pi f} + c_a\:a^2 \Lambda^3, \\
 E^{(\rm sub)}_{\Omega^*_{Q Q}} &=& E^{(\rm sub, 0)} + \Delta_*^{(0)} + c_{{\eta_s}}^{(\rm vv)} \frac{[m_{\eta_s}^{(\rm vv)}]^2-[m_{\eta_s}^{(\rm phys)}]^2}{4\pi f}
 + c_{\pi}^{(\rm ss)} \frac{[m_\pi^{(\rm ss)}]^2}{4\pi f} + c_a\:a^2 \Lambda^3,
\end{eqnarray}
and
\begin{eqnarray}
 E^{(\rm sub)}_{\Omega_{Q Q^\prime}} &=& E^{(\rm sub, 0)} + d_{\eta_s}^{(\rm vv)} \frac{[m_{\eta_s}^{(\rm vv)}]^2-[m_{\eta_s}^{(\rm phys)}]^2}{4\pi f}
 + d_{\pi}^{(\rm ss)} \frac{[m_\pi^{(\rm ss)}]^2}{4\pi f}  + d_a\:a^2 \Lambda^3, \\
 E^{(\rm sub)}_{\Omega^\prime_{Q Q^\prime}} &=& E^{(\rm sub, 0)} + \Delta^{(0)} + c_{{\eta_s}}^{(\rm vv)} \frac{[m_{\eta_s}^{(\rm vv)}]^2-[m_{\eta_s}^{(\rm phys)}]^2}{4\pi f}
 + c_{\pi}^{(\rm ss)} \frac{[m_\pi^{(\rm ss)}]^2}{4\pi f} + c_a\:a^2 \Lambda^3, \\
 E^{(\rm sub)}_{\Omega^*_{Q Q^\prime}} &=& E^{(\rm sub, 0)} + \Delta^{(0)} + \Delta_*^{(0)}
 + c_{{\eta_s}}^{(\rm vv)} \frac{[m_{\eta_s}^{(\rm vv)}]^2-[m_{\eta_s}^{(\rm phys)}]^2}{4\pi f} + c_{\pi}^{(\rm ss)} \frac{[m_\pi^{(\rm ss)}]^2}{4\pi f} + c_a\:a^2 \Lambda^3, 
\end{eqnarray}
respectively. The resulting values of the fit parameters are given in Table \ref{tab:doublyheavyOmega}, and plots of the fits are shown in Fig.~\ref{fig:doublyheavyOmega} and
in the right panel of Fig.~\ref{fig:doublyheavyCB}.

\begin{table}
\begin{tabular}{lrrrrrr}
\hline\hline
                            & & $\{\Omega_{cc}, \Omega^*_{cc}\}$ & & $\{\Omega_{bb}, \Omega^*_{bb}\}$ & & $\{\Omega_{cb}, \Omega^\prime_{cb}, \Omega^*_{cb}\}$     \\
\hline
 $E^{(\rm sub,0)}$ (MeV)    & & 672(21)                    & & 831(29)                    & &   745(28)        \hspace{2ex}                                        \\
 \\[-3ex]
 $\Delta^{(0)}$ (MeV)       & & $\hdots\hspace{4ex}$       & & $\hdots\hspace{4ex}$       & &   34.8(9.8)      \hspace{2ex}                                        \\
 \\[-3ex]
 $\Delta_*^{(0)}$ (MeV)     & & 83.8(1.4)                  & & 35.7(1.3)                  & &   27.4(2.0)      \hspace{2ex}                                        \\
 \\[-3ex]
 $d_{\eta_s}^{(\rm vv)}$    & & $\hdots\hspace{4ex}$       & & $\hdots\hspace{4ex}$       & &   0.379(74)      \hspace{2ex}                                        \\
 \\[-3ex]
 $d_{\pi}^{(\rm ss)}$       & & $\hdots\hspace{4ex}$       & & $\hdots\hspace{4ex}$       & &   $-0.37(45)$    \hspace{2ex}                                        \\
 \\[-3ex]
 $d_a$                      & & $\hdots\hspace{4ex}$       & & $\hdots\hspace{4ex}$       & &   1.5(1.0)       \hspace{2ex}                                        \\
 \\[-3ex]
 $c_{\eta_s}^{(\rm vv)}$    & & 0.325(32)                  & & 0.291(49)                  & &   0.400(69)      \hspace{2ex}                                        \\
 \\[-3ex]
 $c_{\pi}^{(\rm ss)}$       & & $-0.19(34)$                & & $-0.28(46)$                & &   $-0.40(47)$    \hspace{2ex}                                        \\
 \\[-3ex]
 $c_a$                      & & 0.56(76)                   & & 0.4(1.0)                   & &   $1.6(1.0)$     \hspace{2ex}                                        \\
\hline\hline
\end{tabular}
\caption{\label{tab:doublyheavyOmega}Chiral and continuum extrapolation fit parameters for the doubly heavy $\Omega$ baryons.}
\end{table}

\FloatBarrier
\subsection{Triply heavy baryons}
\FloatBarrier

\begin{figure}
\hspace{25pt}\includegraphics[height=18pt]{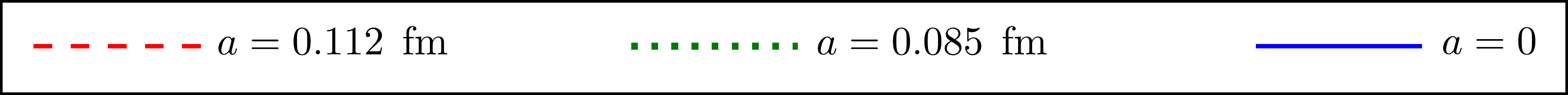}

\includegraphics[width=0.495\linewidth]{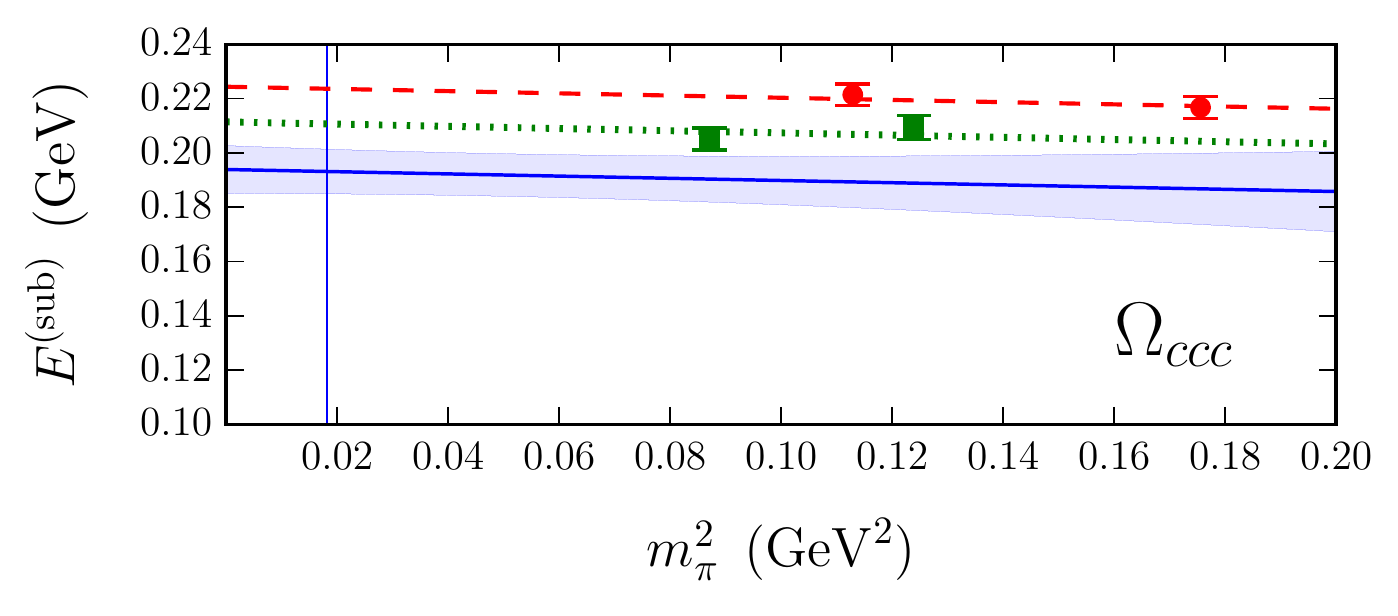}  \hfill \includegraphics[width=0.495\linewidth]{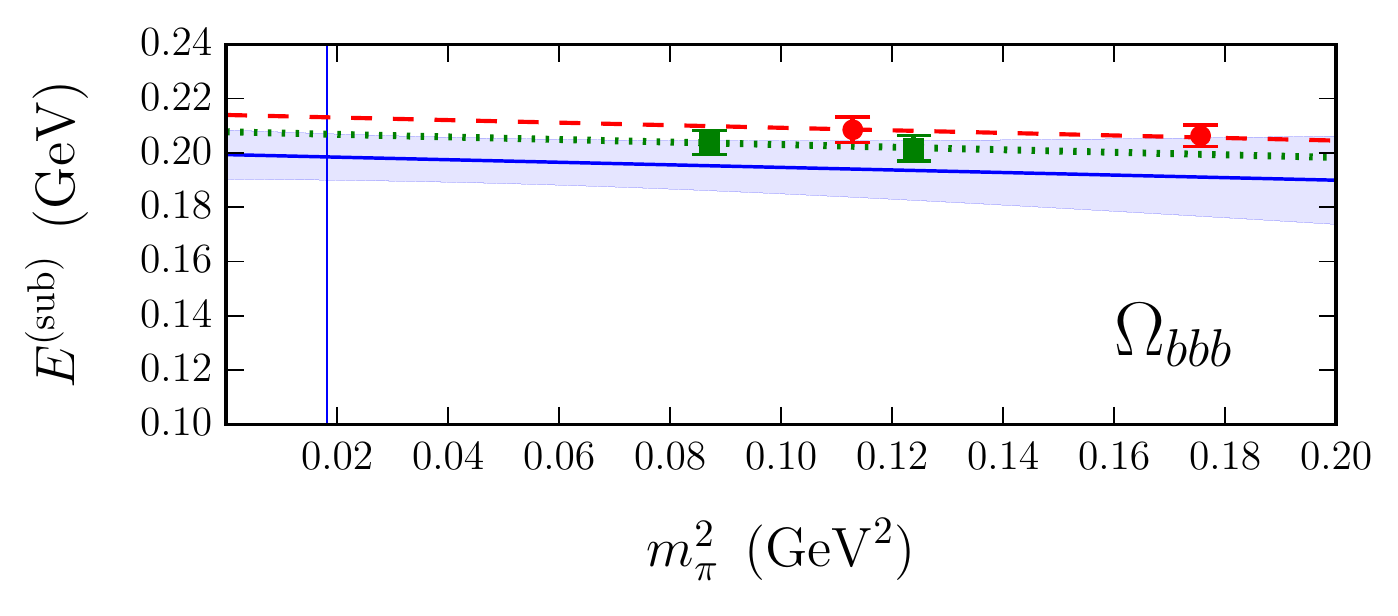}
\caption{\label{fig:OmegaQQQfit}Chiral and continuum extrapolations for the $\Omega_{ccc}$ (left panel) and $\Omega_{bbb}$ (right panel). The curves show the fit functions
for the two different lattice spacing where we have data, and in the continuum limit.
For the continuum curves, the shaded bands indicate the $1\sigma$ uncertainty. Data points at the coarse lattice spacing are plotted with circles,
and data points at the fine lattice spacing are plotted with squares.}
\end{figure}

\begin{figure}
\hspace{25pt}\includegraphics[height=18pt]{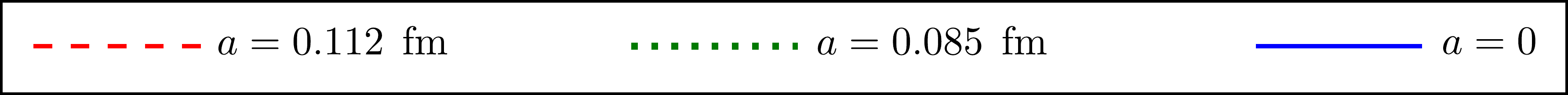}

\includegraphics[width=0.495\linewidth]{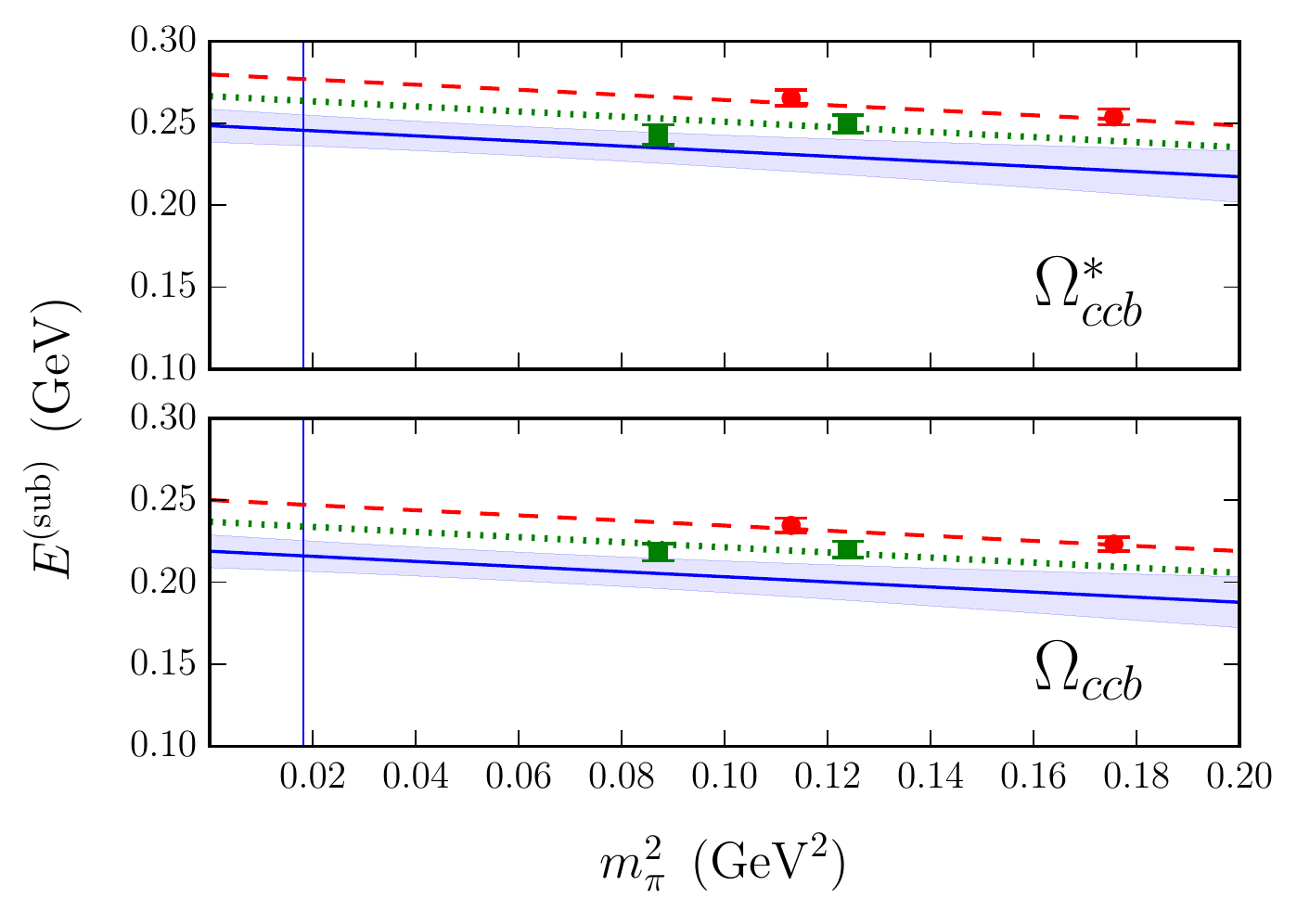}  \hfill \includegraphics[width=0.495\linewidth]{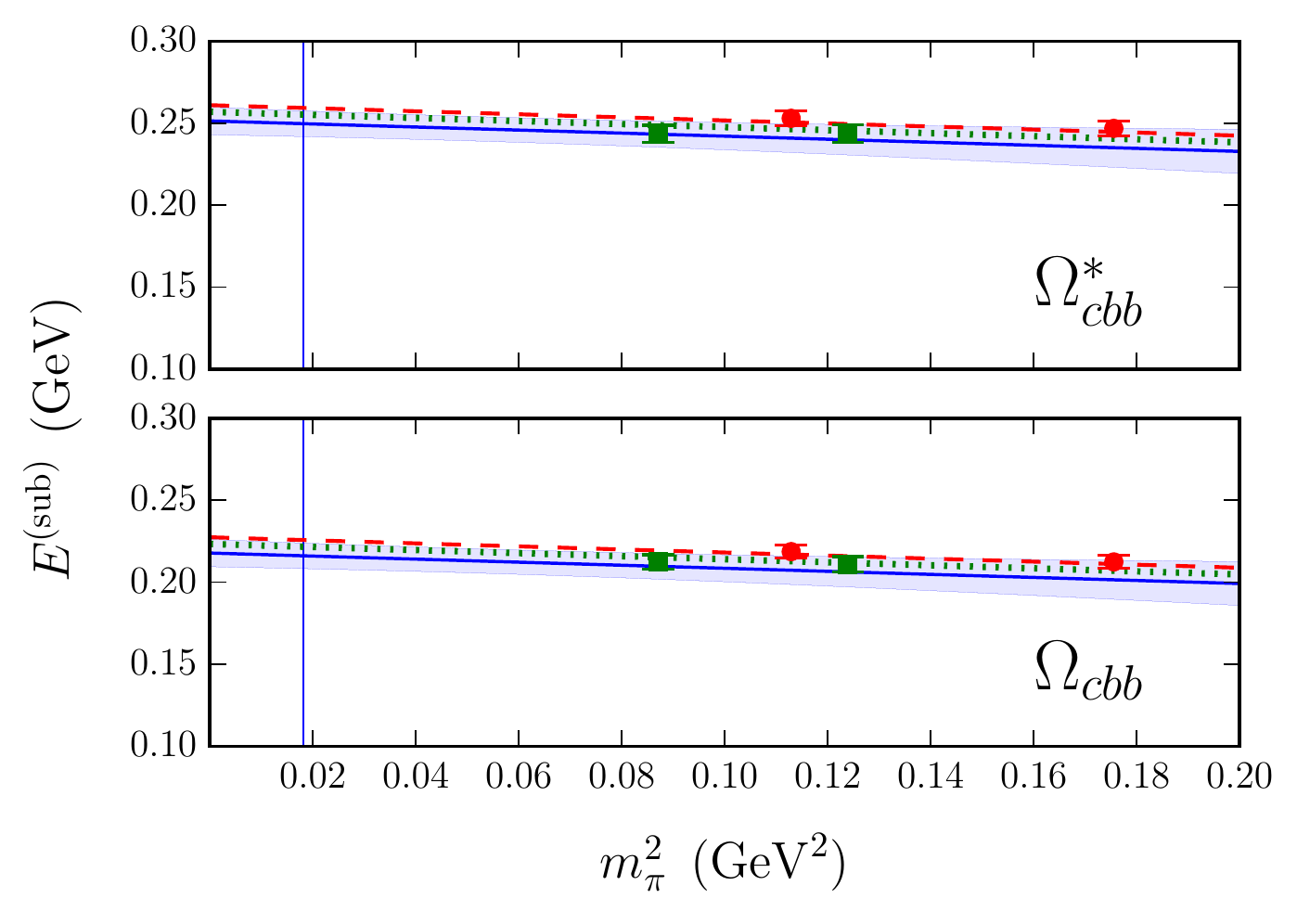}
\caption{\label{fig:OmegaQQQprimefit}Chiral and continuum extrapolations for the $\{ \Omega_{ccb}, \Omega^*_{ccb} \}$ baryons (left panel) and for the
$\{ \Omega_{cbb}, \Omega^*_{cbb} \}$ baryons (right panel). See the caption of Fig.~\ref{fig:OmegaQQQfit} for explanations.}
\end{figure}

With no light or strange valence quarks, the triply heavy baryons represent the simplest systems for the chiral and continuum extrapolations.
Here we allow for a linear dependence on the light sea-quark mass and a quadratic dependence on the lattice spacing. For the case
of three equal-flavor heavy quarks, the Pauli exclusion principle requires the ground-state $\Omega_{QQQ}$ to have $J^P=\frac32^+$. We fit
the subtracted energies of the $\Omega_{ccc}$ and $\Omega_{bbb}$ using the function
\begin{eqnarray}
 E^{(\rm sub)}_{\Omega_{Q Q Q}} &=& E^{(\rm sub, 0)} + c_{\pi}^{(\rm ss)} \frac{[m_\pi^{(\rm ss)}]^2}{4\pi f} + c_a\:a^2 \Lambda^3,
\end{eqnarray}
with parameters $E^{(\rm sub, 0)}$, $c_{\pi}^{(\rm ss)}$, and $c_a$. In the mixed-flavor case, both $J^P=\frac12^+$ and $J^P=\frac32^+$
are possible without requiring orbital angular momentum. Thus, we have the hyperfine multiplets $\{ \Omega_{Q Q Q^\prime}, \Omega^*_{Q Q Q^\prime} \}$,
whose subtracted energies we fit using the form
\begin{eqnarray}
 E^{(\rm sub)}_{\Omega_{Q Q Q^\prime}} &=& E^{(\rm sub, 0)} + c_{\pi}^{(\rm ss)} \frac{[m_\pi^{(\rm ss)}]^2}{4\pi f} + c_a\:a^2 \Lambda^3, \\
 E^{(\rm sub)}_{\Omega^*_{Q Q Q^\prime}} &=& E^{(\rm sub, 0)} + \Delta_*^{(0)}  + c_{\pi}^{(\rm ss)} \frac{[m_\pi^{(\rm ss)}]^2}{4\pi f} + c_a\:a^2 \Lambda^3,
\end{eqnarray}
with the additional hyperfine splitting parameter $\Delta_*^{(0)}$. The resulting fit parameters for all triply heavy baryons are given in Table
\ref{tab:triplyheavy}, and plots of the fits are shown in Figs.~\ref{fig:OmegaQQQfit} and \ref{fig:OmegaQQQprimefit}. Note that the $\mathcal{O}(a^2)$
effects appear to be largest for the systems containing two or more charm quarks.

\begin{table}
\begin{tabular}{lrrrrrrrr}
\hline\hline
                            & & $\Omega_{ccc}$ \hspace{2ex}  & & $\Omega_{bbb}$  \hspace{2ex} & & $\{\Omega_{ccb}, \Omega^*_{ccb}\}$   & & $\{\Omega_{cbb}, \Omega^*_{cbb}\}$         \\
\hline
 $E^{(\rm sub,0)}$ (MeV)    & & 193.9(8.8)                   & & 199.4(9.1)                   & & 218.9(10)        \hspace{0.1ex}      & &  217.8(8.3)       \hspace{0.1ex}           \\
 \\[-3ex]
 $\Delta_*^{(0)}$ (MeV)     & & $\hdots$                     & & $\hdots$                     & & 29.55(74)        \hspace{0.1ex}      & &  33.54(59)        \hspace{0.1ex}           \\
 \\[-3ex]
 $c_{\pi}^{(\rm ss)}$       & & $-0.07(14)$                  & & $-0.08(15)$                  & & $-0.26(15)$      \hspace{0.1ex}      & &  $-0.15(12)$      \hspace{0.1ex}           \\
 \\[-3ex]
 $c_a$                      & & 0.76(30)                     & & 0.36(34)                     & & 0.78(32)         \hspace{0.1ex}      & &  0.24(27)         \hspace{0.1ex}           \\
\hline\hline
\end{tabular}
\caption{\label{tab:triplyheavy} Chiral and continuum extrapolation fit parameters for the triply heavy baryons.}
\end{table}

\FloatBarrier
\section{\label{sec:finalresults} Final results and estimates of systematic uncertainties}
\FloatBarrier

\begin{figure}
\includegraphics[width=0.9\linewidth]{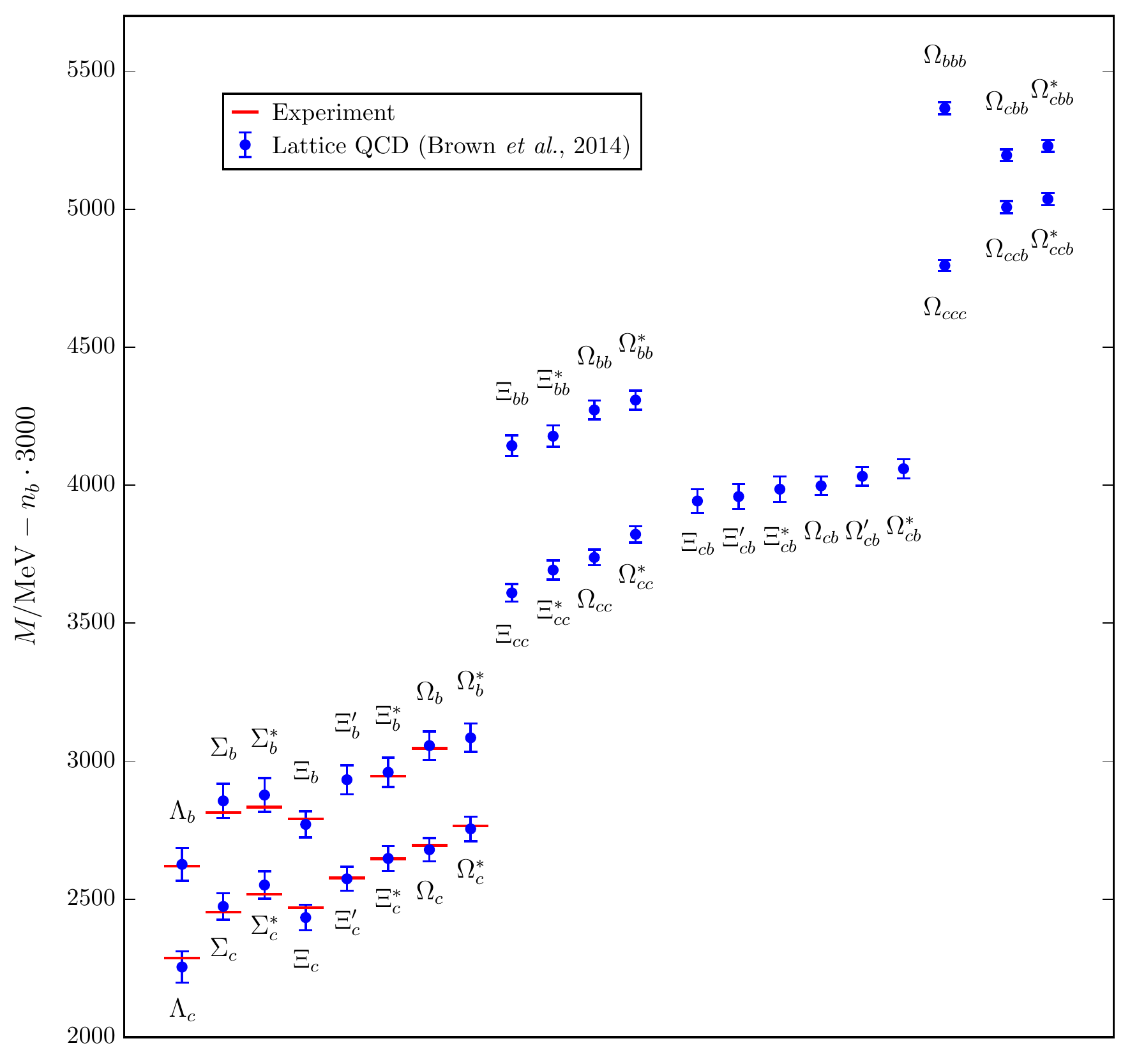}
\caption{\label{fig:spectrumall}Our results for the masses of charmed and/or bottom baryons, compared to the experimental
results where available \cite{Beringer:1900zz, Aaltonen:2009ny, Aaij:2013qja}. The masses of baryons containing $n_b$ bottom quarks
have been offset by $-n_b\cdot (3000\:\:\mathrm{MeV})$ to fit them into this plot. Note that the uncertainties of our results
for nearby states are highly correlated, and hyperfine splittings such as $M_{\Omega_b^*}-M_{\Omega_b}$ can in fact be resolved
with much smaller uncertainties than apparent from this figure (see Table \protect\ref{tab:splittings}).}
\end{figure}

To obtain the subtracted baryon energies at the physical point, $E^{(\rm sub, phys)}$, we evaluated the (real parts of the) fit functions discussed in the previous sections
at $m_\pi^{(\rm vv)}=m_\pi^{(\rm vs)}=m_\pi^{(\rm phys)}$, $m_{\eta_s}^{(\rm vv)}=m_{\eta_s}^{(\rm phys)}$, $a=0$, $L=\infty$, where
$m_\pi^{(\rm phys)}=134.8$ MeV is the pion mass in the isospin limit \cite{Colangelo:2010et}, and $m_{\eta_s}^{(\rm phys)}=689.3$ MeV \cite{Dowdall:2011wh}.
The statistical uncertainties of $E^{(\rm sub, phys)}$ were computed by propagating the uncertainties of all fit parameters in a correlated way, using their
covariance matrices obtained from the second derivatives of $\chi^2$. These statistical uncertainties already include the uncertainties in the lattice spacings (see the discussion
at the beginning of Sec.~\ref{sec:extrap}). To obtain the full baryon masses, we then added the experimental values of
$\frac{n_c}{2}\overline{M}_{c\bar{c}}+\frac{n_b}{2}\overline{M}_{b\bar{b}}$ to $E^{(\rm sub, phys)}$, using \cite{Beringer:1900zz}
\begin{eqnarray*}
 \overline{M}_{c\bar{c}} &=& 3068.61(18)\:\:{\rm MeV}, \\
 \overline{M}_{b\bar{b}} &=& 9444.72(87)\:\:{\rm MeV}.
\end{eqnarray*}
The results for the full baryon masses are given in Table \ref{tab:masses}, and are plotted in Fig.~\ref{fig:spectrumall}. Furthermore, Table \ref{tab:splittings} shows our results for
the mass splittings between baryons with equal quark flavor content, including the hyperfine splittings. The mass splittings have
smaller statistical uncertainties than the baryon masses themselves as a consequence of correlations. For all results, individual estimates of the total systematic uncertainties
are also given in the Tables. These include the uncertainties associated with the assumptions/approximations made in the chiral and continuum extrapolations,
and those associated with the use of lattice NRQCD for the $b$ quarks. In the following, we describe in detail how we obtained these estimates.

\begin{table}
\begin{tabular}{lllll}
\hline\hline
State              & \hspace{2ex} & This work & \hspace{2ex} & Experiment \\
\hline
 $\Lambda_c$         &&  2254(48)(31)       &&  2286.46(14)  \\
 \\[-3ex]
 $\Sigma_c$          &&  2474(41)(25)       &&  2453.79(11)  \\
 \\[-3ex]
 $\Sigma_c^*$        &&  2551(43)(25)       &&  2518.32(42)  \\
 \\[-3ex]
 $\Xi_c$             &&  2433(35)(30)       &&  2468.91(48)  \\
 \\[-3ex]
 $\Xi_c'$            &&  2574(37)(23)       &&  2576.8(2.1)  \\
 \\[-3ex]
 $\Xi_c^*$           &&  2648(38)(25)       &&  2645.90(38)  \\
 \\[-3ex]
 $\Omega_c$          &&  2679(37)(20)       &&  2695.2(1.7)  \\
 \\[-3ex]
 $\Omega_c^*$        &&  2755(37)(24)       &&  2765.9(2.0)  \\
 \\[-3ex]
 $\Xi_{cc}$          &&  3610(23)(22)       &&  $\hdots$     \\
 \\[-3ex]
 $\Xi_{cc}^*$        &&  3692(28)(21)       &&  $\hdots$     \\
 \\[-3ex]
 $\Omega_{cc}$       &&  3738(20)(20)       &&  $\hdots$     \\
 \\[-3ex]
 $\Omega_{cc}^*$     &&  3822(20)(22)       &&  $\hdots$     \\
 \\[-3ex]
 $\Omega_{ccc}$      &&  4796(8)(18)        &&  $\hdots$     \\
 \\[-3ex]
 $\Lambda_b$         &&  5626(52)(29)       &&  5619.4(0.6)  \\
 \\[-3ex]
 $\Sigma_b$          &&  5856(56)(27)       &&  5813.5(1.3)  \\
 \\[-3ex]
 $\Sigma_b^*$        &&  5877(55)(27)       &&  5833.6(1.3)  \\
 \\[-3ex]
 $\Xi_b$             &&  5771(41)(24)       &&  5790.6(2.0)  \\
 \\[-3ex]
 $\Xi_b'$            &&  5933(47)(24)       &&  $\hdots$     \\
 \\[-3ex]
 $\Xi_b^*$           &&  5960(47)(25)       &&  5945.5(2.3)  \\
 \\[-3ex]
 $\Omega_b$          &&  6056(47)(20)       &&  6046.8(2.1)  \\
 \\[-3ex]
 $\Omega_b^*$        &&  6085(47)(20)       &&  $\hdots$     \\
 \\[-3ex]
 $\Xi_{bb}$          &&  10143(30)(23)      &&  $\hdots$     \\
 \\[-3ex]
 $\Xi_{bb}^*$        &&  10178(30)(24)      &&  $\hdots$     \\
 \\[-3ex]
 $\Omega_{bb}$       &&  10273(27)(20)      &&  $\hdots$     \\
 \\[-3ex]
 $\Omega_{bb}^*$     &&  10308(27)(21)      &&  $\hdots$     \\
 \\[-3ex]
 $\Omega_{bbb}$      &&  14366(9)(20)       &&  $\hdots$     \\
 \\[-3ex]
 $\Xi_{cb}$          &&  6943(33)(28)       &&  $\hdots$     \\
 \\[-3ex]
 $\Xi_{cb}'$         &&  6959(36)(28)       &&  $\hdots$     \\
 \\[-3ex]
 $\Xi_{cb}^*$        &&  6985(36)(28)       &&  $\hdots$     \\
 \\[-3ex]
 $\Omega_{cb}$       &&  6998(27)(20)       &&  $\hdots$     \\
 \\[-3ex]
 $\Omega_{cb}'$      &&  7032(28)(20)       &&  $\hdots$     \\
 \\[-3ex]
 $\Omega_{cb}^*$     &&  7059(28)(21)       &&  $\hdots$     \\
 \\[-3ex]
 $\Omega_{ccb}$      &&  8007(9)(20)        &&  $\hdots$     \\
 \\[-3ex]
 $\Omega_{ccb}^*$    &&  8037(9)(20)        &&  $\hdots$     \\
 \\[-3ex]
 $\Omega_{cbb}$      &&  11195(8)(20)       &&  $\hdots$     \\
 \\[-3ex]
 $\Omega_{cbb}^*$    &&  11229(8)(20)       &&  $\hdots$     \\
 \\[-3ex]
\hline\hline
\end{tabular}
\caption{\label{tab:masses}Final results for the full baryon masses (in MeV). The first uncertainty is statistical
and the second uncertainty is systematic. Where available, we also show the experimental averages from the Particle Data Group \cite{Beringer:1900zz}.
Where experimental results were available for multiple isospin states, we show the isospin-averaged mass. The experimental value for the $\Omega_b$
mass given here is our average of the CDF \cite{Aaltonen:2009ny} and LHCb \cite{Aaij:2013qja} results. }
\end{table}

\begin{table}
\begin{tabular}{lllll}
\hline\hline
Splitting                       & \hspace{2ex} & This work & \hspace{2ex} & Experiment \\
\hline
 $\Sigma_c - \Lambda_c$         &&  219(36)(43)      && 167.33(18)    \\
 \\[-3ex]
 $\Sigma_c^* - \Lambda_c$       &&  297(33)(43)      && 231.86(44)    \\
 \\[-3ex]
 $\Sigma_c^*-\Sigma_c$          &&  78(7)(11)        && 64.53(43)     \\
 \\[-3ex]
 $\Xi_c'-\Xi_c$                 &&  140(16)(38)      && 107.9(2.2)    \\
 \\[-3ex]
 $\Xi_c^*-\Xi_c$                &&  214(16)(39)      && 176.99(61)    \\
 \\[-3ex]
 $\Xi_c^*-\Xi_c'$               &&  73.7(5.0)(8.7)   && 69.1(2.2)     \\
 \\[-3ex]
 $\Omega_c^*-\Omega_c$          &&  75.3(1.9)(7.6)   && 70.7(2.6)     \\
 \\[-3ex]
 $\Xi_{cc}^*-\Xi_{cc}$          &&  82.8(7.2)(5.8)   &&  $\hdots$     \\
 \\[-3ex]
 $\Omega_{cc}^*-\Omega_{cc}$    &&  83.8(1.4)(5.3)   &&  $\hdots$     \\
 \\[-3ex]
 $\Sigma_b-\Lambda_b$           &&  230(47)(40)      && 194.1(1.4)    \\
 \\[-3ex]
 $\Sigma_b^*-\Lambda_b$         &&  251(46)(40)      && 214.2(1.5)    \\
 \\[-3ex]
 $\Sigma_b^*-\Sigma_b$          &&  21.2(4.9)(7.3)   && 20.1(1.9)     \\
 \\[-3ex]
 $\Xi_b'-\Xi_b$                 &&  162(29)(33)      &&  $\hdots$     \\
 \\[-3ex]
 $\Xi_b^*-\Xi_b$                &&  189(29)(33)      &&  154.41(0.79) \\
 \\[-3ex]
 $\Xi_b^*-\Xi_b'$               &&  27.0(3.2)(8.6)   && $\hdots$      \\
 \\[-3ex]
 $\Omega_b^*-\Omega_b$          &&  28.4(2.2)(7.7)   && $\hdots$      \\
 \\[-3ex]
 $\Xi_{bb}^*-\Xi_{bb}$          &&  34.6(2.5)(7.4)   && $\hdots$      \\
 \\[-3ex]
 $\Omega_{bb}^*-\Omega_{bb}$    &&  35.7(1.3)(5.5)   && $\hdots$      \\
 \\[-3ex]
 $\Xi_{cb}'-\Xi_{cb}$           &&  16(18)(38)       && $\hdots$      \\
 \\[-3ex]
 $\Xi_{cb}^*-\Xi_{cb}$          &&  43(19)(38)       && $\hdots$      \\
 \\[-3ex]
 $\Xi_{cb}^*-\Xi_{cb}'$         &&  26.7(3.3)(8.4)   && $\hdots$      \\
 \\[-3ex]
 $\Omega_{cb}'-\Omega_{cb}$     &&  35(9)(25)        && $\hdots$      \\
 \\[-3ex]
 $\Omega_{cb}^*-\Omega_{cb}$    &&  62(9)(25)        && $\hdots$      \\
 \\[-3ex]
 $\Omega_{cb}^*-\Omega_{cb}'$   &&  27.4(2.0)(6.7)   && $\hdots$      \\
 \\[-3ex]
 $\Omega_{ccb}^*-\Omega_{ccb}$  &&  29.6(0.7)(4.2)   && $\hdots$      \\
 \\[-3ex]
 $\Omega_{cbb}^*-\Omega_{cbb}$  &&  33.5(0.6)(4.1)   && $\hdots$      \\
 \\[-3ex]
\hline\hline
\end{tabular}
\caption{\label{tab:splittings}Mass splittings (in MeV) between baryons with equal flavor. The first uncertainty is statistical
and the second uncertainty is systematic. Where available, we also show the experimental averages from the Particle Data Group \cite{Beringer:1900zz}
(the $\Xi_b^{*0}-\Xi_b^-$ splitting was taken from Ref.~\cite{Chatrchyan:2012ni}). Where experimental results were available for multiple
isospin states, we show the isospin-averaged mass splitting.}
\end{table}

\FloatBarrier
\subsection{Chiral and continuum extrapolation systematic uncertainties}

The chiral and continuum extrapolations
were performed at next-to-leading-order in the chiral expansion, and included quadratic dependence on the lattice spacing.
To estimate the uncertainty associated with this truncation, we added higher-order analytic
terms to the fit functions and redid the fits. For example, in the case of the $\{\Lambda_Q, \Sigma_Q, \Sigma_Q^*\}$, we added the terms
\begin{eqnarray}
\nonumber E^{(\rm sub,HO)}_{\Lambda_Q} &=& d_{\pi}^{(\rm vv,vv)}\frac{[m_\pi^{(\rm vv)}]^4}{(4\pi f)^3} + d_{\pi}^{(\rm ss,ss)}\frac{[m_\pi^{(\rm ss)}]^4}{(4\pi f)^3}
+ d_{\pi}^{(\rm vv,ss)}\frac{[m_\pi^{(\rm vv)}]^2 [m_\pi^{(\rm ss)}]^2}{(4\pi f)^3} \\
 &&     + d_{a,\pi}^{(\rm vv)}\frac{[m_\pi^{(\rm vv)}]^2\, a^2 \Lambda^2}{4\pi f}  + d_{a,\pi}^{(\rm ss)}\frac{[m_\pi^{(\rm ss)}]^2\, a^2 \Lambda^2}{4\pi f}
 + d_a^{(3)} a^3 \Lambda^4, \label{eq:ELambdaQHO}
\end{eqnarray}
\begin{eqnarray}
\nonumber E^{(\rm sub,HO)}_{\Sigma_Q} &=& c_{\Delta,\pi}^{(\rm vv)} \frac{[m_\pi^{(\rm vv)}]^2}{(4\pi f)^2} \Delta^{(0)}
+ c_{\Delta,\pi}^{(\rm ss)} \frac{[m_\pi^{(\rm ss)}]^2}{(4\pi f)^2} \Delta^{(0)} + c_{\Delta,a}\: a^2\Lambda^2 \Delta^{(0)}  \\
\nonumber && + c_{\pi}^{(\rm vv,vv)}\frac{[m_\pi^{(\rm vv)}]^4}{(4\pi f)^3} + c_{\pi}^{(\rm ss,ss)}\frac{[m_\pi^{(\rm ss)}]^4}{(4\pi f)^3}
+ c_{\pi}^{(\rm vv,ss)}\frac{[m_\pi^{(\rm vv)}]^2 [m_\pi^{(\rm ss)}]^2}{(4\pi f)^3} \\
 &&     + c_{a,\pi}^{(\rm vv)}\frac{[m_\pi^{(\rm vv)}]^2\, a^2 \Lambda^2}{4\pi f}
 + c_{a,\pi}^{(\rm ss)}\frac{[m_\pi^{(\rm ss)}]^2\, a^2 \Lambda^2}{4\pi f} + c_a^{(3)} a^3 \Lambda^4, \label{eq:ESigmaQHO}
\end{eqnarray}
\begin{eqnarray}
\nonumber E^{(\rm sub,HO)}_{\Sigma^*_Q} &=& c_{\Delta,\pi}^{(\rm vv)} \frac{[m_\pi^{(\rm vv)}]^2}{(4\pi f)^2} \Delta^{(0)}
+ c_{\Delta,\pi}^{(\rm ss)} \frac{[m_\pi^{(\rm ss)}]^2}{(4\pi f)^2} \Delta^{(0)} + c_{\Delta,a}\: a^2\Lambda^2 \Delta^{(0)}  \\
\nonumber && +  c_{\Delta_*,\pi}^{(\rm vv)} \frac{[m_\pi^{(\rm vv)}]^2}{(4\pi f)^2} \Delta_*^{(0)}
+ c_{\Delta_*,\pi}^{(\rm ss)} \frac{[m_\pi^{(\rm ss)}]^2}{(4\pi f)^2} \Delta_*^{(0)} + c_{\Delta_*,a}\: a^2\Lambda^2 \Delta_*^{(0)}  \\
\nonumber && + c_{\pi}^{(\rm vv,vv)}\frac{[m_\pi^{(\rm vv)}]^4}{(4\pi f)^3} + c_{\pi}^{(\rm ss,ss)}\frac{[m_\pi^{(\rm ss)}]^4}{(4\pi f)^3}
+ c_{\pi}^{(\rm vv,ss)}\frac{[m_\pi^{(\rm vv)}]^2 [m_\pi^{(\rm ss)}]^2}{(4\pi f)^3} \\
 &&     + c_{a,\pi}^{(\rm vv)}\frac{[m_\pi^{(\rm vv)}]^2\, a^2 \Lambda^2}{4\pi f}  + c_{a,\pi}^{(\rm ss)}\frac{[m_\pi^{(\rm ss)}]^2\, a^2 \Lambda^2}{4\pi f}
 + c_a^{(3)} a^3 \Lambda^4 \label{eq:ESigmaQstarHO}
\end{eqnarray}
to Eqs.~(\ref{eq:ELambdaQ}), (\ref{eq:ESigmaQ}), and (\ref{eq:ESigmaQstar}). At this order, the energy splitting parameters from the original fit
also need to be expanded in powers of the quark masses and lattice spacing, leading to the terms with
products of $\Delta^{(0)}$ or $\Delta_*^{(0)}$ with $[m_\pi^{(\rm vv)}]^2$, $[m_\pi^{(\rm ss)}]^2$, or $a^2$. The terms proportional to $a^3$ may arise from heavy-quark discretization
errors. We followed a Bayesian approach and constrained the additional parameters in
Eqs.~(\ref{eq:ELambdaQHO}), (\ref{eq:ESigmaQHO}), and (\ref{eq:ESigmaQstarHO}) to be natural-sized. Because we have
introduced appropriate powers of the relevant energy scales in the definitions of the fit functions,
the new parameters are dimensionless, and we used Gaussian priors with central value 0 and width 3 for each one.
We then recomputed $E^{(\rm sub, phys)}$ for each baryon from the new higher-order fits. A good measure for the systematic uncertainty due to the higher-order effects is
the resulting increase in the uncertainty of $E^{(\rm sub, phys)}$, computed in quadrature,
\begin{equation}
 \sigma_{\rm syst.,HO} = \sqrt{\sigma_{\rm NLO+HO}^2-\sigma_{\rm NLO}^2} \, ,
\end{equation}
where $\sigma_{\rm NLO}$ is the uncertainty obtained from the original fit and $\sigma_{\rm NLO+HO}$ is the uncertainty of the fit including
the higher-order analytic terms. We applied the same procedure to the baryon mass splittings and their uncertainties. Using the increase in the uncertainty is
far more robust than using the change in the central value, because the change in the central value may be close to zero with our
choice of priors for the higher-order terms. 

We separately estimated and added the uncertainties associated with our choices made for $\Delta$ and $\Delta_*$ in the evaluation of the chiral functions $\mathcal{F}$.
As discussed in Sec.~\ref{sec:extrap}, for the larger, well-resolved splittings, we used the results of linear extrapolations of the lattice results
to the chiral limit. To estimate the effect of this choice, we repeated the analysis with $\Delta$ and $\Delta_*$ set equal to constant fits of the lattice results instead,
and we took the resulting changes in the central values of $E^{(\rm sub, phys)}$ as our estimates for this particular source of systematic uncertainty. The smallest
of the splittings (such as the hyperfine splittings $\Delta_*$ in the bottom sector), for which we already used the results of constant fits to the lattice data, have very little
effect on the values of $\mathcal{F}$ in the first place.

\subsection{NRQCD systematic uncertainties: baryon masses via $E^{(\rm sub,phys)}$ }

For baryons containing $b$ quarks, the uncertainties associated with the use of lattice NRQCD enter in
Eq.~(\ref{eq:Esub}) both through the baryon energies themselves, and through the subtraction term $-\frac{n_b}{2}\overline{E}_{b\bar{b}}$.
We estimate the NRQCD uncertainties using power counting \cite{Lepage:1992tx}.

For the subtraction term, the relevant expansion
parameter is the typical velocity of the $b$-quark inside bottomonium, $v_b^2 \approx 0.1$ \cite{Lepage:1992tx}. The NRQCD
action used here only includes terms up to order $v^4$. The effect of the missing $v^6$ corrections on the subtraction
term is of order $\frac{n_b}{2} m_b v_b^6 \approx \frac{n_b}{2} (5\:\:{\rm MeV})$. The matching coefficients of the order-$v^4$
operators in the NRQCD action were set to their tadpole-improved tree-level values (except for $c_4$, which was determined to one loop),
introducing an additional systematic uncertainty of order $ \frac{n_b}{2} \alpha_s m_b v_b^4 \approx \frac{n_b}{2} (10\:\:{\rm MeV})$.
Furthermore, the NRQCD action used here did not include four-fermion operators, whose effect is expected to be of order
$\frac{n_b}{2} \alpha_s^2\:m_b\:v_b^3 \approx \frac{n_b}{2} (6\:\:{\rm MeV})$.

For the $b q q^\prime$ baryons containing a single $b$ quark and no charm quarks, we need to use heavy-light power counting with expansion parameter $\Lambda/m_b$.
In this case, the operators $-c_4\:\frac{g}{2 m_b}\:\bss{\sigma}\cdot\mathbf{\widetilde{B}}$ and $H_0=-\frac{\Delta^{(2)}}{2 m_b}$ are both of first order. While $H_0$ does not require
a matching coefficient, the matching coefficient $c_4$ was computed only through one loop. We estimate the uncertainty in $E_{b q q^\prime}$ resulting from this truncation
to be of order $\alpha_s^2\Lambda^2/m_b \approx 2$ MeV. The matching coefficients of the $\mathcal{O}(\Lambda^2/m_b^2)$ operators were computed at tree level, and most of the
$\mathcal{O}(\Lambda^3/m_b^3)$ operators are missing altogether, which introduces systematic uncertainties in $E_{b q q^\prime}$ of order
$\alpha_s\Lambda^3/m_b^2 \approx 1$ MeV and $\Lambda^4/m_b^3 \approx 0.5$ MeV, respectively. For heavy-light systems, the effect of the missing four-quark operators
(containing products of two heavy and two light quark fields) on the energies is expected to be of order $\alpha_s^2 \Lambda^3/m_b^2$; a more detailed
study shows that the energy shifts caused by the four-quark operators in heavy-light systems are around 3 MeV \cite{Blok:1996iz}.

For the $b c c$ baryons, we use heavy-heavy power counting; there, the typical velocity of the $b$ quark is expected to be comparable to that in a $B_c$ meson,
$v^2_{b(c)}\approx 0.05$ \cite{Gregory:2009hq}. Thus, we estimate the systematic uncertainties associated with the missing $v^6$ terms, the missing
radiative corrections in the matching coefficients of the $v^4$ terms, and the missing four-quark operators to be of order $m_b v_{b(c)}^6 \approx 1\:\:{\rm MeV}$,
$\alpha_s m_b v_{b(c)}^4 \approx 3\:\:{\rm MeV}$, and $\alpha_s^2\:m_b\:v_{b(c)}^3 \approx 3\:\:{\rm MeV}$, respectively.

For the $b c q$ baryon energies, we conservatively add the power-counting estimates obtained in the previous
two paragraphs (for the $b q q^\prime$ and $b c c$ baryons) in quadrature.

For the triply-bottom $\Omega_{bbb}$, heavy-heavy power counting applies, and the NRQCD expansion
converges with the same rate as in bottomonium (this was demonstrated numerically in Refs.~\cite{Meinel:2010pw, Meinel:2012qz}). Here we expect a partial
cancellation of the NRQCD uncertainties between $E_{\Omega_{bbb}}$ and $-\frac{n_b}{2}\overline{E}_{b\bar{b}}$. Therefore, instead of adding
the uncertainties from these two terms in quadrature, we estimate the total NRQCD systematic uncertainty in $E^{(\rm sub, phys)}_{\Omega_{bbb}}$ to be equal to $(1/2)$
times the NRQCD systematic uncertainty in $-\frac{n_b}{2}\overline{E}_{b\bar{b}}$ (with $n_b=3$).

The $bbc$ and $bbq$ baryons are also similar to bottomonium,
and we again assume a 50\% cancellation of the NRQCD systematic uncertainty from $-\frac{n_b}{2}\overline{E}_{b\bar{b}}$. Because of the
presence of a charm or light valence quark, we estimate the total NRQCD systematic uncertainty in $E_{bbc}^{(\rm sub, phys)}$ or $E_{bbq}^{(\rm sub, phys)}$
to be the quadratic sum of $(1/2)$ times the uncertainty in $-\frac{n_b}{2}\overline{E}_{b\bar{b}}$ (with $n_b=2$) and our above estimate of the NRQCD
uncertainty in $E_{b c c}$ or $E_{b q q^\prime}$, respectively.

\subsection{NRQCD systematic uncertainties: baryon mass splittings}

In the mass splittings between different baryon states with the same valence quark content, given in Table \ref{tab:splittings}, the subtraction term
$- \frac{n_c}{2} \overline{E}_{c\bar{c}}  - \frac{n_b}{2} \overline{E}_{b\bar{b}}$
cancels. Furthermore, in the hyperfine splittings, the leading contributions from the spin-independent operators in the NRQCD action cancel.

For the heavy-light ${b q q^\prime}$ baryons, our estimates of the NRQCD systematic uncertainties in $E_{b q q^\prime}$, as discussed
in the previous subsection, were in fact dominated by spin-dependent effects, and hence we assign the same estimates also for the mass splittings in this sector.

In the $\Omega^*_{b c c}-\Omega_{b c c}$ hyperfine splitting, the effects of the spin-independent order-$v^4$ operators are expected to cancel to a large extent,
and this splitting is expected to primarily originate from the operator $-c_4\:\frac{g}{2 m_b}\:\bss{\sigma}\cdot\mathbf{\widetilde{B}}$. Thus,
the NRQCD systematic uncertainties originate from the one-loop matching of $c_4$, from the missing spin-dependent $v^6$ terms, and from the missing
spin-dependent four-quark operators. We estimate the sizes of these uncertainties to be $\alpha_s^2\: m_b\: v_{b(c)}^4 \approx 1\:\:{\rm MeV}$,
$m_b\: v_{b(c)}^6 \approx 1\:\:{\rm MeV}$, and $\alpha_s^2\:m_b\:v_{b(c)}^3 \approx 3\:\:{\rm MeV}$, respectively, using the same power counting as for $B_c$ mesons.
For the $\Omega^*_{b b c}-\Omega_{b b c}$ hyperfine splitting, we note that (in the limit of large $m_b$) the bottom diquark has spin 1 in both states. Therefore,
it is appropriate to apply the $B_c$ power counting also to this splitting, and we assign the same NRQCD uncertainty.

Similarly, the $\Xi^*_{b b}-\Xi_{b b}$ and $\Omega^*_{b b}-\Omega_{b b}$ mass splittings predominantly arise through the hyperfine interaction of a spin-1 
bottom diquark with the spin of the light or strange valence quark. Therefore, we use heavy-light power counting for these splittings, and assign
the same NRQCD uncertainties as for the mass splittings of singly bottom $b q q^\prime$ baryons.

Finally, for the mass splittings in the mixed $bcq$ sector, where the power counting is less trivial, we add in quadrature our estimates of the NRQCD systematic
uncertainties for the mass splittings in the $bcc$ and $bqq$ sectors.

\subsection{Other systematic uncertainties}

In this subsection we briefly comment on other sources of systematic uncertainties in the baryon masses. First, recall that our
calculation was performed in the isospin limit, setting $m_u=m_d$ and neglecting QED effects. Isospin breaking effects caused by $m_u-m_d$
and by QED are typically of the order of a few MeV \cite{Borsanyi:2014jba}. The electromagnetic contribution to $E^{(\rm sub, phys)}_{\Omega_{bbb}}$
was estimated using a potential model to be $5.1\pm 2.5$ MeV \cite{Meinel:2010pw}; an effect of similar size can be expected for the $\Omega_{ccc}$
(the charge of the charm quark is twice as large, but the average interquark distance is also expected to be larger than in the $\Omega_{bbb}$).
Uncertainties associated with the tuning of the charm and bottom quark masses are expected to be negligible in our results,
because this tuning was performed with high precision, and, more importantly, the leading dependence on the heavy-quark masses cancels in the subtracted
energies $E^{(\rm sub, phys)}$. When computing the spin-averaged quarkonium masses for the subtracted energies, we neglected the annihilation
contributions, which predominantly affect the $\eta_c$ and the $\eta_b$. A perturbative estimate for the resulting mass shift was given
in Eq.~(\ref{eq:annihilation}). This evaluates to about $-3$ MeV for the $\eta_c$ mass and about $-1$ MeV for the $\eta_b$ mass, and these masses
enter only with a factor of $1/4$ in the spin averages. Finally, we note that our chiral and continuum extrapolations already removed the leading finite-volume
effects from the baryon masses. Given that these leading finite-volume effects were at most 2 MeV (see Tables \ref{tab:volumeshiftssinglyheavy}
and \ref{tab:volumeshiftsdoublyheavy}), we expect that higher-order finite-volume effects are negligible.

\FloatBarrier
\section{\label{sec:conclusions}Conclusions}
\FloatBarrier

We have presented a comprehensive lattice QCD calculation of the masses of baryons containing one or more heavy quarks. We have extrapolated all
results to the continuum limit and to the physical light-quark mass (in the isospin limit), and we have carefully estimated the remaining
systematic uncertainties. For the singly charmed and singly bottom baryons that have already been observed in experiments, our results
for the masses agree with the experimental values within the uncertainties, as can bee seen in Fig.~\ref{fig:spectrumall}. In the case of the $\Omega_b$, our calculation
agrees with the CDF \cite{Aaltonen:2009ny} and LHCb \cite{Aaij:2013qja} measurements, but deviates from the D$\slashed{0}$ measurement \cite{Abazov:2008qm}
by 2 standard deviations.

Our results for the heavy-baryon hyperfine splittings (see Table \ref{tab:splittings}) have smaller uncertainties than our results for the baryon masses themselves.
Combining our lattice QCD determinations of the $\Xi_b^*-\Xi_b^\prime$ and $\Omega_b^*-\Omega_b$ splittings with the experimental
values of the $\Xi_b^{*0}$ \cite{Chatrchyan:2012ni, Beringer:1900zz} and $\Omega_b$ \cite{Aaltonen:2009ny, Aaij:2013qja} masses, we obtain more precise predictions
for the masses of the as yet undiscovered $\Xi_b^{\prime 0}$ and $\Omega_b^*$:
\begin{eqnarray}
 m_{\Xi_b^{\prime 0}} &=& 5918.5(3.2)(8.6)(2.3)\:\:{\rm MeV}, \\
 m_{\Omega_b^*} &=& 6075.2(2.2)(7.7)(2.1)\:\:{\rm MeV}.
\end{eqnarray}
Here, the first two uncertainties are from our lattice QCD calculation (statistical and systematic), and the third uncertainty is experimental.
We assumed that the baryon discovered by the CMS Collaboration \cite{Chatrchyan:2012ni} is indeed the $\Xi_b^{*0}$ (and not the $\Xi_b^{\prime 0}$).
For the $\Omega_b$ mass, we used the average of the CDF \cite{Aaltonen:2009ny} and LHCb \cite{Aaij:2013qja} measurements.

\begin{figure}
 \includegraphics[width=\linewidth]{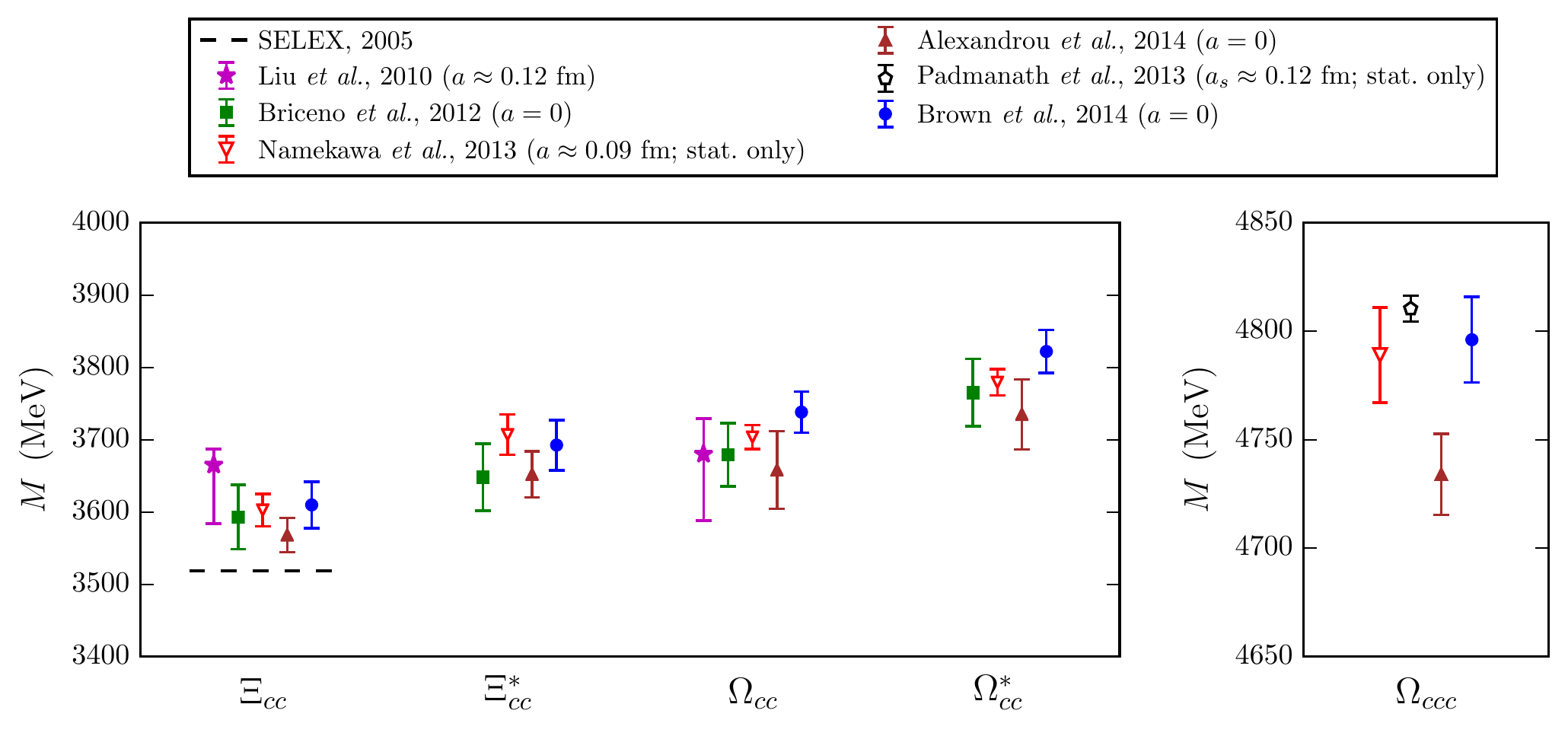}
 \caption{\label{fig:charmedcomp}Comparison of lattice QCD results for the doubly and triply charmed baryon masses
 \cite{Liu:2009jc, Briceno:2012wt, Namekawa:2013vu, Padmanath:2013zfa, Alexandrou:2014sha}; our results are labeled as ``Brown \textit{et al.}, 2014''.
 Only calculations with dynamical light quarks are included; for the doubly charmed baryons, we further required that the calculations were performed at or
 extrapolated to the physical pion mass.
 Results without estimates of systematic uncertainties are labeled ``stat.~only''. The lattice spacing values used in the calculations are also given;
 $a=0$ indicates that the results have been extrapolated to the continuum limit.
 Reference \cite{Padmanath:2013zfa} (Padmanath \textit{et al.}, 2013) gives results for
 $m_{\Omega_{ccc}}-\frac32 m_{J/\psi}$ for two different values of the Sheikholeslami-Wohlert coefficient; here we took the result with $c_{\rm SW}=1.35$ and added
 the experimental value of $\frac32 m_{J/\psi}$ \cite{Beringer:1900zz}.
 In the plot of the doubly charmed baryons, the unconfirmed experimental result for the $\Xi_{cc}^+$ mass from SELEX \cite{Mattson:2002vu, Ocherashvili:2004hi}
 is shown with a dashed line.
 Note that the lattice QCD calculations consistently predict a $\Xi_{cc}$ mass higher than the SELEX result.}
\end{figure}

A comparison of our results for the doubly and triply charmed baryons with other unquenched lattice calculations is shown in Fig.~\ref{fig:charmedcomp}.
Of particular interest is the lightest doubly-charmed baryon, the $\Xi_{cc}$. The SELEX collaboration reported signals interpreted as the $\Xi_{cc}^+$ at $3518.7(1.7)$ MeV
\cite{Mattson:2002vu, Ocherashvili:2004hi}, but subsequent searches by FOCUS \cite{Ratti:2003ez}, BaBar \cite{Aubert:2006qw},
Belle \cite{Chistov:2006zj}, and LHCb \cite{Aaij:2013voa} did not confirm the existence of this structure. As can be seen in Fig.~\ref{fig:charmedcomp}, all
recent lattice QCD determinations of the $\Xi_{cc}$ mass in the isospin limit are consistent with each other and give masses around 100 MeV higher than
the SELEX result; our own calculation deviates from the SELEX measurement by $91\pm32$ MeV, corresponding
to 2.8 standard deviations. Note that the isospin splitting $m_{\Xi_{cc}^{++}}-m_{\Xi_{cc}^+}$ was recently computed in lattice QCD+QED to be
$2.16(11)(17)$ MeV \cite{Borsanyi:2014jba}. 

Regarding the $\Omega_{ccc}$ mass, we note that our result is higher than the recent
result from Alexandrou \textit{et al.} \cite{Alexandrou:2014sha} by 2.3 standard deviations, but agrees with earlier calculations \cite{Namekawa:2013vu, Padmanath:2013zfa}
(note, however, that the earlier calculations of the $\Omega_{ccc}$ mass did not include continuum extrapolations, and lack estimates of the systematic uncertainties).
While our lattice calculation was based on the mass difference $M_{\Omega_{ccc}}-\frac32\overline{M}_{c\bar{c}}$, Ref.~\cite{Alexandrou:2014sha} calculated
$M_{\Omega_{ccc}}$ directly and may therefore be more susceptible to a slight mistuning of the charm-quark mass.

\begin{figure}
 \includegraphics[width=0.56\linewidth]{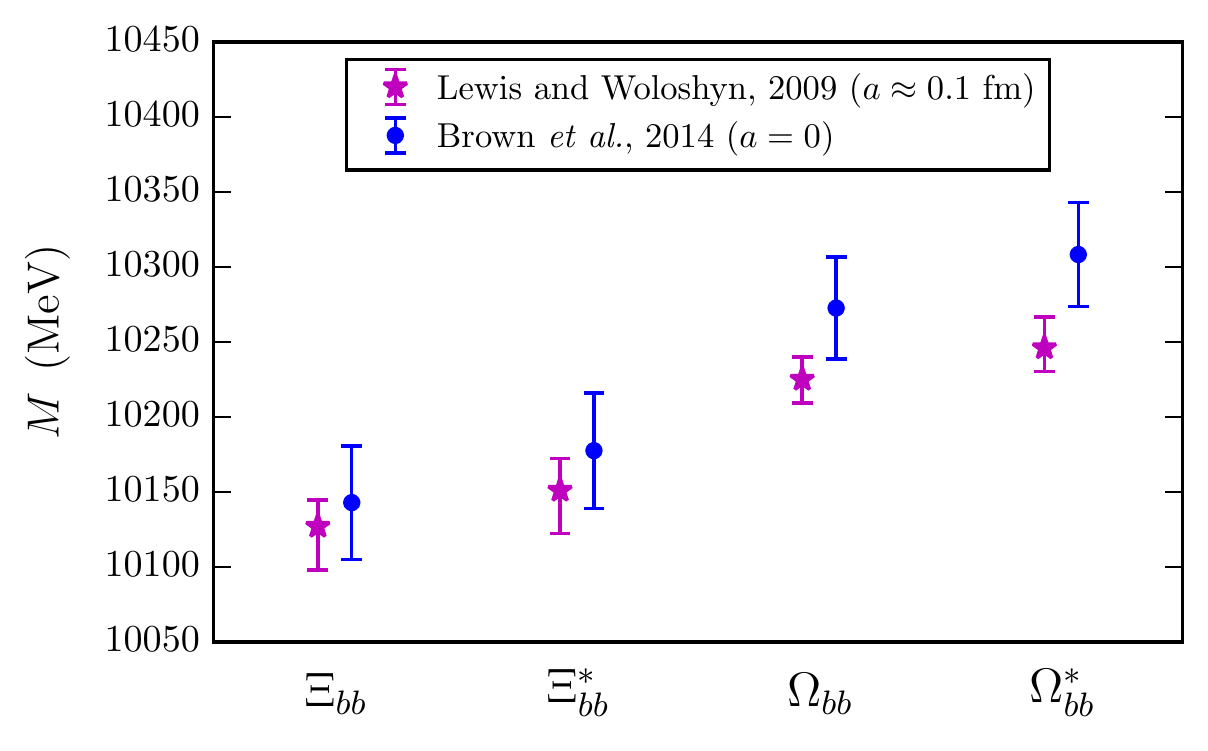}
 \caption{\label{fig:doublybottomcomp}Comparison of lattice QCD results for the doubly bottom baryon masses. The only other published unquenched calculation
 is the one of Ref.~\cite{Lewis:2010xj}. Our results have larger statistical uncertainties, but our calculation
 was performed with closer-to-physical pion masses and included a combined chiral and continuum extrapolation.}
\end{figure}

For the doubly bottom baryons, we compare our results to those from Ref.~\cite{Lewis:2010xj} in Fig.~\ref{fig:doublybottomcomp}. Our results
are consistent with Ref.~\cite{Lewis:2010xj} but have larger statistical uncertainties. This may be because we performed
our numerical calculations with lighter (closer to physical) up and down-quark masses were the two-point correlation functions are exponentially noisier \cite{Lepage:1991ui},
and because our continuum extrapolations amplified the statistical uncertainties. For the triply bottom $\Omega_{bbb}$ baryon, our present result is not
completely independent from the result obtained by one of us in earlier work \cite{Meinel:2010pw}, and we refer the reader to Ref.~\cite{Meinel:2010pw} for further
discussions.

It is interesting to compare our lattice QCD results for the hyperfine splittings of the doubly heavy baryons to the hyperfine splittings
of the corresponding heavy-light mesons. This comparison is shown in Table \ref{tab:doublyheavyHFS}, where we use the experimental results of the
heavy-light meson hyperfine splittings (preliminary lattice results for the heavy-light meson hyperfine splittings from the same data sets as used for
the baryons are consistent with the experimental results). Heavy quark-diquark symmetry \cite{Savage:1990di} predicts that the ratio of
these hyperfine splittings approaches the value $3/4$ in the heavy-quark limit \cite{Brambilla:2005yk}. We do indeed see some evidence
that the ratios in the bottom sector are closer to this value than the ratios in the charm sector.

\begin{table}
\begin{tabular}{ccccccccc}
\hline\hline
  Splitting                     & & This work (MeV) & & Splitting             & & Experiment (MeV)  & & Ratio           \\
\hline
 $\Xi_{cc}^*-\Xi_{cc}$          & & $82.8(9.2)$     & &  $D^{*0} - D^0$       & &  $142.12(7)$      & & $0.58(6)\nb$    \\
 $\Omega_{cc}^*-\Omega_{cc}$    & & $83.8(5.5)$     & &  $\:\, D_s^* - D_s$   & &  $143.8(4)\nb$    & & $0.58(4)\nb$    \\
 $\Xi_{bb}^*-\Xi_{bb}$          & & $34.6(7.8)$     & &  $B^* - B$            & &  $\:45.78(35)$    & & $0.76(17)$      \\
 $\Omega_{bb}^*-\Omega_{bb}$    & & $35.7(5.7)$     & &  $\:\, B_s^* - B_s$   & &  $48.7(2.3)$      & & $0.73(12)$      \\
\hline\hline
\end{tabular}
\caption{\label{tab:doublyheavyHFS} Hyperfine splittings of doubly heavy baryons calculated in this work, compared to experimental results \cite{Beringer:1900zz} for the hyperfine
splittings of mesons related by heavy quark-diquark symmetry. The ratio of these hyperfine splittings is expected to approach the value $3/4$ in the heavy-quark limit \cite{Brambilla:2005yk}.}
\end{table}

\begin{figure}
 \includegraphics[width=\linewidth]{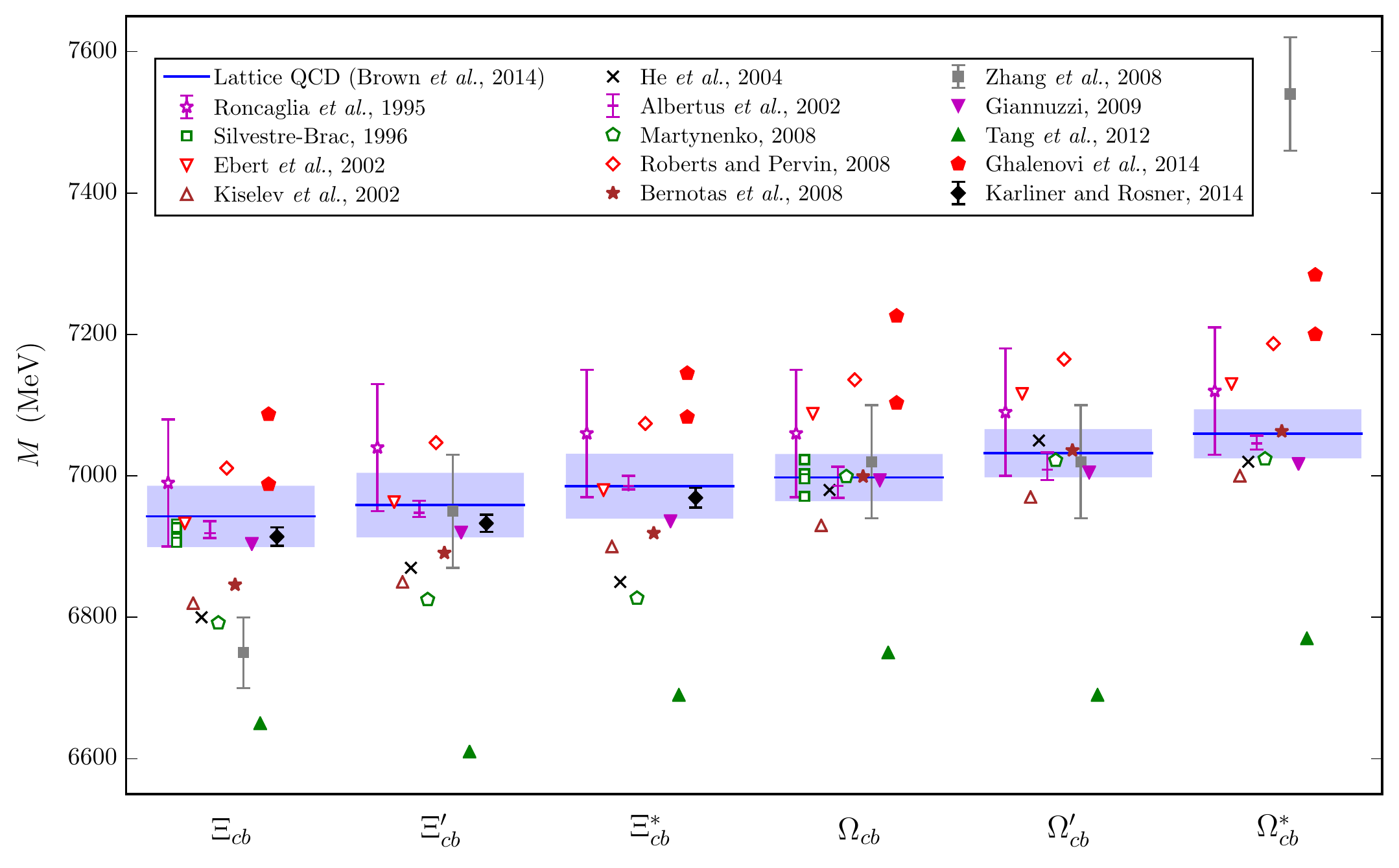}
 \caption{\label{fig:mixeddoublyheavycomp}Comparison of our lattice QCD results for the
 $\Xi_{cb}$, $\Xi_{cb}^\prime$, $\Xi_{cb}^*$, $\Omega_{cb}$, $\Omega_{cb}^\prime$, and $\Omega_{cb}^*$ baryon masses with estimates
 based on continuum methods, including quark models and QCD sum rules
 \cite{Roncaglia:1995az, SilvestreBrac:1996bg, Ebert:2002ig, Kiselev:2002iy, He:2004px, Albertus:2006ya, Martynenko:2007je, Roberts:2007ni,
 Bernotas:2008fv, Zhang:2008rt, Giannuzzi:2009gh, Tang:2011fv, Ghalenovi:2014swa, Karliner:2014gca}.
 From Refs.~\cite{SilvestreBrac:1996bg} (Silvestre-Brac, 1996) and \cite{Ghalenovi:2014swa} (Ghalenovi \textit{et al.}, 2014), we show results for multiple different choices of the
 interquark potentials. Note that the bag-model calculation of Ref.~\cite{He:2004px} (He \textit{et al.}, 2004) predicts $m_{\Xi_{cb}^*} < m_{\Xi_{cb}^\prime}$ and
 $m_{\Omega_{cb}^*} < m_{\Omega_{cb}^\prime}$, and the QCD sum rule calculation of Ref.~\cite{Tang:2011fv} (Tang \textit{et al.}, 2012)
 predicts $m_{\Xi_{cb}^\prime} < m_{\Xi_{cb}}$ and $m_{\Omega_{cb}^\prime} < m_{\Omega_{cb}}$, both rather unusual. The sum-rule calculation
 of Ref.~\cite{Zhang:2008rt} (Zhang \textit{et al.}, 2008) gives extremely large hyperfine splittings
 $m_{\Xi_{cb}^*}-m_{\Xi_{cb}^\prime} \approx 1\:{\rm GeV}$ and $m_{\Omega_{cb}^*}-m_{\Omega_{cb}^\prime}\approx 0.5\:{\rm GeV}$
 [our results for the hyperfine splittings are $m_{\Xi_{cb}^*}-m_{\Xi_{cb}^\prime}=26.7(3.3)(8.4)\:{\rm MeV}$,
 $m_{\Omega_{cb}^*}-m_{\Omega_{cb}^\prime}=27.4(2.0)(6.7)\:{\rm MeV}$];  the $\Xi_{cb}^*$ mass from Ref.~\cite{Zhang:2008rt} is beyond the upper limit of the plot.}
\end{figure}

\begin{figure}

 \includegraphics[width=\linewidth]{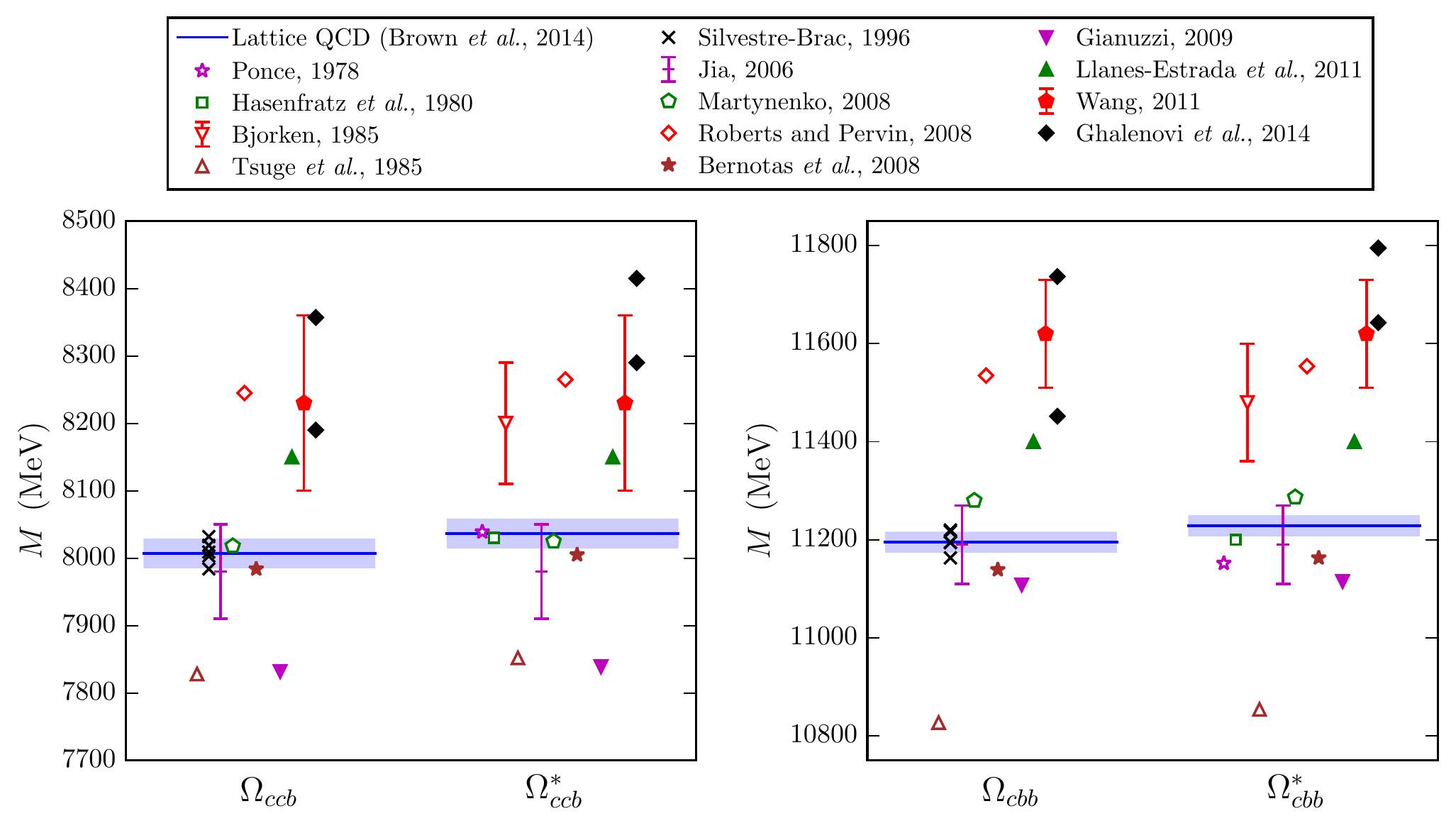}
 \caption{\label{fig:triplyheavymixedcomp}Comparison of our lattice QCD results for the masses of triply heavy charm-bottom baryons with
 estimates based on continuum methods, including quark models, QCD sum rules, and perturbative QCD
 \cite{Ponce:1978gk, Hasenfratz:1980ka, Bjorken:1985ei, Tsuge:1985ei, SilvestreBrac:1996bg, Jia:2006gw,  Martynenko:2007je, Roberts:2007ni,
       Bernotas:2008bu, Giannuzzi:2009gh, LlanesEstrada:2011kc, Wang:2011ae, Ghalenovi:2014swa}.
       From Refs.~\cite{SilvestreBrac:1996bg} (Silvestre-Brac, 1996) and \cite{Ghalenovi:2014swa} (Ghalenovi \textit{et al.}, 2014),
       we show results for multiple different choices of the  interquark potentials. 
       From Ref.~\cite{LlanesEstrada:2011kc}, which used the static three-quark potential from perturbative QCD, we show the NNLO results from Table 16.
       Not shown in this plot are the results of the QCD sum-rule calculation of Ref.~\cite{Zhang:2009re}, which are far lower than all other results.}
\end{figure}

No other dynamical lattice QCD calculations have been published so far for mixed charm-bottom baryons 
(results of a quenched lattice calculation can be found in Ref.~\cite{Mathur:2002ce}). We therefore compare our
lattice QCD results for the masses of these baryons to predictions from potential models, QCD sum rules, and other continuum-based
approaches. These comparisons are shown in Fig.~\ref{fig:mixeddoublyheavycomp} for the $\Xi_{cb}$, $\Xi_{cb}^\prime$, $\Xi_{cb}^*$, $\Omega_{cb}$, $\Omega_{cb}^\prime$, and $\Omega_{cb}^*$,
and in Fig.~\ref{fig:triplyheavymixedcomp} for the $\Omega_{ccb}$, $\Omega_{ccb}^*$, $\Omega_{cbb}$, and $\Omega_{cbb}^*$. It is evident
that the mass predictions in the literature cover ranges far wider than our uncertainties. We hope that our lattice QCD results provide a useful benchmark for
future studies of these interesting systems.

\appendix

\FloatBarrier
\section{\label{sec:F}The chiral function $\mathcal{F}$}
\FloatBarrier

The chiral function $\mathcal{F}\left(m,\delta,\mu\right)$ results from the evaluation of a one-loop self-energy diagram,
where the internal baryon has a mass difference $\delta$ from the external baryon, in heavy-hadron chiral perturbation theory.
We write it as the sum of the infinite-volume part and a finite-volume correction,
\begin{equation}
\mathcal{F}\left(m,\delta,\mu\right) = \mathcal{F}^{(\rm IV)}\left(m,\delta,\mu\right) + \mathcal{F}^{(\rm FV)}\left(m,\delta\right).
\end{equation}
The infinite-volume part is given by
\begin{equation}
\label{eqn:ChiralF}
\mathcal{F}^{(\rm IV)}\left(m,\delta,\mu\right) = \left(m^2 - \delta^2\right) m\:R\left(\frac{\delta}{m}\right)
- \left( \frac32 m^2 - \delta^2\right) \delta\, \log\left(\frac{m^2}{\mu^2}\right) -\delta^3\,\log\left(\frac{4\,\delta^2}{\mu^2}\right),
\end{equation}
where
\begin{equation}
 R(x) = \sqrt{x^{2}-1}\:\left [ \log\left (  x - \sqrt{x^{2}-1+i\epsilon} \right )- \log\left (  x + \sqrt{x^{2}-1+i\epsilon} \right ) \right ] \nonumber.
\end{equation}
Here, we use a renormalization scheme in which the real part of $\mathcal{F}^{(\rm IV)}$ vanishes in the chiral limit \cite{WalkerLoud:2006sa}. Our definition of
$\mathcal{F}^{(\rm IV)}$ differs from the one used in Ref.~\cite{Tiburzi:2004kd} by the term $-\delta^3\,\log\left(\frac{4\,\delta^2}{\mu^2}\right)$.

An approximate expression for the finite-volume correction is given by \cite{Arndt:2004bg, Detmold:2005pt}
\begin{equation}
\mathcal{F}^{(\rm FV)}\left(m,\delta\right) = -m^2 \pi \sum_{\vec{u} \neq \vec{0}} \frac{e^{-umL}}{uL} \mathcal{A}\, ,
\end{equation}
where $\vec{u}=(u_1,u_2,u_3)$, $u_i\in {\mathbb{Z}}$, $u \equiv|\vec{u}|$, and 
\begin{eqnarray}
\label{eq:calA_def}
 {\mathcal{A}} &=& e^{(z^{2})} \big [ 
1 - {\mathrm{Erf}}(z)\big ]
+ \frac{1}{u m L} \bigg [
 \frac{1}{\sqrt{\pi}} \left ( \frac{9z}{4} - 
\frac{z^{3}}{2}\right )
 + \left(\frac{z^{4}}{2}-2\,z^2\right)e^{(z^{2})} 
 \big [ 1 - {\mathrm{Erf}}(z)\big ]
\bigg ]\\
&&
-\frac{1}{(u m L)^2} \bigg [
\frac{1}{\sqrt{\pi}}\left ( -\frac{39z}{64} + 
\frac{11z^{3}}{32}
  -\frac{9z^{5}}{16} + \frac{z^{7}}{8} \right )
-\left ( -\frac{z^{6}}{2} + \frac{z^{8}}{8}\right )
e^{(z^{2})} \big [ 1 - {\mathrm{Erf}}(z)\big ]
\bigg ]
+\mathcal{O}\left ( \frac{1}{(u m L)^{3}}\right ) ,
\nonumber
\end{eqnarray}
with
\begin{equation}
 z =  \frac{\delta}{m} 
  \sqrt{\frac{u m L}{2}} .
\end{equation}

\begin{acknowledgments}
We thank the RBC and UKQCD collaborations for making their gauge field configurations publicly available.
SM was supported by the U.S.~Department of Energy under cooperative research agreement Contract Number DE-FG02-94ER40818.
KO and ZSB were supported by the U.S.~Department of Energy through Grant Number DE-FG02-04ER41302 and
through Grant Number DE-AC05-06OR23177 under which JSA operates the Thomas Jefferson National Accelerator Facility.
ZSB also acknowledges support by the JSA Jefferson Lab Graduate Fellowship Program.
WD was supported by the U.S.~Department of Energy Early Career Research Award D{E}-S{C001}0495 and the Solomon Buchsbaum Fund
at MIT. This work made use of high-performance computing resources provided by XSEDE (supported by National
Science Foundation Grant Number OCI-1053575) and NERSC (supported by U.S.~Department of Energy Grant Number DE-AC02-05CH11231). 
\end{acknowledgments}

\FloatBarrier

\end{document}